\documentclass[twocolumn,aps,prb,notitlepage,10pt]{revtex4-1}
\RequirePackage{snapshot}
\usepackage{graphicx,color,amsmath,amssymb}
\usepackage{soul}
\usepackage{multirow}
\usepackage{dcolumn}
\usepackage{array}
\usepackage[normalem]{ulem}

\newcommand*\RR{{\mathcal R}}

\begin{document}

\title{Machine learning a general purpose interatomic potential for silicon}
\author{Albert P. Bart\'ok}
\affiliation{Scientific Computing Department, Science and Techology Facilities Council, Rutherford Appleton Laboratory, Didcot, OX11 0QX, United Kingdom}
\author{James Kermode}
\affiliation{Warwick Centre for Predictive Modelling, School of Engineering, University of Warwick, Coventry, CV4 7AL, United Kingdom}
\author{Noam Bernstein}
\affiliation{Center for Materials Physics and Technology, U.~S. Naval Research Laboratory, Washington, DC, 20375}
\author{G\'abor Cs\'anyi}
\affiliation{Engineering Laboratory, University of Cambridge, Trumpington Street, Cambridge, CB2 1PZ, United Kingdom}
\date{\today} 

\begin{abstract}
The success of first principles electronic structure calculation for
predictive modeling in chemistry, solid state physics, and materials
science is constrained by the limitations on simulated length and time
scales due to computational cost and its scaling.  Techniques based on
machine learning ideas for interpolating the Born-Oppenheimer potential
energy surface without explicitly describing electrons have recently
shown great promise, but accurately and efficiently fitting the
physically relevant space of configurations has remained a challenging
goal.  Here we present a Gaussian Approximation Potential for silicon
that achieves this milestone, accurately reproducing density functional
theory reference results for a wide range of observable properties,
including crystal, liquid, and amorphous bulk phases, as well as point,
line, and plane defects.  We demonstrate that this new potential enables
calculations that would be very expensive with a first principles
electronic structure method, such as finite temperature phase boundary
lines, self-diffusivity in the liquid, formation of the amorphous by
slow quench, and dynamic brittle fracture.  We show that the uncertainty
quantification inherent to the Gaussian process regression framework
gives a qualitative estimate of the potential's accuracy for a given
atomic configuration.  The success of this model shows that it is indeed
possible to create a useful machine-learning-based interatomic potential
that comprehensively describes a material, and serves as a template for
the development of such models in the future.
\end{abstract}

\maketitle
\section{Introduction}
\label{sec:intro}

\subsection{Background}

First principles molecular simulation, based on various approximations
of electronic structure theory, is the workhorse of materials
modelling. For example, density functional theory (DFT), a prominent
method, is indicated as the topic of about $19000$ papers published
in 2017 according to the Web of Science database. However, due to the
combination of its computational expense and unfavourable scaling,
simulations that require thousands of atoms and/or millions of energy
or force evaluations are carried out not using electronic structure
methods, but {\em empirical analytical potentials} (also known as {\em force fields} in the chemistry literature). These are parametrised approximations of the Born-Oppenheimer
potential energy surface (PES), the electronic ground state energy viewed as
a function of the nuclear positions\cite{Finnis:2004da}.

The functional form of analytical potentials is typically simple, partly
based on a combination of physical and chemical intuition, and partly
on convenience. The parameters are usually optimised  so that the model
reproduces, as best as it is able, either some macroscopic observables, or
microscopic quantities such as total energies and/or forces, corresponding
to selected configurations and calculated separately using an electronic
structure method. Unsurprisingly,  while it is easy to find parameter
sets that reproduce individual observables (e.g. the melting point
corresponding to a particular composition, or the binding energy of a
crystalline structure), the simple functional forms postulated are not
flexible enough to allow matching many properties simultaneously, and the potential energy surface is not accurate.
This suggests
that when empirical analytical potentials are successful,
there is a risk that this is due to a fortuitous cancellation of errors, and is then a case of
getting the right answer for the wrong reasons. The practitioner is forced
to select parameters subject to a tradeoff between maximizing the accuracy for a
few selected properties and transferability, i.e.\ avoiding large or even qualitative error for a wide range
of configurations and observables of interest\cite{tersoff_phys_rev_b_1988b,Brenner:2000uh}. This severely limits the
predictive power of empirical analytical potentials when the very parameters that
give sufficient accuracy for some known observables result in wildly varying predictions for new phases and properties
whose prediction is the ultimate goal of the simulation.

Machine learning (ML) methods have provided a systematic approach to fitting
functions in high dimensional spaces without employing simply parametrised functional forms,\cite{Bishop:2016ui} thus opening up the possibility
of creating interatomic potentials for materials with unprecedented
accuracy for a wide range of configurations.   The development in the last 10 years required
exploring a variety of ways to describe the chemical environment of atoms,
the basis functions used to construct the potential, e.g. by various kernels or artificial neural network models, and the way such fits
can be regularised, either by linear algebra or by various protocols of
neural network optimisation\cite{Behler:2016bp,rbpmk2017q}. 
The general approach is to define an atomic energy as a function of its
local environment, and fit this function in a 60-100 dimensional space corresponding
to the relative positions of 20-30 nearby atoms.  The challenge we take on
here is to use this approach to develop a {\em general purpose} interatomic
potential, neither restricted to a narrow range of configurations or
observables, nor compromising accuracy in reproducing the reference
potential energy surface.  This requires adequately sampling enough
of the space that is relevant to a wide range of atomistic simulations to interpolate it accurately, and
doing so in a computationally tractable manner.

The Achilles heel of machine learning models,
directly related to their flexibility, is their naturally much-reduced transferability: the flexible
function representation is informed by a large training database, leading to a good fit for configurations
nearby the database (in the space of the chosen representation), and progressively poorer away from it.  This is often summarised by
saying that high dimensional fits are good at {\em interpolation}, but less good at {\em extrapolation}, and
this can be viewed as another manifestation of the ``curse of dimensionality''.

When first encountering the ML potential approach, one might wonder
why such high dimensional fits work at all, given that it is impractical
to thoroughly sample a 60-100 dimensional space (e.g. on a grid)? It is an empirical observation
that they often do, so the real question is what are the special
properties of potential energy surfaces that make them amenable to such
approximations? {\em Regularity} is almost certainly one of these, the
mathematical concept encompassing the colloquial idea of a potential
varying smoothly as a function of atomic position. Indeed the
regular kernels that are used (and the corresponding regular activation
functions in artificial neural networks) define the length scales over
which predictions are interpolated, rather than extrapolated. This can also
point towards explaining why some methods work better than others:
kernels that better capture the inherent regularity of the underlying
function will interpolate better and extrapolate farther. Another
property is that the configurations that are likely
to arise in an atomistic simulation actually occupy a volume of configuration space
that is much smaller than the full space.  Consequently, a database derived from configurations
found in reference atomistic simulations is sufficient for fitting an
interatomic potential, so long as it includes not only the low energy
configurations, but also nearby high energy ones to constrain the
potential at the {\em boundary} of the region that will be explored when
it is used.

Thus, the  tradeoff made by empirical analytical potentials (viz. between accuracy
and transferability), is now replaced by another tradeoff: that between transferability and
database size, because high accuracy is possible, but only near the training set.
In order to achieve the promised wider impact in materials modelling,
it would be desirable to explore this tradeoff. In particular, is it possible to create a training
database of manageable size that covers almost all relevant configurations of a material, and thus a potential for future larger length scale simulations? Or
will such models always be confined to a narrow set of atomic configurations,
with every new scientific question necessitating a new fit trained on a
problem-specific database of first principles calculations? Since the notion of nearness is
intimately tied to the representation, in this paper we will explore this question for a particular case, a kernel-based fit using the
previously introduced Smooth Overlap of Atomic Positions (SOAP) kernel~\cite{Bartok:2013cs,Bartok:2013ks} and the
Gaussian Approximation Potential (GAP) framework\cite{Bartok:2015iw,Bartok:2016bq}. (Everywhere in this paper when we refer to GAP models, we mean a Gaussian process regression model using the SOAP kernel, although of course other kernels and also combinations of different kernels can be used within the GAP framework and have indeed been used for other systems.\cite{Bartok:2013gf,Gillan:2013hq,Deringer:2016uf,Rowe:2018ct})

An obvious alternative approach to the
transferability problem is to give up on it entirely, and accept that an interatomic potential will
always be extremely narrowly confined to its training database. One can then develop algorithms that
actively {\em adapt/grow} the training database during the course of a simulation.\cite{Li:2015eb,Glielmo:2017dj,Podryabinkin:2017jpa,
gps2018q,
sy2018q,
jjh2018q,
mo2017q,doyt2017q} The obvious
disadvantage is that an electronic structure method always has to be part of the simulation,
to be called upon to calculate new target data as and when necessary. The efficacy of this approach
then depends on what algorithm is used to detect that the simulation has strayed into parts
of configuration space not sufficiently well covered by the current database, and
precisely what subsequent action is taken.

\subsection{State of the art}
\label{sec:stateoftheart}

The following is a brief review of the recent works in the emergent
field of interatomic potential construction using high dimensional
non-parametric fits. Although fits to the potential energy
surface of molecules and small molecule clusters have a much longer
history\cite{Blank:1995dy,Brown:2003jo,Lorenz:2006eb,Braams:2009eb,Hawe:2010cb,Mills:2011iw,Babin:2013fs,Babin:2014bn,Medders:2014bp,Manzhos:2014jb,Hughes:2015bp,lm2017q,um2017q,um2017q,pktmttd2017q},
here we limit our scope to only include efforts that model strongly
bound materials in the condensed phase. On the one hand, material
models generally have a number of critical requirements that differentiate them
from molecular models: (i) the potential must be reactive, i.e. need
to describe the forming and breaking of many covalent bonds, often
simultaneously; (ii) a wide range of neighbour configurations need to
be covered, including radical changes in neighbour count. Comprehensive
models also need to (iii) cover multiple phases, e.g. metallic and
insulating, solid and liquid, etc.  On the other hand, many works cited
below (and the present work) consider only one type of element, which allows the consideration of only relatively short range
interactions, because the absence of charge transfer obviates the need
to describe long range electrostatic effects.

Modelling the short range interactions with artificial neural networks
(NN) really took off about a decade ago\cite{Behler:2017ku}, starting with the bulk
phases of silicon\cite{Behler:2007fe,Behler:2008ft},
with many more to follow: describing some silicon defects\cite{Sanville:2008jt},
the graphite-diamond transition\cite{Khaliullin:2010el}, bulk zinc
oxide\cite{Artrith:2011em}, copper with some defects\cite{Artrith:2012fw},
the phase change material GeTe in its various phases\cite{Sosso:2012dv},
various ionic solids\cite{Ghasemi:2015ch,fgrrsga2017q}, Li-Si alloys\cite{auc2018q,Onat:2018cp,cmowk2017q}, bulk
TiO$\mbox{}_2$\cite{Artrith:2016gc}, alloys\cite{Kobayashi:2017cy,hsk2017q},  Ta$_2$O$_5$~\cite{law2017}, Li$_3$PO$_4$~\cite{lamw2017q}, gold clusters\cite{jcb2017q}, graphene\cite{rmlht2018q} and various surfaces\cite{sbmk2017q,qhb2017q,nb2016bq,bk2017q,klzjg2017q}.
Fitting NN potentials is beginning to be combined with combinatorial structure
search\cite{Dolgirev:2016fz,Deringer:2017ck,Deringer:2018eh}.

Kernel fitting is a different approach to high dimensional interpolation,
with origins in statistics (c.f. kriging\cite{Stein:2012tt} and Gaussian
process regression (GPR)\cite{Rasmussen:2006vz}) and widely applied in numerical
analysis and machine learning\cite{Vapnik:1998uq}. The key to its success is
the choice of kernel, and through it the basis functions employed. In the
context of atomistic potentials, a significant step was the introduction
of rotationally and permutationally invariant descriptors that also
varied smoothly with coordination number, based on the spherical Fourier
transform and the bispectrum constructed from it\cite{Bartok:2010fj}. This
was later simplified to the SOAP
descriptor and kernel\cite{Bartok:2013cs,Bartok:2013ks} and applied
to tungsten\cite{Szlachta2014}, amorphous carbon\cite{Deringer:2016uf,cdklc2018},
iron\cite{Dragoni:2018je}, graphene\cite{Rowe:2018ct} and boron\cite{dpc2018q}. Retaining the
original spherical bispectrum as a descriptor was used to make a potential for tantalum with linear
\cite{Thompson:2015dw} and quadratic\cite{Wood:2018da} regression,
molybdenum\cite{Chen:2017ih}, and with a nonlinear kernel for bulk
LiBH\cite{Miwa:2017kj}. Linear regression using yet another class of
basis functions was introduced by Shapeev\cite{Shapeev:2016kn} and
used to make a potential for Li\cite{Podryabinkin:2017jp}. Others
used GPR with different descriptors to fit forces directly
without constructing a potential, starting with a test for
silicon\cite{Li:2015eb}, and more comprehensive potentials for
aluminium\cite{Botu:2015kb,Botu:2015kc,Botu:2016ih,Kruglov:2017ju}.

Machine learning methods and novel molecular descriptors have
also been used for other regression tasks for molecules, using a variety
of approaches to predict e.g. atomization energies, atomic charges, NMR shifts
etc.\cite{Rupp:2007dy, Rupp:2012kx, Montavon:2013cn,Hansen:2013dp,Ramakrishnan:2014ij, Hansen:2015jb, Bereau:2015jv,Dral:2015hd,Ramakrishnan:2015if,Rupp:2015cr,Faber:2016jc,Huang:2016gh,Faber:2017cs,Schutt:2017kd,
bsr2018q,
gwcc2017q,
msvp2018q,
fhb2017q
}, constructing molecular force fields\cite{Smith:2017jd,Smith:2017jd,Zhang:2018kz,
ssktm2018q,
fchl2018q,
lsb2017q,
bdtl2017q,
um2018,
ffmrb2018q
} and even in combination with QM/MM.\cite{wsw2017q}

\subsection{A general potential}

Here we demonstrate that, using ML techniques, it is indeed possible to develop an accurate
potential that spans a wide range of physically important structures
and properties.  Using silicon as an example, we create a potential and demand that it give reasonably
accurate predictions for all configurations relevant to  scientific
questions within a wide temperature and pressure range, including
surfaces, point and line defects, cracks, etc. Silicon is a good
material for such a study for a number of reasons. Firstly, it has a
rich phase diagram with many stable and metastable crystal structures,
as well as a wide variety of point and line defects, and surface
reconstructions. Secondly, the simulation community has extensive experience
in understanding many aspects of its
potential energy surface. Finally, there are a large number of empirical
analytical potentials that have been constructed over the past decades, whose
successes and failures give a detailed picture of what is it about the
potential energy surface that is relatively easy to get right, and what
are the more difficult aspects.

\begin{figure*}
    \centerline{\includegraphics[width=15cm]{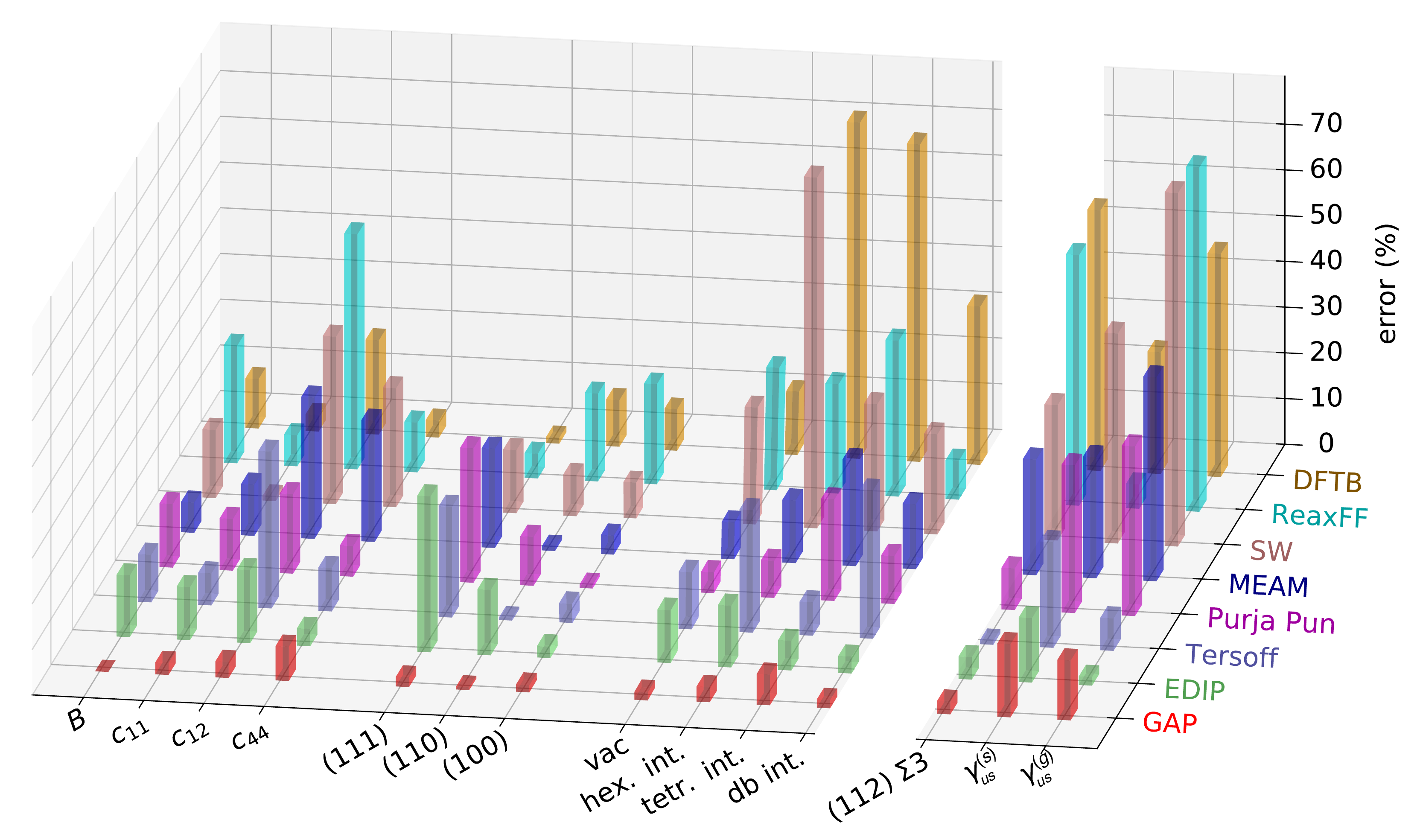}}
    \begin{tabular}{l|rrrr|rrr|rrrr|rrr}
Model & \multicolumn{4}{c|}{Elastic props. / GPa} & \multicolumn{3}{c|}{Surfaces / J/m$^2$} & \multicolumn{4}{c|}{Point defects / eV} & \multicolumn{3}{c}{Planar defects / J/m$^2$}\\ & $B$
 & $c_{11}$
 & $c_{12}$
 & $c_{44}$
 & $(111)$
 & $(110)$
 & $(100)$
 & $\mathrm{vac}$
 & $\mathrm{hex.\ int.}$
 & $\mathrm{tetr.\ int.}$
 & $\mathrm{db\ int.}$
 & $(112)\;\Sigma3$
 & $\gamma^{(s)}_\mathrm{us}$
 & $\gamma^{(g)}_\mathrm{us}$ \\
\hline
DFT reference &  88.6 & 153.3 &  56.3 &  72.2 &  1.57 &  1.52 &  2.17 &  3.67 &  3.72 &  3.91 &  3.66 &  0.93 &  1.61 &  1.74 \\ \multicolumn{1}{c}{} \\
\multicolumn{15}{c}{Relative error [\%]} \\ \hline
GAP &     0 &    -3 &     4 &    -8 &    -2 &    -1 &    -2 &    -2 &    -3 &    -7 &    -2 &     3 &   -16 &    13 \\
EDIP &    14 &    12 &    16 &    -4 &   -34 &   -14 &    -3 &   -12 &    14 &     6 &    -4 &     5 &   -14 &    -2 \\
Tersoff &    10 &    -7 &    34 &   -10 &   -24 &    -0 &     4 &    13 &    27 &    -7 &    32 &    -1 &   -23 &    10 \\
Purja Pun &    14 &    11 &    17 &     7 &   -29 &   -11 &     1 &     5 &     8 &   -22 &   -10 &     9 &   -32 &    37 \\
MEAM &     7 &   -11 &    31 &   -26 &   -22 &    -1 &     4 &    -8 &   -14 &   -23 &   -14 &    25 &   -26 &    45 \\
SW &    14 &    -1 &    36 &   -26 &   -14 &     9 &     8 &   -27 &    77 &    28 &    22 &    30 &   -46 &    77 \\
ReaxFF &    26 &     7 &    51 &   -11 &    -5 &    19 &   -23 &    28 &    24 &    34 &     8 &    55 &     5 &    75 \\
DFTB &    11 &     4 &    21 &    -4 &     1 &    10 &    10 &    15 &    74 &    69 &    35 &    57 &    27 &    49 \\
\end{tabular}
    \caption{
    Comparison of percentage errors made by a range of interatomic potentials for selected properties, with respect to our DFT reverence.
    Those on the left of the break in the axis are interpolative, i.e. well represented within training set of the GAP model:
     elastic constants (bulk modulus $B$, stiffness tensor components $c_{ij}$),
    unreconstructed (but relaxed) surface energies ($(111)$, $(110)$, and $(100)$ low-index surfaces), point defect formation
    energies (vacancy, and hexagonal, tetrahedral, and dumbbell interstitials); while the planar defects to the right are
    extrapolative: $(112) \Sigma 3$
    symmetric tilt grain boundary, and unstable stacking fault energies on shuffle plane $\gamma^{(s)}_\mathrm{us}$
    and glide plane $\gamma^{(g)}_\mathrm{us}$).
    The first row in the corresponding table shows reference quantities computed with DFT (units indicated in header row).
    \label{fig:big_fractional_error_plot}}
\end{figure*}

\begin{figure}
  \centerline{\includegraphics[width=\columnwidth]{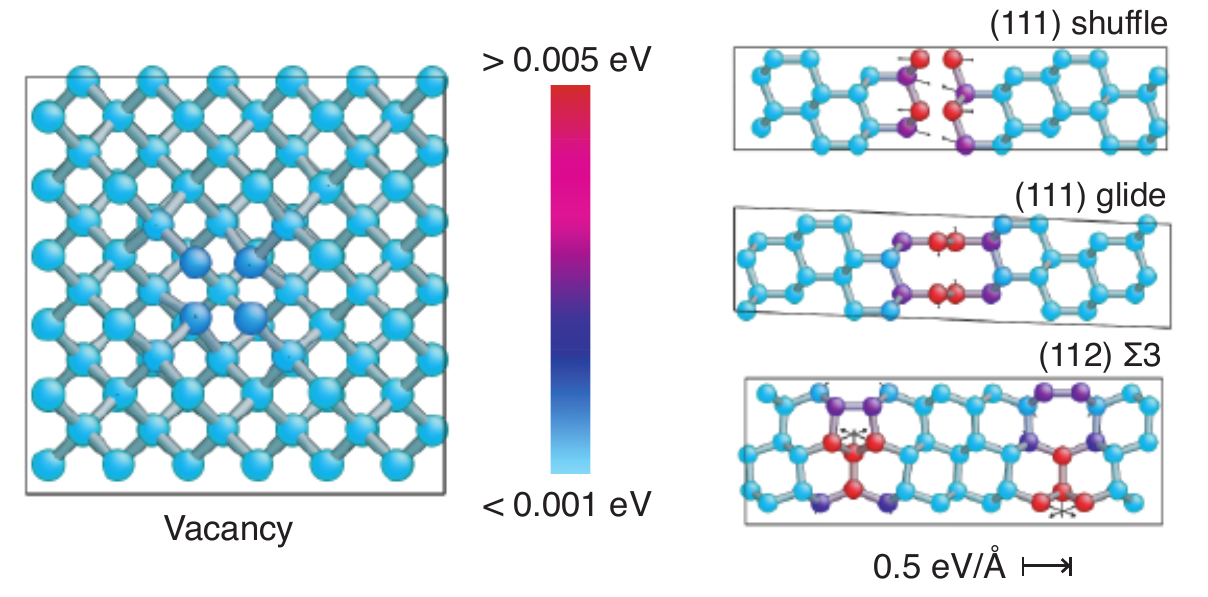}}
\caption{
    Visualisation of vacancy, $(111)$ shuffle and glide unstable stacking faults and $(112)~\Sigma3$
    grain boundary configurations. Atoms are coloured by the per-atom energy errors predicted by the GAP on the DFT-relaxed configurations, and arrows show the non-zero GAP forces.
    \label{fig:pred-err}
}

\end{figure}

Indeed, many advances in materials simulation methodology over
the past decades have been demonstrated first using silicon.
Some of these were new approaches where silicon was used as a test
system, including the Car-Parrinello method for {\em ab initio}
molecular dynamics (MD)~\cite{Car:1985ix}, maximally-localized Wannier
functions for analyzing electronic structure~\cite{Marzari:1997co},
concurrent coupling of length scales combining different simulation
method~\cite{Broughton:1999wq,Bernstein:2003gz}, and the learn-on-the-fly method for
extending the time scale of {\em ab initio} MD~\cite{Csanyi:2004dh}.
Others used these new methods to explain experimentally observed
phenomena in silicon, for example using density functional theory
(DFT) to study the $7 \times 7$
dimer-adatom-stacking-fault reconstruction of the Si (111) surface~\cite{Stich:1992hh}.
Silicon was also extensively used as a model system to understand fracture, and in particular
the interplay between brittle and ductile failure~\cite{Zhu:2004ds,buehler_phys_rev_lett_2006,Holland1998,Holland1998b,Kermode:2008gh}.

A large number of interatomic potentials have been developed
for silicon with the intent of describing its bulk phases and
defects.  While there are too many publications to thoroughly
review here, we discuss the most widely used and successful ones,
to motivate our choices for comparison models in this work.
By far the two most commonly used are those of Stillinger and
Weber~\cite{stillinger_phys_rev_b_1985,stillinger_phys_rev_b_1986_erratum}
(SW), and
Tersoff~\cite{tersoff_phys_rev_lett_1986,tersoff_phys_rev_b_1988a,tersoff_phys_rev_b_1988b,tersoff_phys_rev_b_1989,tersoff_phys_rev_b_1990_erratum}.
Both include pair terms and three body terms, the former defined
in terms of bond lengths and bond angles, the latter in terms of a
repulsive core and a bond-order dependent attractive bonding interaction.
Many other functional forms have also been used, as reviewed for example
by Balamane {\em et al.}~\cite{balamane_phys_rev_b_1992}, and more
recently by Purja Pun and Mishin~\cite{purja_pun_phys_rev_b_2017}.
While none produced sufficient improvement to lead to significant
adoption by the simulation community,  recent attempts to add
terms that depend on more than 3-body interactions have been at least
somewhat successful.  These include the environment dependent interatomic
potential~\cite{justo_phys_rev_b_1998} (EDIP), modified embedded atom
method~\cite{baskes_phys_rev_b_1992,lenosky_model_sim_mater_sci_eng_2000},
ReaxFF~\cite{buehler_phys_rev_lett_2006,ReaxFF_personal_comm}, and
screened Tersoff~\cite{Pastewka2008,pastewka_phys_rev_b_2013}. EDIP uses the local
coordination of each atom to approximate a bond order (a chemical concept
that is also integral to the Tersoff potential), and change the preferred
bond length, strength, and bond angles correspondingly.  MEAM is an angle
dependent functional form that evolved out of the simpler embedded atom
method, mainly used for metals, and we use the parameterization due to
Lenosky {\em et al.}~\cite{lenosky_model_sim_mater_sci_eng_2000}
The ReaxFF form was originally developed in the context of computational
chemistry to describe reactions of molecules, and the silicon potential we
use~\cite{ReaxFF_personal_comm} was previously used to simulate brittle
fracture\cite{buehler_phys_rev_lett_2006}.  The screened Tersoff form
(TersoffScr) was developed by Pastewka {\em et al.}, who modifed
the Tersoff functional form with a screening term to improve its
performance for fracture properties~\cite{Pastewka2008,pastewka_phys_rev_b_2013},
where bonds are broken and formed.  Finally, Purja Pun and
Mishin took the modified Tersoff form developed by Kumagai {\em
et al.}~\cite{kumagai_comp_mat_sci_2007} and optimized it for
a wide range of properties~\cite{purja_pun_phys_rev_b_2017}.
We compare the results of GAP to these interatomic potential
models (EDIP, MEAM, Purja Pun, ReaxFF, SW, Tersoff, and
TersoffScr), and also to the density-functional tight-binding (DFTB)
method~\cite{porezag_phys_rev_b_1995,frauenheim_phys_rev_b_1995,DFTB_web}.

The inclusion of a tight binding (TB) model in
the above list is essential because TB represents a middle ground between
DFT and interatomic potentials.  The TB approach is a minimal
description of electronic structure\cite{Finnis:2004da}, significantly
cheaper than DFT, yet still carrying the essentially quantum mechanical
nature of the electrons, giving a qualitatively robust description of
their behaviour in solids in a wide range of materials. Like interatomic
potential models, TB can be easily implemented with a cost that is linear in the
number of atoms. A lot of effort has gone into making accurate TB models~\cite{Gehrmann:2015dc,Lu:2015ja,Bernstein:2000ut,Wang:1999eq,Cohen:1997jm,Lenosky:1997is,Bernstein:1997wp,frauenheim_phys_rev_b_1995,Menon:1994fa,Kwon:1994dk,JamesLMercer:1994ji,Goodwin:1989bs},
and if they were clearly more accurate or transferable than conventional
interatomic potentials, they might present the same tradeoff
between speed and accuracy as ML models, which are also significantly
more computationally expensive than conventional interatomic potentials.
However, this does not appear to be the case: the widely used TB model included
here does not perform better on the whole than analytical potentials.
For the DFTB calculations we used the ``pbc'' parameter set and a k-point density of 0.007~\AA$^{-1}$ for
bulk configurations and 0.04~\AA$^{-1}$ for others. To reduce the computational
cost, no charge-self-consistency iterations were performed, since they are not
expected to lead to substantial differences for this monatomic covalently
bonded material.

Our focus is on creating {\em useful}
models, and therefore the guiding principle was to create a dataset and
fitting protocol that is the {\em least} specific to the material and
the observables as possible, while still achieving the aims.
On the one hand, we tried to create tests
that are as relevant to the materials modeller as possible, focusing
on observables that are either directly comparable to experiments, or at
least generally agreed to be important for the understanding of material
behaviour. On the other hand, we think of the ML potential as an
interpolation scheme for the reference DFT method, so with a few exceptions,
we compare the interatomic potential results to DFT, rather
than to experiment, for example.

It is worth noting that the comparisons with analytical potentials we
show in this paper are not meant as a definitive evaluation of their accuracy.  Since the
analytical potentials were fit to different sets of properties from
different sources, their performance for any particular observable could
very well be improved somewhat by refitting them to our DFT data; we did
not see the relevance of doing that here, and worked with the published
parametrisations, since the analytical potentials' main advantages are simplicity,
computational efficiency, and some transferability, rather than ultimate accuracy.
In the case of a machine learned model, the unique selling point
{\em is} the accuracy with which the target potential energy surface is
matched, and this is best demonstrated by comparing to DFT results. The
route to improved agreement with experimental observables is to
improve the target, i.e. using a more accurate description of electronic
structure.

In contrast to many earlier works on materials modelling and machine
learning, we do not emphasize learning outcomes in the statistical sense,
using splits of the data set into training, testing and validation. This
is partly because this has been done many times before for the same
kernel, descriptor and approach to parameter choice that we use here,
and partly because the SOAP kernel  does not have hyperparameters
that are worth optimising: they are dictated
by the length and energy scales inherent in atomic interactions, which
are well known. (The other parameters in the regression correspond to accuracy targets on  a few classes of configurations.)
Ultimately, the paper is about the validation on
material properties, and all those are based on atomic configurations
which were themselves not in the training set.

Figure~\ref{fig:big_fractional_error_plot} provides an overview of many of the
verification and validation tests carried out for our new silicon GAP model in
comparison to the empirical analytical models mentioned above. While the individual tests are
discussed in more detail below, we present an overview here.
The first three groups of quantities in the figure are verification tests, in
the sense that they require accuracy on configurations which are directly
represented in the training set. These are split into three classes of test:
bulk properties, surfaces and point defects. Bulk properties, namely the bulk
modulus $B$ and diamond cubic elastic constants $C_{11}$, $C_{12}$ and $C_{44}$,
are well reproduced by the GAP model with fractional errors relative to DFT of
less than 10\%; none of the other interatomic potentials reach this accuracy,
although in many cases they were fit to different training data (e.g.
experiment, or simply other exchange-correlation functionals). The largest
relative errors in bulk properties are typically made in the softest elastic
constant $C_{12}$, with the EDIP model being the next most accurate after our
new GAP model. The second class of verification tests demonstrates that the GAP
model performs consistently at describing surface energies of the $(111)$,
$(110)$ and $(100)$ cleavage planes, with errors of around 2\% with respect to
our reference DFT calculations. Here, the scatter across the various other
models is smaller than for bulk properties. For example, the $(100)$ surface energy is in
general well described by most models. For the third class of verification tests, formation
energies of vacancy and interstitial point defects, we  see a wide range of errors
across the models evaluated. The new GAP model again predicts all these quantities within 10\%
of the reference DFT results. In general, for any particular property there is
often a model that provides an accurate description but apart from GAP, we
are not aware of any model that provides uniform accuracy across the whole range of properties
in Fig.~\ref{fig:big_fractional_error_plot}.

Moving to more stringent tests of the new model, we considered a set of planar
defects which were not represented in the training set (right hand group in
Fig.~\ref{fig:big_fractional_error_plot}), namely the $(112)\Sigma 3$ symmetric
tilt grain boundary, and unstable stacking fault energies on the $(111)$ shuffle plane
$\gamma_\mathrm{us}^{(s)}$ and $(111)$ glide plane  $\gamma_\mathrm{us}^{(g)}$. For these tests  the
accuracy of the GAP model is reduced, but still within 20\% of DFT,
comparing favourably with all other models, some of which included stacking
fault values in their training sets (e.g. EDIP). Moreover, the ability of the
GAP model to provide an estimated error along with its predictions allows us to
qualitatively assess the expected reliability of the model for particular
classes of configurations. Figure~\ref{fig:pred-err} shows the predicted errors for
each atom in the vacancy, shuffle, glide and grain boundary
configurations. For the vacancy, the confidence of the model is high on all
atoms (blue colour), and the corresponding accuracy with respect to DFT is
high. The reduced confidence close to the planar defects (red atoms) is
consistent with the larger errors made for these configurations and the fact that
the database does not include any similar atomic environments.

The rest of the paper is organised as follows. In section~\ref{sec:method} we give
an overview of the potential fitting methodology and the construction of the database.
In section~\ref{sec:verification} we report on extensive tests that serve to
verify that those properties which the database is explicitly designed
to capture are indeed correctly predicted. This includes equations of
state, average structural properties of liquid and amorphous states, point
defect energetics, surface reconstructions, and crack tip geometries. In
section~\ref{sec:validation}, we validate the model by showing predictions for properties that
are deemed fundamental for modeling this material, but for which the
database makes no special provision. This includes further crystal
structures, thermal expansion, di-interstitials, grain boundaries and
random structure search. We finally give a  brief outlook in section~\ref{sec:conclusion}.

\section{Methodology}
\label{sec:method}

\subsection{Potential fitting}

The interatomic potential, even after assuming a finite interaction radius, is a relatively
high dimensional function, with dozens of atoms affecting the energy and force on any
given atom at the levels of tolerances we are interested in (around a meV/atom). However,
much of the interaction energy (in absolute magnitude) is captured by a simple pair
potential, describing exchange repulsion of atoms at close approach and potentially the chemical bonding in an average sense farther out. In anticipation of the kernel approach for fitting the interatomic potential, the
pair potential also serves a useful purpose from the numerical efficiency point of view, because
the exchange repulsion it takes care of is a component of the potential that is very steep, in comparison
which the bonding region, and such disparate energy scales are difficult to capture with a single
kernel in high dimensions.

In the present case we chose a purely repulsive pair potential, given by cubic splines that were fitted to the interaction of a pair of Si atoms, computed using DFT. This leaves the description of the attractive part entirely for the many-body kernel fit.

We start by giving a concise account of the Gaussian Approximation Potential kernel fitting approach, as we use it here. The total GAP model energy for our system is a sum of the pre-defined pair potential and a many body term which is given by a linear sum over kernel basis functions\cite{Rasmussen:2006vz},
\begin{equation}
E = \sum_{i<j} V^{(2)}(r_{ij}) +  \sum_i \sum_s^{M} \alpha_s K(\RR_i,\RR_s),
\label{eq:gapE}
\end{equation}
where $i$ and $j$ range over the number of atoms in the system, $V^{(2)}$ is the pair potential, $r_{ij}$ is the distance between atoms $i$ and $j$, $K$ is a kernel basis function defined below, and $\RR_i$ is the collection of relative position vectors corresponding to the neighbours of atom $i$ which we call a {\em neighbourhood}. The last sum runs over a set of $M$  {\em representative} atoms, selected from the input data set, whose environments have been chosen to serve as a basis in which the potential is expanded; more on this below.

The value of the kernel quantifies the similarity between neighbourhoods (in the Gaussian process literature it is a covariance between values of the unknown function at different locations), which is largest when its two arguments are equal, and smallest for maximally different configurations. The degree to which the kernel is able to capture the variation of the energy with neighbour configuration will determine how efficient the above fit is. The better the correspondence, the fewer representative configurations are needed to achieve a given accuracy. It also helps tremendously if exact symmetries of the function to be fitted are already built into the form of the kernel. For an interatomic potential, we  need a kernel that is invariant with respect to permutation of like atoms, and 3D rotations of the atomic neighbourhood. Note that translational invariance is already built in, because the kernel fit is applied to each atom individually---this very natural decomposition of the total energy is customary when fitting interatomic potentials, and is directly analogous with the spatial decomposition of convolutional neural networks\cite{Goodfellow:2016}.

Here we use the SOAP kernel\cite{Bartok:2013cs,Bartok:2013ks,Bartok:2015iw,Bartok:2016bq}. We start by representing the neighbourhood $\RR_i$ of atom $i$ by
its {\em neighbour density},
\begin{equation}
\rho_i({\bf r}) = \sum_{i'} f_\text{cut}(r_{ii'})e^{-({\bf r}-{\bf r}_{ii'})/2\sigma^2_\text{atom}}
\end{equation}
where the sum ranges over the neighbours $i'$ of atom $i$ (including itself), $f_\text{cut}$ is a cutoff function that smoothly goes to zero beyond a cutoff radius $r_\text{cut}$, and $\sigma_\text{atom}$ is a smearing parameter, typically 0.5~\AA. Invariance to rotations is achieved by constructing a Haar integral over the SO(3) rotation group\cite{Bartok:2013cs,Bartok:2013ks}. The SOAP kernel between two neighbour environments is the integrated overlap of the neighbour densities, squared, and then also integrated over all possible 3D rotations,
\begin{equation}
\tilde K(\RR_i,\RR_j) = \int_{\hat R \in SO_3} d\hat R \left| \int d{\bf r} \rho_i({\bf r}) \rho_j (\hat R {\bf r}) \right|^2
\end{equation}
To obtain the final kernel, we normalise and raise to a small integer power,
\begin{equation}
K(\RR_i,\RR_j) = \delta^2 \left|\frac{\tilde K(\RR_i,\RR_j) }{\sqrt{\tilde K(\RR_i,\RR_i) \tilde K(\RR_j,\RR_j) }}\right|^\zeta
\end{equation}
with $\zeta=4$ in the present case. The $\delta$ hyperparameter corresponds to the energy scale of the many body term, and we use $\delta=3$~eV, commensurate with typical atomization energy/atom. The accuracy of the fit is not particularly sensitive to this parameter.

In practice, we do not evaluate the above integrals directly, but expand the neighbour density in a basis of spherical harmonics $Y_{lm}({\bf \hat r})$ and radial  functions $g_n(r)$ (we use equispaced Gaussians, but the formalism works with any radial basis),
\begin{equation}
\label{eq:density_expansion}
\rho_i({\bf r}) = \sum_{nlm} c^i_{nlm} Y_{lm}({\bf \hat r}) g_n(r).
\end{equation}
The following spherical power spectrum vector (henceforth termed the ``SOAP vector'') is a unique, rotationally and permutationally invariant description of the neighbour environment,
\begin{align}
{\tilde p}^i_{nn'l} &= \sum^l_{m=-l} c^{i*}_{nlm} c^i_{n'lm}\\
{\bf p}^i &= {\bf\tilde p}^i/|{\bf\tilde p}^i |
\end{align}
and the SOAP kernel can be written as its scalar product,
\begin{equation}
\label{eq:kernel}
K(\RR_i,\RR_j) = \delta^2 | {\bf p}^i \cdot {\bf p}^j |^\zeta,
\end{equation}
The coefficients $\alpha_s$ in Eq.~\ref{eq:gapE} are determined by solving a linear system
that is obtained when available data are substituted into the equation, as we detail below, as we detail beloww. In the present case these data take the form of total energies and gradients (forces and stresses) corresponding to small and medium sized periodic unit cells, calculated using density functional theory.

We also need an algorithm to select the set of representative environments over which the sum in  Eq.~\ref{eq:gapE} is taken. This could be done by simple random sampling, but we find it advantageous to use this freedom to optimise interpolation accuracy. One approach to this is to maximise the dissimilarity between the elements of the representative set\cite{De:2016iaa}, such that the small number of environments best represent the variety of the entire set. Here we use a matrix reconstruction technique called CUR\cite{Mahoney:2009fz} and apply it to the rectangular matrix formed by the concatenation of SOAP vectors corresponding to all the neighbour environments appearing in the input data.  The CUR decomposition leads  to a low rank approximation of the full kernel matrix using only a subset of its rows and columns\cite{Imbalzano:2018er}.

There are two factors that complicate the determination of the vector of linear expansion coefficients, $\boldsymbol{ \alpha}$. The first is that atomic energies are not directly available from density functional theory, and the second is the presence of gradients in the input data. The following treatment addresses both of these.  We denote the number of atoms in the input database with $N$, and define
$\boldsymbol{y}$ as the vector with $D$ components containing the input data: all total energies, forces and virial stress components in the training database,  and $\boldsymbol{y}'$ as the vector with $N$ components containing the \textit{unknown}  atomic energies of the $N$ atomic environments in the database,  and  $\mathbf{L}$ as the linear differential operator of size $N\times D$  which connects $\boldsymbol{y}$ with $\boldsymbol{y}'$ such that $\boldsymbol{y}=\mathbf{L}{^\mathrm{T}}\boldsymbol{y}'$. After selecting $M$ representative atomic environments (with $M \ll N$), the regularised least-squares solution for the coefficients in Eq.~\ref{eq:gapE} is given by\cite{Snelson:2005vi,QuinoneroCandela:2005wpb}
\begin{equation}
\label{eq:alphas}
    \boldsymbol{\alpha} = {\big[ \mathbf{K}_{MM} + \mathbf{K}_{MN} \mathbf{L} \mathbf{\Lambda}^{-1} \mathbf{L}{^\mathrm{T}} \mathbf{K}_{NM}  \big]}^{-1} \mathbf{K}_{MN} \mathbf{L} \mathbf{\Lambda}^{-1} \boldsymbol{y} ,
\end{equation}
where $K_{MM}$ is the kernel matrix corresponding to the $M$
representative atomic environments (with matrix elements from Eq.~\ref{eq:kernel}), $K_{MN}$ is the kernel
matrix corresponding to the representative set and all of the
$N$ environments in the training data, and the elements of the diagonal matrix
$\mathbf{\Lambda}^{-1}$ represent weights for the input data values. 
The Bayesian interpretation of the inverse weights are expected
errors in the fitted quantities. While taking $\mathbf{\Lambda}=\sigma_{\nu}^2 \mathbf{I}$ with an empirical value for $\sigma_\nu$
would be sufficient to carry out the fit, this interpretation makes it straightforward to set sensible values.
The expected errors are not just due to lack of numerical
convergence in the electronic structure calculations, but also include
the \emph{model error} of the GAP representation, e.g.\ due to the
finite cutoff of the local environment. Our informed choices for these
parameters are reported in Table~\ref{table:sidatabase}.

For several systems below, we include results on the {\em predicted
error}, the  measure of uncertainty intrinsic to our interpolated
potential energy surface. These come from the Bayesian view of the
above regression procedure, in which the data (and the predicted values)
are viewed as samples from a Gaussian process whose covariance function
is the chosen kernel function\cite{Rasmussen:2006vz}. The mean of this Gaussian
process is of course just the second term of the predicted energy, Eq.~\ref{eq:gapE},
and the predicted variance of the atomic energy for atom $i$ is given by
\begin{equation}
K(\RR_i,\RR_i) - \mathbf{k}^T (\mathbf{K}_{MM}+\sigma_e \mathbf{I})^{-1} \mathbf{k}
\label{eq:gpvar}
\end{equation}
where the element $s$ of the vector $\mathbf{k}$ is given by
$K(\RR_i,\RR_s)$, the covariance between
the environment of atom $i$ and the environments of the representative atoms $s$ in
the database.  The above is a simplified error estimate, in which we
regularise using the parameter $\sigma_e$, typically set to 1~meV (equal
to the value used for the per-atom energy data components  of $\mathbf \Lambda$ for most of the
database in Eq.~\ref{eq:alphas}), rather than using the more complicated
regularisation as in Eq.~\ref{eq:alphas}. We interpret this variance as
the (square of the) ``one sigma'' error bar for the atomic energies.

\subsection{Database}

The database of atomic configurations (periodic unit cells) is described in Table~\ref{table:sidatabase}. It was built over an extended period, using multiple computational facilities. The kinds of configurations that we included were chosen using intuition and past experience to guide what needs to be included to obtain good coverage pertaining to a range of properties. The number of configuration in the final database is a result of somewhat ad-hoc choices, driven partly by the varying computational cost of the electronic structure calculation, and partly by observed success in predicting properties, signalling sufficient amount of data.
Each configuration yields a total energy, six components of the stress tensor and 3 force components for each atom. The database therefore has a total of 531710 pieces of electronic structure data. We represent the diversity of atomic neighbourhoods using $M=9000$ representatives, and the number of these picked from each of the structure types by the CUR algorithm is also shown in the table.

We used the Castep software package\cite{castep} as our density functional theory implementation, and manual cross-checking was done to ensure that the calculations are consistent between different computers. The main parameters of the electronic structure calculation were as follows: PW91\cite{Perdew:1992jd} exchange-correlation functional (the choice was motivated by the existence of large scale simulation of the melting point with this functional), 250~eV plane wave cutoff (with finite basis corrections), Monkhorst-Pack k-point grids with 0.03~\AA$^{-1}$ spacing, ultrasoft pseudopotentials, and 0.05~eV smearing of the electronic band filling. The remaining numerical error is dominated by the finite k-point grid, leading to errors on the order of a few meVs. The reference data for testing purposes was calculated with the parameters kept the same, except for: bulk energy-volume curves, which used a k-point spacing of 0.015~\AA$^{-1}$; the re-optimisation of IP minima of amorphous configurations (Table~\ref{table:amorph_coord}) which used a k-point spacing of 0.07~\AA$^{-1}$; and molecular dynamics of the liquid, whose parameters are given further below.

\begin{figure}
    \centerline{\includegraphics[width=0.9\columnwidth]{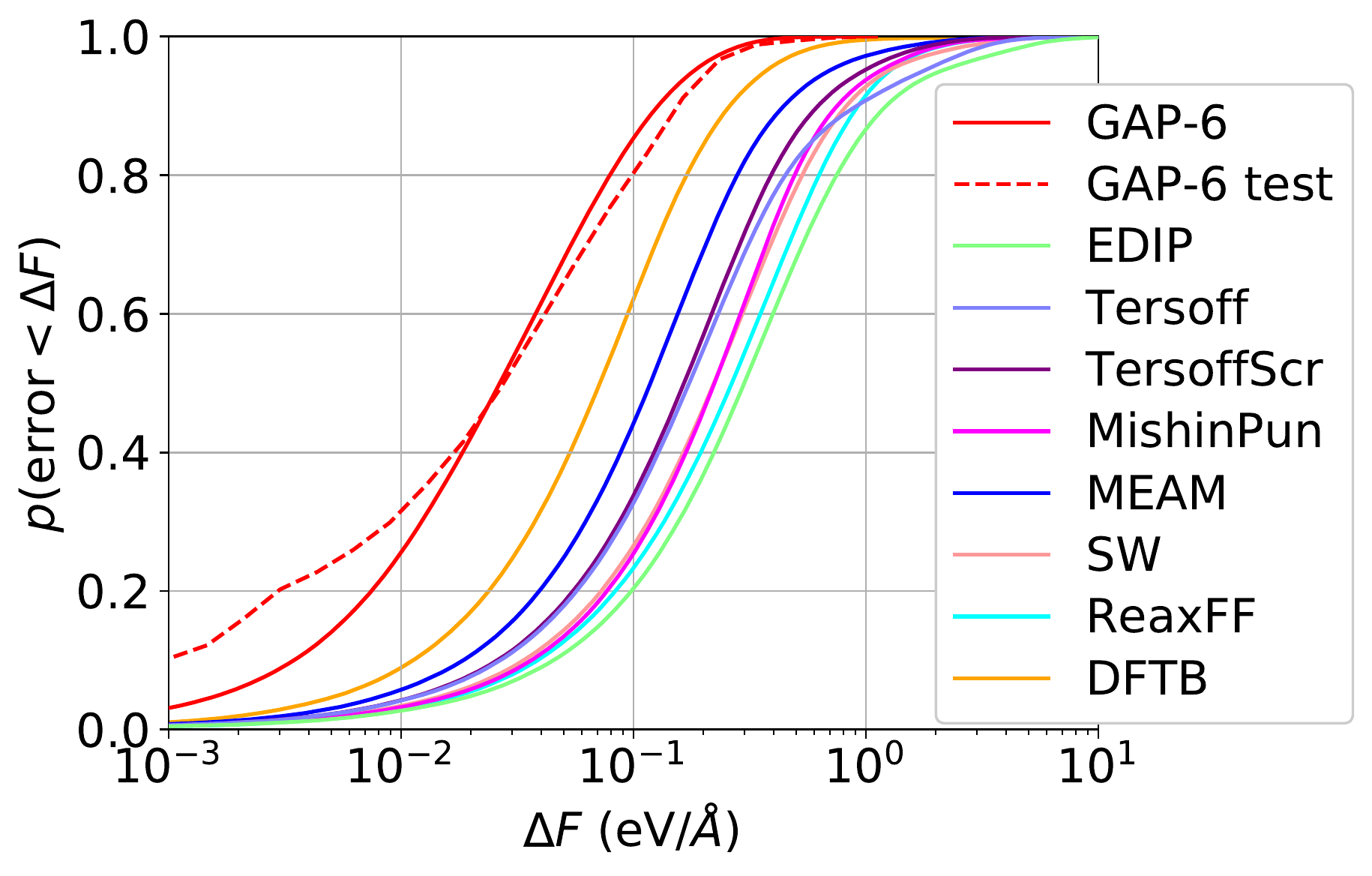}}
    \caption{
    Cumulative probability distribution of force component errors (relative to
    reference DFT calculations) for all potentials evaluated on the GAP model
    fitting database (solid lines), and for the GAP model only on
    a separate testing database (dashed line).
    \label{fig:testing_force_error}
    }
\end{figure}

While we focus our efforts here on testing the GAP for its predictions
for scientifically interesting observables, we have also evaluated the global
 distribution of force errors relative to DFT calculations.
The results for all the potentials evaluated on the GAP fitting
database, as well as for the GAP on a simple testing database
(distinct from the fitting database) are shown in Fig.~\ref{fig:testing_force_error}.
The GAP shows much lower force errors than any other potential
tested, with a median of about 0.025~eV/{\AA}, an order of magnitude smaller
than for the analytical potentials. The testing database,
which consists of a grain boundary, 6 di-interstitials, the unrelaxed
and relaxed shuffle and glide generalized stacking fault paths,
and an amorphous configuration, shows very similar distribution of force
error, although the actual errors are strongly dependent on the
type of geometry, so changing the proportions of each could
change the resulting distribution somewhat.

Note that the testing database for Fig.~\ref{fig:testing_force_error} is not the result of a usual random split into
training and test sets, but represents extrapolation into configurations entirely
different from those in the training database. This is a more stringent test
than the usual split. Since the empirical analytical potentials have not been fit to our database,
the latter serves as a test for the potentials. It is remarkable how good the analytical
potentials' predictions are for macroscopic properties, which are mostly energy differences,
given the large force errors shown here.

\newcommand{\centercol}[1]{\multicolumn{1}{>{\centering}p{2cm}}{#1}}
\begin{table*}
%
%
    \centering
    \begin{tabular}{p{3.5cm}|r r r r r r r}
    \hline\hline
        Structure type &
        \centercol{\# atoms} & \centercol{\# structures} & \centercol{\#~environments} & \centercol{\#~representative~atoms} & $\sigma_{\rm energy}$ & $\sigma_{\rm force}$ & $\sigma_{\rm virial}$\\
        &&&&&\multicolumn{3}{c}{default values:}\\
        &&&&&0.001&0.1&0.05\\
    \hline\hline
        isolated atom & 1 & 1 & 1 & 1  & &  &  \\\hline
        \multirow{4}{*}{diamond}%
                & 2& 104& 208 & 6& &&\\
               & 16&220& 3520 &53 & & &\\
               & 54&110& 5940 & 58 & & &\\
               & 128&55& 7040 & 92 & & &\\\hline
        \multirow{4}{*}{$\beta$-Sn}%
                & 2& 60& 120 & 32 & &&\\
               & 16&220& 3520 & 51& & &\\
               & 54&110& 5940 & 66& & &\\
               & 128&55&  7040 & 157& & &\\\hline
        \multirow{4}{*}{simple hexagonal}%
                & 1& 110& 110 & 13 & &&\\
               & 8&30&240 & 15 &  & &\\
               & 27&30& 810 & 42 & & &\\
               & 64&53& 3392 & 89 & & &\\\hline
        hexagonal diamond & 4& 49 & 196 & 7 & & &\\\hline
        bcc & 2 & 49 & 98 & 40& & &\\\hline
        bc8 & 8 & 49 & 392 & 66 & & &\\\hline
        fcc & 4 & 49 & 196 & 46 & & &\\\hline
        hcp & 2 & 49 & 98 & 28 & & &\\\hline
        st12 & 12 & 49 & 588 & 94 & & &\\\hline
        \multirow{2}{*}{liquid}%
           &64& 69& 4416& 1114 & \multirow{2}{*}{0.003} & \multirow{2}{*}{0.15}& \multirow{2}{*}{0.2}\\
        & 128 & 7 & 896 & 323 & & &\\\hline
        \multirow{2}{*}{amorphous}%
        &64  & 31  & 1984  & 231  & \multirow{2}{*}{0.01}  & \multirow{2}{*}{0.2} & \multirow{2}{*}{0.4}\\
        &216 & 128 & 27648 & 1719 & & &\\\hline
        diamond surface (001)%
                &144 & 29 & 4176 & 514 & &&\\
                \quad decohesion & 32 & 11 & 352 &28&&&\\\hline
        diamond surface (110)%
                &108 & 26& 2808 & 338 & &&\\
                \quad decohesion & 16 & 11 & 176 &8&&&\\\hline
        diamond surface (111)&&&&&&\\
            \quad decohesion & 24 & 11 & 264 &10&&&\\
            \quad unreconstructed& 96 & 47 & 4512 & 573 &&&\\
            \quad adatom &146 & 11 & 1606 & 62 &&&\\
            \quad Pandey reconstruction &96 & 50 & 4800 & 632 &&&\\
           \quad DAS $3 \times 3$ unrelaxed& 52& 1& 52 & 6 &&&\\\hline
        \multirow{2}{*}{diamond vacancy}%
                &63 & 100& 6300 & 168 & &&\\
                &215 & 111& 23865 & 405 &   &  & \\\hline
        diamond divacancy%
                &214  &78 & 16692 & 416 & && \\\hline
        diamond interstitial%
                &217 & 115 & 24955 & 605 & &&\\\hline
        small (110) crack tip%
                &200 & 7 & 1400 & 130 & &&\\\hline
        small (111) crack tip%
                &192 & 10 & 1920 & 185 & &&\\\hline
        screw dislocation core%
                &144 &19  & 2736 & 124 & &&\\\hline
        sp$^2$ bonded%
                &8   & 51 & 408 & 61 & &&\\\hline
        sp bonded%
                &4   & 100 & 400 & 392 & 0.01 & 0.2 & 0.4\\\hline\hline
        Total &  & 2475 &  171815& 9000& &\\
        \hline        \hline
    \end{tabular}
    \caption{Summary of the database for the silicon model. The first
    column shows the number of atoms in the periodic unit cells, the
    second column shows the number of such unit cells in the database,
    while the third column is the product of the first two, and thus
    shows the number of atoms (and therefore atomic environments) in the
    database for each structure type. The fourth column shows the number
    of representative atoms picked automatically from each structure type
    by the CUR algorithm (see text). The last three columns show the
    regularisation we used in the linear system (empty rows correspond
    to using the defaults, given at the top).
    }
    \label{table:sidatabase}
\end{table*}

\subsection{Convergence}

Since the principal goal of machine learned interatomic potentials is to
enable the prediction material properties by fitting the Born-Oppenheimer
potential energy surface, it is interesting to consider the {\em
convergence} of such a potential. The expectation is that a closer
match of the potential energy surface will result in more accurate
predictions. While a comprehensive convergence study is beyond the scope
of this work, there are simple convergence parameters in the SOAP/GAP framework that
directly control the tradeoff between computational cost and accuracy of
the fit. One is the number $M$ of representative environments (effectively
the number of basis functions in the regression), the other is the truncation of the
spherical harmonic and radial basis expansion of the atomic neighbour density (Eq.~\ref{eq:density_expansion}).
\begin{figure*}
    \centerline{\includegraphics[width=15cm]{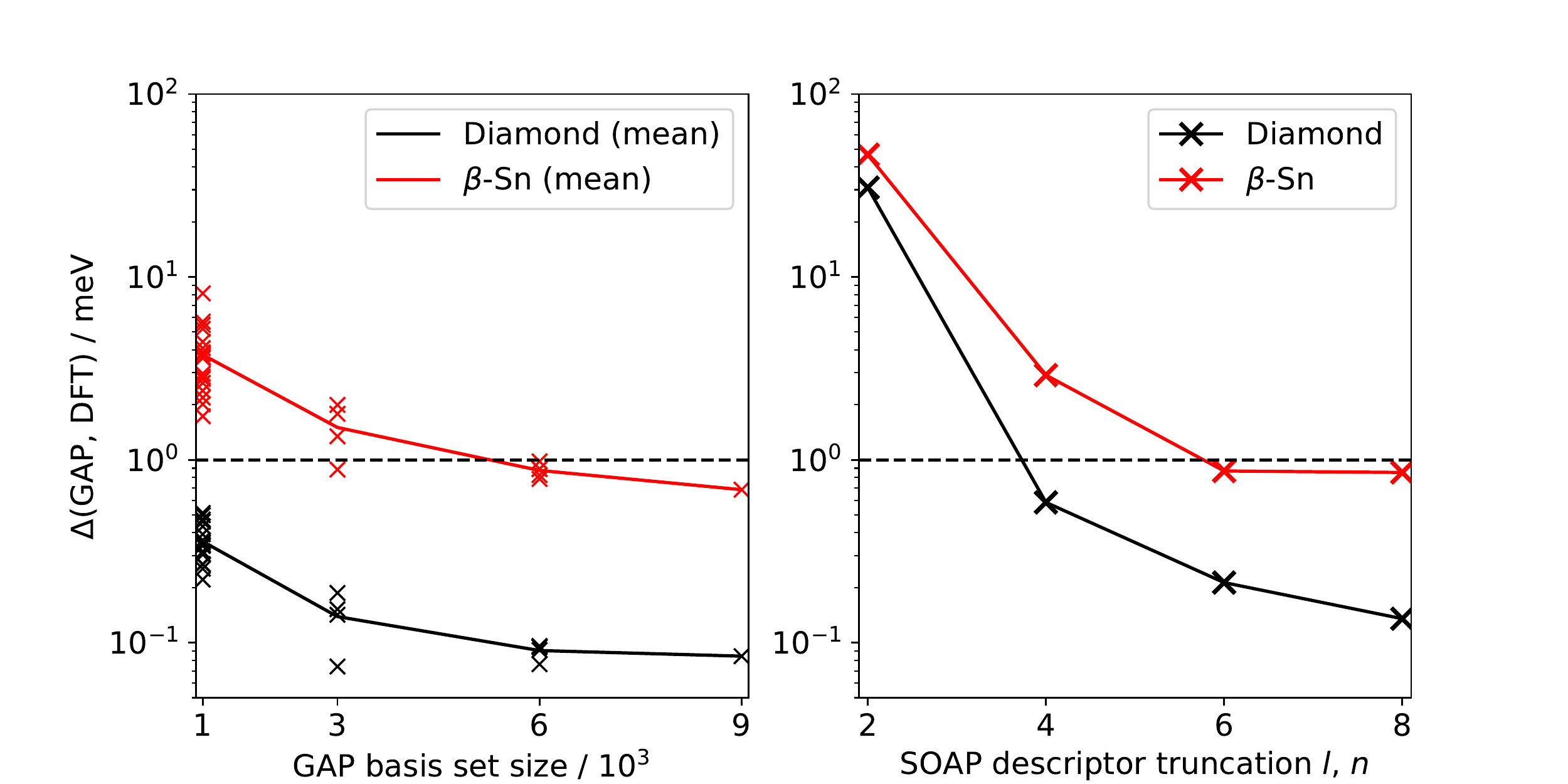}}
    \caption{
    Error of the SOAP/GAP model based on the $\Delta$-value of Ref.~\onlinecite{Lejaeghere2016} with respect to DFT for diamond (black) and $\beta$-Sn (red) structures. The left  panel shows the error as a function of the number of basis functions used in the Gaussian process regression. Because we use a stochastic algorithm to select which basis functions to use, multiple models are shown, which only differ in the pseudo-random seed. The right panel shows the error as a function of the length of the SOAP descriptor vector, which in our case is controlled by the truncation of the radial and spherical harmonic expansion of the atomic neighbour density. The horizontal dashed line corresponds to $1$~meV/atom error, our default ``energy'' accuracy target.
    \label{fig:error_convergence}}
\end{figure*}
Figure~\ref{fig:error_convergence} shows the convergence of the SOAP/GAP model with respect to these. We use the $\Delta$-value of \citet{Lejaeghere2016} to compute the error in the energy-volume curves for diamond and $\beta$-Sn with respect to our DFT reference, defined as
\begin{equation*}
  \Delta = \sqrt{
    \frac{\int_{0.94 V_0}^{1.06 V_0} \left[
      E^\mathrm{GAP}(V) - E^\mathrm{DFT}(V)
      \right]^2\, \mathrm{d}V}{0.12 V_0}
  }
\end{equation*}
where $E^\mathrm{GAP}$ and $E^\mathrm{DFT}$ denote GAP and DFT energies relative to the diamond energy minimum to allow comparison, $V_0$ is the DFT minimum-energy volume for each phase and the integral is computed numerically by fitting cubic splines to 12 $(E, V)$ pairs for each model. Good convergence can be seen with respect to both basis set size and the accuracy of the expansion of the atomic density, with a precision of the order of a meV (for $\beta$-Sn, for diamond another order of magnitude better), indicating that GAP reproduces the target DFT energy surface better than the typical variability between DFT codes of order $\Delta=1$~meV reported in Ref.~\onlinecite{Lejaeghere2016}.

In principle, a Gaussian process regression model should be able to
converge to a given target function with arbitrary accuracy as the
database size grows. However, in this case the only remaining physical
approximation is the finite cutoff of the interatomic potential, which
means that the force on an atom that is computed using our DFT engine is
not strictly a function of the finite neighbourhood of the atom.  From
the point of view of a model with finite cutoff, the target function
appears to have an finite amount of uncertainty, and this  uncertainty
is taken into account when fitting the model, as mentioned above. Indeed,
previous investigations have shown that with a cutoff of 5~\AA, an error
of 0.1~eV/\AA\ on the forces is about what is to be expected for the
diamond structure.\cite{Csanyi2005,Bernstein2009,Peguiron2015} Note that
it is possible to estimate the expected force error due to the finite
cutoff directly from the DFT engine because forces are themselves local
quantities, as opposed to site energies and virial stress components,
which are not observable directly.

It is noteworthy how much more accurate the potential is for the diamond structure than for $\beta$-Sn. Two factors contribute to this: first, there are many more diamond-like configurations in the database, particularly the configurations associated with various defects, and second, the locality error is expected to be significantly larger for the $\beta$-Sn structure due to its metallic electron density of states.

We do not claim that our database in the present work is complete in a mathematical sense (even within the restriction of the given cutoff), but that for any particular application whose relevant configurations are well represented in the database, errors can be improved only by choosing a larger cutoff, which in turn might lead to the need to enlarge the database further.

\subsection{Testing}

A software testing framework was built
to run tests of the potential using the Atomic Simulation Environment
(ASE)~\cite{Larsen2017}.
Each model and test is implemented as an
independent Python module, allowing all tests to be run with each model (similarly
to the design of the OpenKIM project~\cite{OpenKIM}).  The model modules
are simple, consisting of calls to existing ASE interfaces to QUIP~\cite{QUIP} (GAP, DFTB,
Stillinger-Weber, Tersoff, and MEAM), LAMMPS~\cite{lammps} (EDIP, Purja Pun, and ReaxFF),
and Atomistica~\cite{Atomistica} (TersoffScr).  Reference DFT
results were obtained using the same tests with a model based on the ASE
interface to Castep, and using the parameters discussed above.
One advantage of this automated approach is that it ensures consistency in
starting configurations, minimization algorithms, and the final test results that
are shown in our figures.
Another is that it enables automated re-running of tests
when changes are made, e.g.\ to the GAP training database, allowing incremental
improvements to be assessed. The framework is available for download.\cite{testing-framework}

\section{Results: verification}
\label{sec:verification}

In this section, we report on a series of basic tests which the GAP model was
designed to pass, because they correspond to configurations that were selected
for inclusion in the database for the purpose of describing those very
observables.  We refer to these as ``verification'', by analogy to the usage of the term in software
engineering, where it refers to confirmation that the software implements the
specifications correctly.

It is important to note that by the very nature of such data-driven models, in
some sense the database (and the corresponding models) will never be deemed
completely final and definitive.  By designating some tests as part of
``verification,'' we mean to be open about the fact that the database was
amended qualitatively and quantitatively until these tests were passed to our
satisfaction, and therefore these tests are in some sense merely the achievement of a good
fit. This is in contrast to the next section, ``validation'' (again by analogy to the use of the term in software engineering where it refers to confirmation that the specifications
describe a method that achieves the desired goal), in which we collect tests for
which the database was not explicitly designed, but concern observables that a good model for the material ought to be able to describe. We made no attempt to augment or modify the database in order to improve the results of those tests, and this could, indeed should, be done in future work.

\subsection{Bulk crystals}

As an initial test we calculated the energy vs.\ volume for a number of
bulk crystal structures for silicon, including the ground state diamond
structure, closely related hexagonal diamond, known high pressure
structures $\beta$-Sn, simple hexagonal (sh), bc8 and st12 structures,
as well as even higher pressure phases, hexagonal close packed (hcp), body
centered cubic (bcc), and face centered cubic (fcc).  When calculating
these curves with DFT as well as DFTB and each interatomic potential we
deform the lattice to the target volume and relax it with respect to
unit cell shape and atomic position while approximately constraining
the volume, and also constraining the symmetry (using spglib~\cite{spglib})
to remain that of the initial
structure.   We find that the hcp structure has two minima, the conventional
one with $c/a \approx \sqrt{3/2}$, and another we label hcp' which has
a much lower $c/a < 1$.

\begin{figure}
    \centerline{\includegraphics[width=\columnwidth]{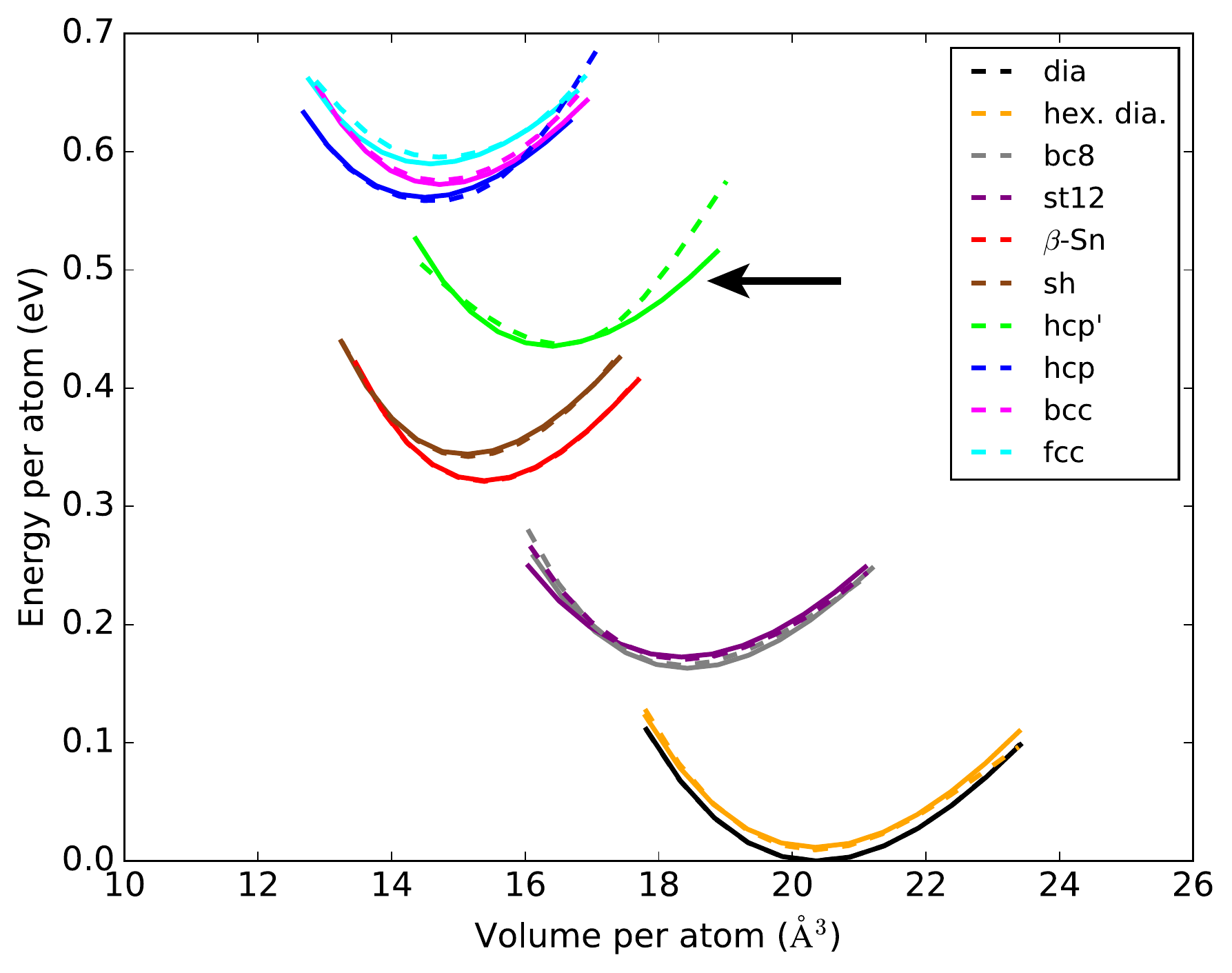}}
    \caption{Energy per atom vs.\ volume per atom for various
    bulk crystal lattice structures computed using DFT (solid lines) and
    GAP (dashed lines).  The hcp' structure (indicated by an arrow),
    which is not in the fitting database, has a substantially larger
    discrepancy between DFT and GAP than any of the other structures,
    all of which are in the database.
    }
    \label{fig:bulk_phases_GAP}
\end{figure}

\begin{figure}
    \centerline{\includegraphics[width=\columnwidth]{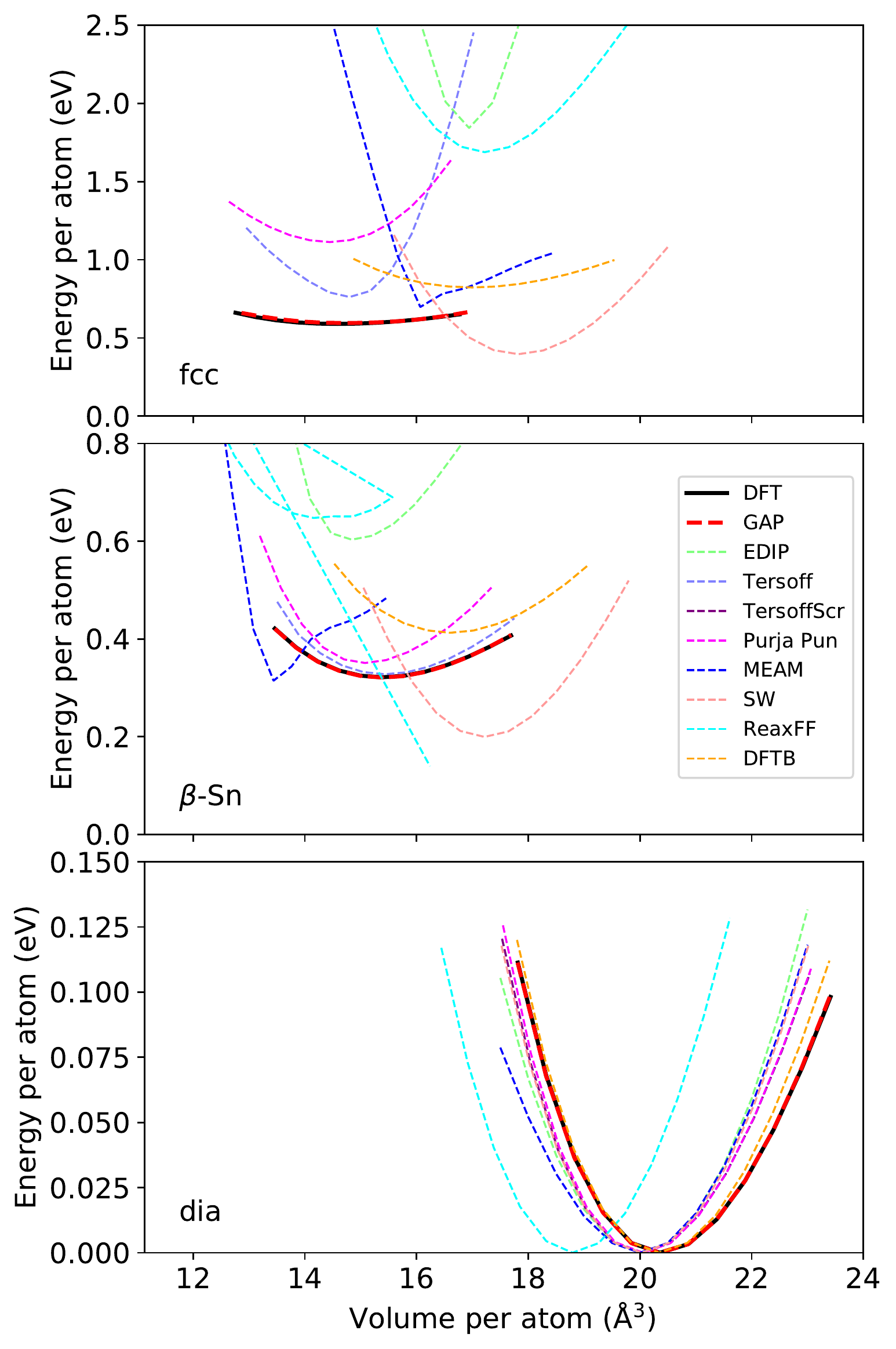}}
    \caption{Energy per atom relative to diamond structure vs.\ volume per atom for
    fcc (top panel), $\beta$-Sn (middle panel), and diamond structure (bottom panel),
    computed with DFT (black, solid line), GAP (red dashed line),
    and all other models (various colors, dashed lines).  Note different y-axis
    range on each panel. }
    \label{fig:bulk_dia_bsn_fcc_ALL}
\end{figure}

The resulting $E(V)$ curves for each crystal structure calculated
with GAP and compared to our reference DFT calculations are shown in
Fig.~\ref{fig:bulk_phases_GAP}.  The results are in excellent agreement
for all structures tested, including minima positions (volume), depths
(cohesive energy relative to the ground state), and curvatures (bulk
modulus).  The hcp' structure, which is not in the fitting database, has a larger
discrepancy than the other structures, although it is still in good agreement.
A comparison of all the models for a few selected crystal
lattices (diamond structure, $\beta$-Sn, and fcc), are shown in
Fig.~\ref{fig:bulk_dia_bsn_fcc_ALL}.  Only GAP is even qualitatively reproducing
all three selected structures, and many of the models fail to reproduce
even the first structure seen experimentally under applied pressure, $\beta$-Sn.

\subsection{Liquid}

To simulate the structure of liquid silicon with each interatomic
potential and DFTB we used constant pressure ($P=0$~GPa) molecular dynamics
as implemented in the QUIP package through the quippy Python
interface~\cite{QUIP}.  A $2 \times 2 \times 2$ supercell of the 8-atom
diamond cubic cell (64 atoms total) was heated from $T=0$~K to $T=5000$~K
for rapid melting over 20000 0.5~fs time steps, then equilibrated at
$T=2000$~K for 10000 0.25~fs time steps.  Structural data was gathered
over an additional 5000 0.25~fs time steps.  Reference DFT results
were obtained from a similar MD simulation using the Castep software,
averaging over 9700 0.25~fs time steps at $T=2000$~K. For the electronic structure
calculations, a 200~eV plane-wave energy cutoff and a $2 \times 2 \times 2 $
Monkhorst-Pack\cite{Monkhorst:1976cv} k-point grid was used (equivalent to a k-point density of about 0.05 \AA$^{-1}$).
The radial distribution function (RDF) and angular distribution
function (ADF) were calculated and averaged using the tools included in QUIP.

\begin{figure}
    \centerline{\includegraphics[width=\columnwidth]{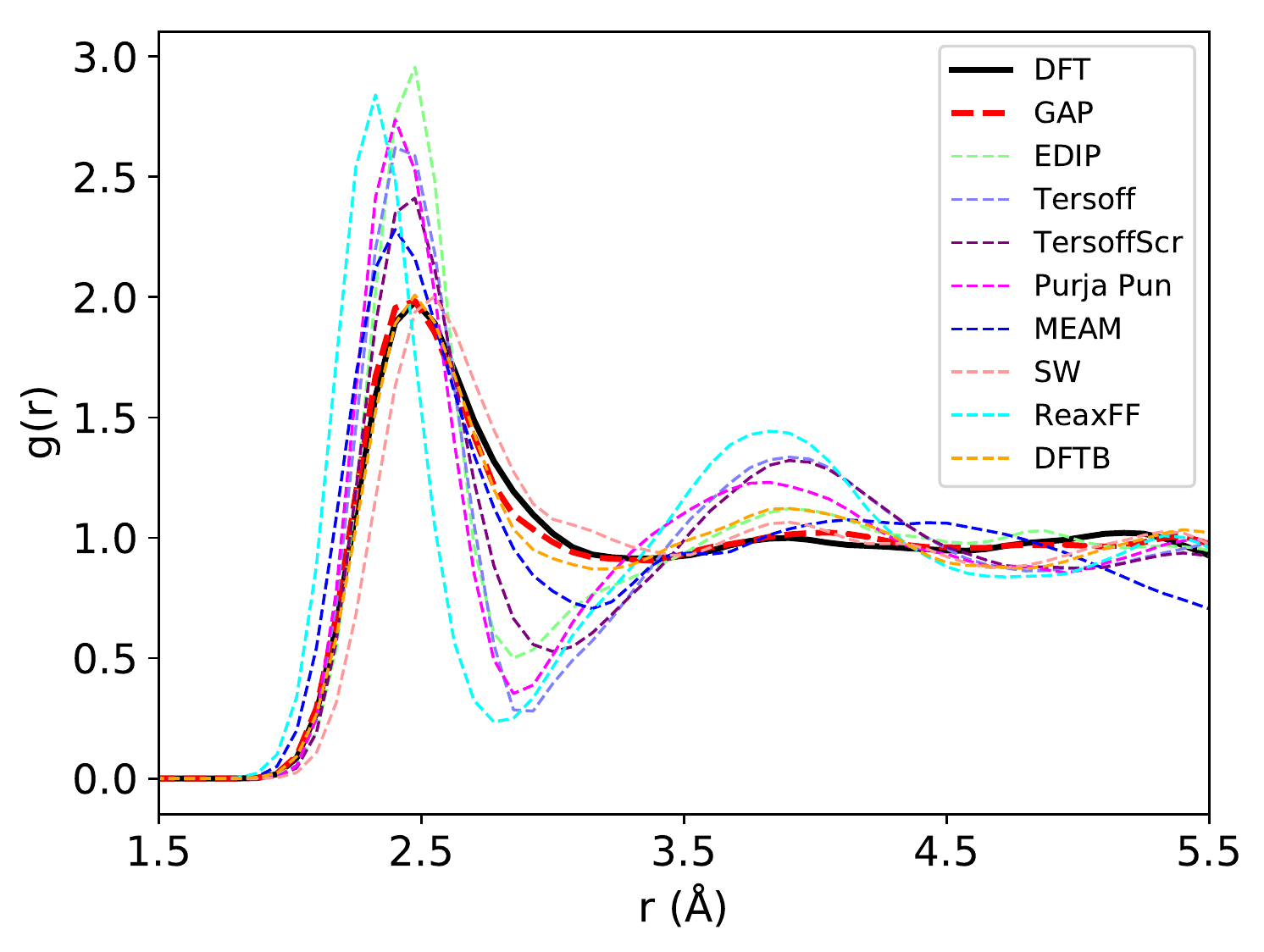}}
    \centerline{\includegraphics[width=\columnwidth]{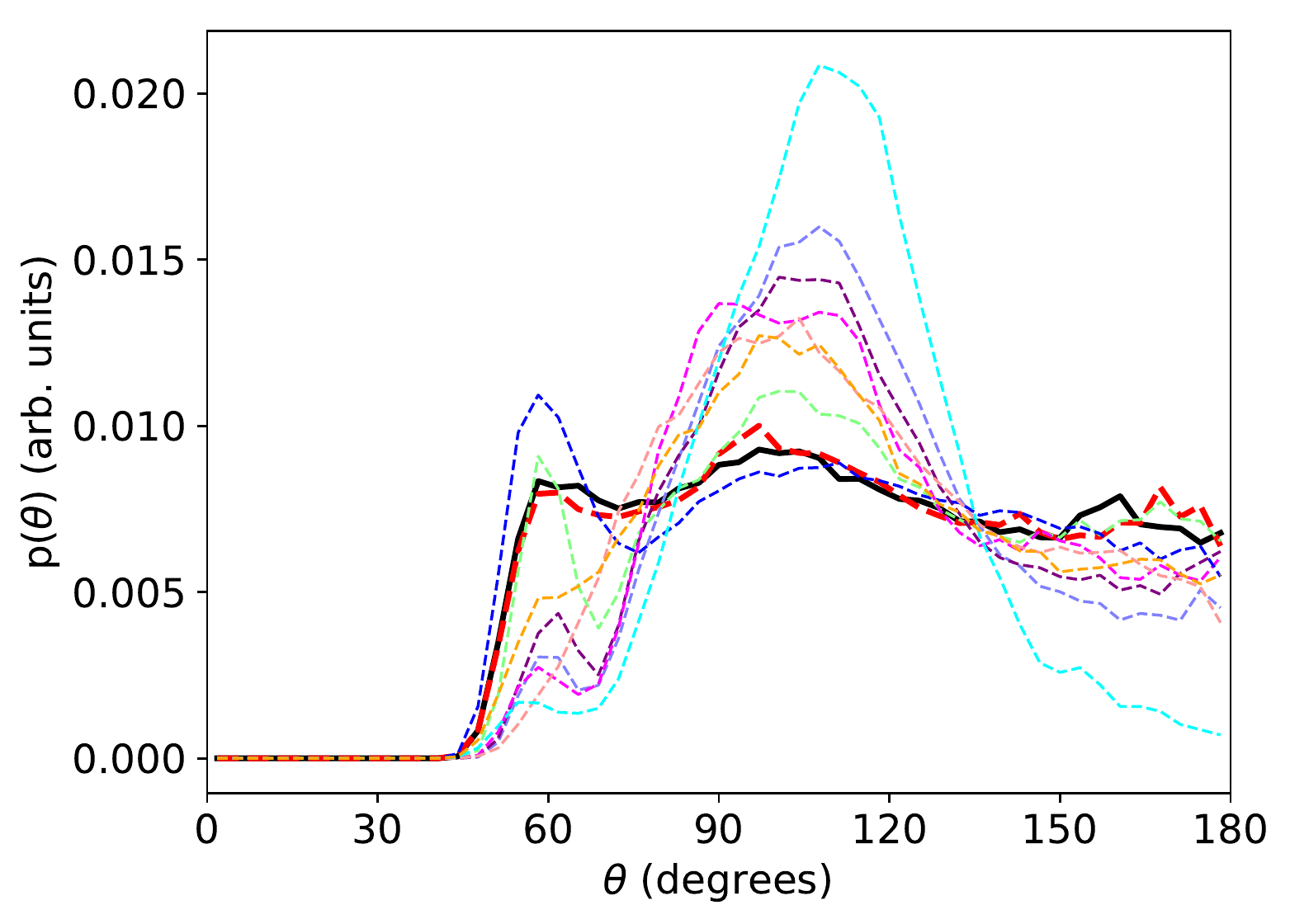}}
    \caption{
        Liquid silicon radial and angular structure from well equilibrated constant temperature
        and pressure 64 atom samples at $P=0$~GPa and $T=2000$~K.  Top: radial distribution function (RDF). Bottom: angular distribution
        function (ADF). Black solid line indicates DFT results, red dashed line and symbols
        indicate GAP results, results, and dashed lines (various colours) indicate DFTB and other
        interatomic potentials.}
    \label{fig:liquid_struct}
\end{figure}

The resulting structural quantities are shown in Fig.~\ref{fig:liquid_struct}.
The GAP RDF is in excellent agreement with the DFT result, including
both peak heights and radii at all distances captured in the simulation
cell.  DFTB is in comparably good agreement on this structural quantity,
and the various interatomic potentials are in much worse agreement,
with signficant variation among them.  The ADF proves to be an even more
stringent test.  Again the GAP results are in excellent agreement with
DFT, showing a narrow peak at about 60$^\circ$, and a broader peak with
similar height at about 100$^\circ$.  Most of the potentials greatly
underestimate the height of the small angle peak and overestimate the
height of the large angle peak.  The only two that are qualitatively
correct are EDIP and MEAM, but those both overestimate the depth
of the trough separating the two peaks. 
Several issues with the analytical interatomic potentials may be the source
of the differences.  Some, for example Tersoff~\cite{tersoff_phys_rev_b_1988b}, greatly overestimate
the melting point and are therefore strongly undercooled at $T=2000$~K rather than
an equilibrium liquid.  In other cases it's possible that
the wide
variety of curves observed is consistent with the hypothesized
liquid-liquid phase transition between a high coordination, high density
metallic phase and a low coordination, low density semiconductor-like phase.~\cite{Ganesh:2009ks}
Some of the potentials may simply be incorrectly predicting the low coordination
phase to be present at $T=2000$~K and zero pressure, leading to a
predominantly tetrahedral-like bond angle distribution.

\begin{figure}
    \centerline{\includegraphics[width=\columnwidth]{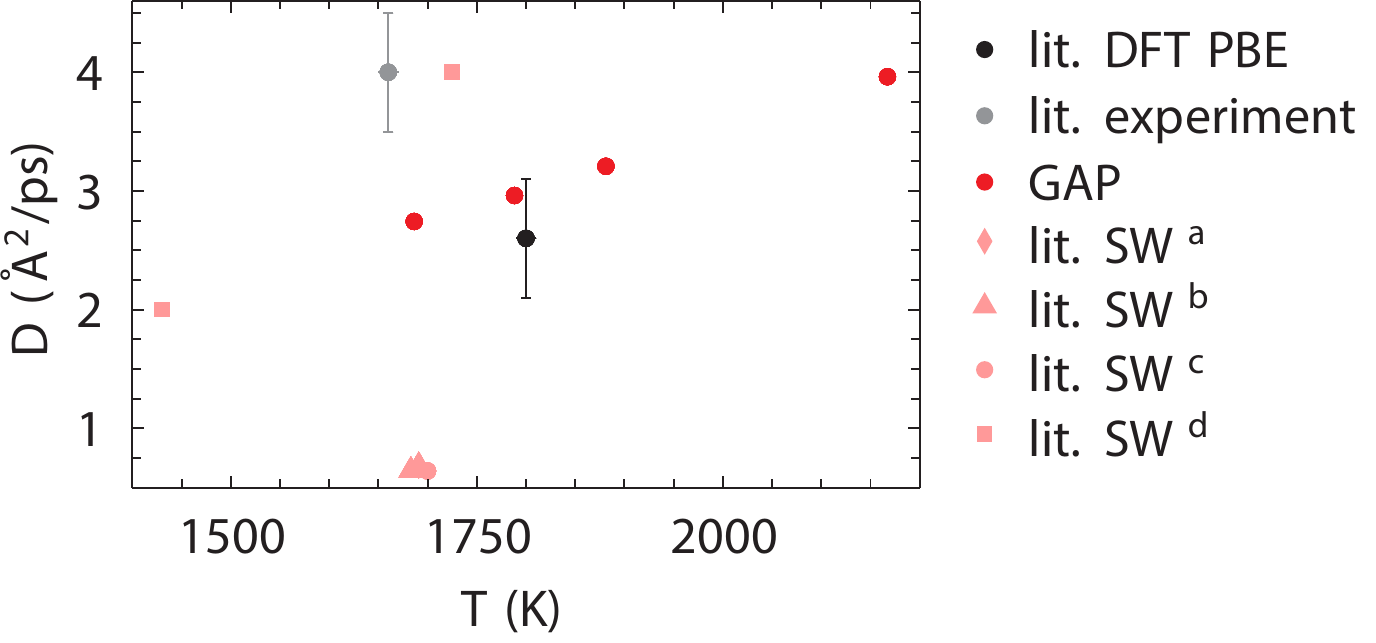}}
    \caption{Diffusivity of liquid silicon from literature DFT simulations~\cite{remsing_phys_rev_b_2017} (black),
    literature experiment~\cite{sanders_j_appl_phys_1999} (grey), GAP (red), and literature SW potential
    (Refs.~\onlinecite{broughton_phys_rev_b_1987}, \onlinecite{kakimoto_j_appl_phys_1998}, \onlinecite{yu_phys_rev_b_1996},
    and \onlinecite{sastry_nat_mater_2003} for $a$-$d$, respectively, pink). Error bars for GAP simulations
    are smaller than symbols on this scale, and were not available for literature SW results. }
    \label{fig:liquid_diffusivity}
\end{figure}

In addition to the two structural quantities we evaluated a
dynamical quantity, the diffusivity of liquid Si, by carrying out
variable cell size constant enthalpy MD simulations using the LAMMPS
software~\cite{plimpton_j_comp_phys_1995,lammps} on a 512 atom cell for 10$^5$ 1~fs time steps
at temperatures ranging from about 1700~K to 2200~K.
The resulting diffusivity  as a function of temperature
is shown in Fig.~\ref{fig:liquid_diffusivity}, and compared to the experimental results~\cite{sanders_j_appl_phys_1999},
DFT results~\cite{remsing_phys_rev_b_2017} (using the PBE GGA
exchange-correlation functional, which is somewhat
different from the PW91 functional we used to generate our fitting
database), and previously published SW potential
results~\cite{broughton_phys_rev_b_1987,kakimoto_j_appl_phys_1998,yu_phys_rev_b_1996,sastry_nat_mater_2003}.
The GAP results are in excellent agreement with DFT, and so both underestimate
the experimental diffusivity.  This difference relative to experiment has
previously been ascribed to the tendency for DFT to exaggerate the structure
of the liquid,~\cite{remsing_phys_rev_b_2017} and so the similar diffusivities of GAP and DFT are consistent
with the similarities in their liquid RDF and ADF.

\subsection{Amorphous phase}

Amorphous silicon is an interesting tetrahedrally coordinated phase that
forms upon various forms of processing, including ion implantation,
low temperature deposition, and rapid quenching from the melt.  The
last of these is commonly used in simulations, but it is challenging to
reach experimentally relevant cooling rates using
accurate methods such as DFT.  We therefore carried out zero pressure
variable cell volume (hydrostatic strain) simulations of
the quenching of a 216 atom sample of liquid Si, cooled at 10$^{12}$~K/s
from 2000~K to 500~K with a 1~fs time step ($1.5 \times 10^6$ steps) using the LAMMPS software,
and then relaxed to the local energy minimum with respect to atomic positions
and cell size and shape.  The initial configuration for all quenches
was from a GAP equilibrated liquid at $T=1800$~K, which was further equilibrated
with each potential at $T=2000$~K for an additional 10$^5$ time steps before cooling.
As for the liquid above, for some potentials this initial thermodynamic
state may be a strongly undercooled liquid due to their overestimation of
the melting temperature.

\begin{figure}
    \centerline{\includegraphics[width=\columnwidth]{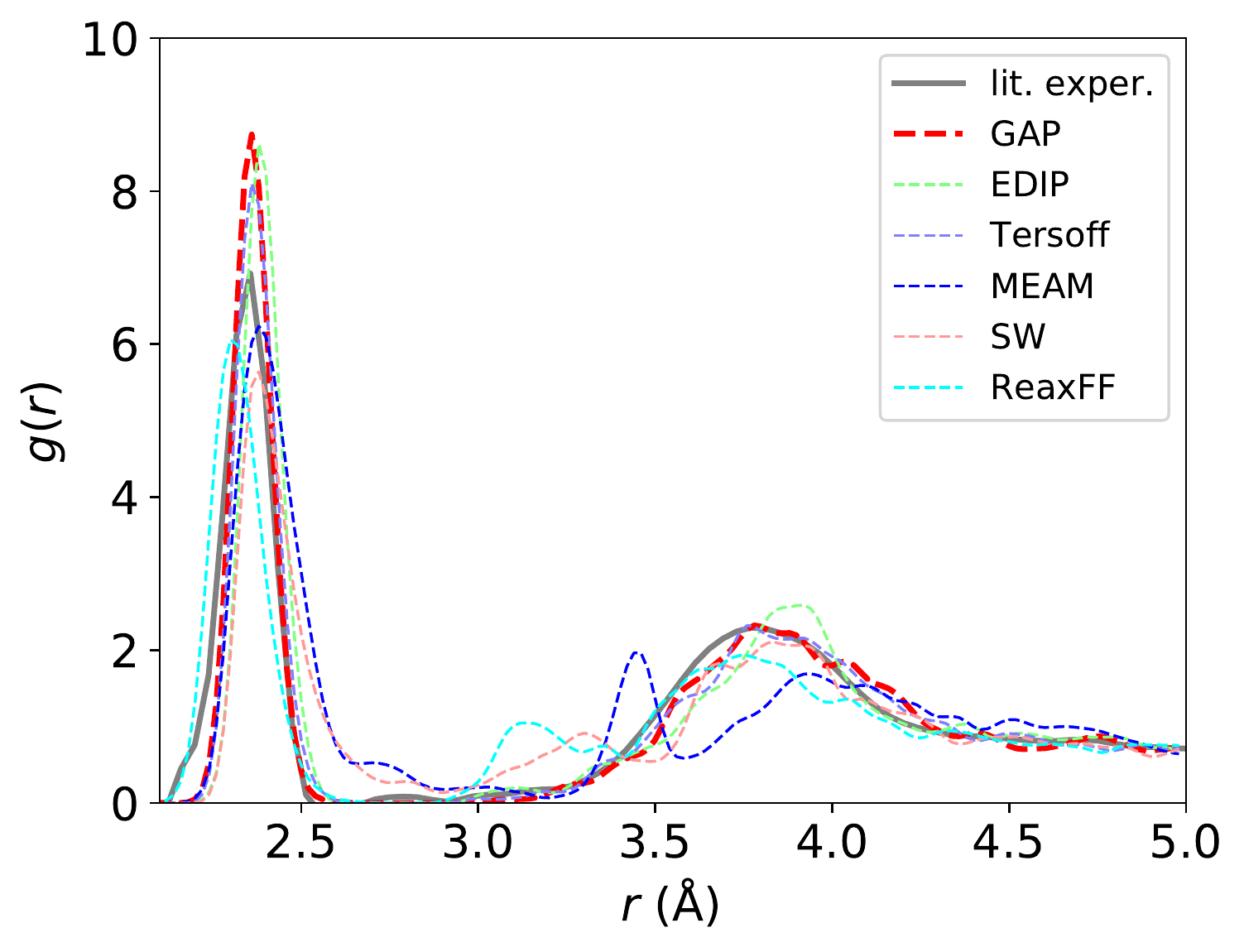}}
    \caption{Radial distribution function (RDF) for 216 atom amorphous
    configuration generated by cooling at 10$^{12}$~K/s from 2000~K to 500~K and then
    minimized, for GAP (red) and other interatomic potentials (other colors).
    Experimental results (grey) generated by ion implantation from
    Ref.~\onlinecite{laaziri_phys_rev_lett_1999} are shown for comparison.  }
    \label{fig:amorphous_rdf}
\end{figure}

The RDFs of the resulting structures are shown in Fig.~\ref{fig:amorphous_rdf},
in comparison with experimental results~\cite{laaziri_phys_rev_lett_1999}
(since DFT results for comparable sizes or quench rates are not computationally
feasible).  The various interatomic models vary widely in the overall shape
of their RDF, with GAP, EDIP and Tersoff in best agreement with experiment,
showing a sharp first neigbor peak at about 2.35~\AA, and a broad second
peak at about 3.8~\AA.  These three models have essentially no atoms between the
two peaks ($2.5$~{\AA}~$\lesssim r \lesssim 3.25$~\AA).   The other models show various
qualitative problems, including smaller peaks between the two expected ones,
or an excess of atoms throughout the entire distance range between the first and
second neighbour peaks.  The corresponding coordination statistics (using $r=2.75$~{\AA}
as the nearest neighbor distance cutoff) are shown in
Table~\ref{table:amorph_coord}.  The GAP and Tersoff models have the
lowest coordination defect concentration, significantly lower than the
next best model, EDIP, and closest to the experimental estimates of
$\leq 1$\%\cite{nygren_j_appl_phys_1991}.

Table~\ref{table:amorph_coord} also lists the amorphous-crystal energy difference $\Delta
E_{ac}$ relative to the diamond structure.  The obvious way to evaluate
the energy difference for each structure is to use the same interatomic
potential that was used to generate the structure, i.e. a calculation
that is entirely self-consistent for that potential.  This $\Delta
E_{ac}^\mathrm{IP}$ listed in the table shows GAP with the closest
value to experiment~\cite{laaziri_phys_rev_lett_1999} (excluding
MEAM, which has a very unphysical structure), while other interatomic
potentials result in higher energy differences.  However, using the
potential to evaluate the energy difference risks mixing up errors in
the structure with errors in the energy difference given the structure,
with the possibility of exaggerating or understating the stability of the
amorphous structure, depending on the sign of the energy error.  For an
independent evaluation of the quality of the quenched a-Si structures,
we evaluated their energies with DFT ($\Delta E_{ac}^\mathrm{eD}$),
and also further relaxed them with DFT ($\Delta E_{ac}^\mathrm{rD}$).
Note that these calculations were done with a lower k-point density,
0.07~\AA$^{-1}$, due to computational expense of the 216 atom cells.
In general the unrelaxed DFT energy shows a similar trend to the IP
energy, except for SW and MEAM, where the IP energy greatly underestimates
the (more reliable) DFT energy difference.  Relaxing the structure leads
to a small energy reduction for GAP as well as EDIP and Tersoff, indicating
a structure that is relatively close to the nearest DFT local minimum,
but much larger reductions for the other potentials.

All these DFT results show that quenching a liquid with GAP produces
the most stable a-Si structure with the lowest energy difference
relative to the diamond structure crystal as compared with the
other interatomic potentials, and that the GAP evaluated
energy of this structure is in good agreement with DFT.
Further work at lower quench rates will be required to generate structures that can be reasonably
argued to be directly comparable to experiment.\cite{aSi-preprint}

\begin{table}
    \caption{Coordination statistics $c_i$ (fraction of atoms with number of
    neighbors $i$ within $r_c = 2.75$~\AA, in percent) and energy per atom relative to diamond
    structure ($\Delta E_{ac}$, in eV) for amorphous structures resulting
    from quenching of the liquid.  Energy difference evaluated using interatomic
    potential is $\Delta_{ac}^\mathrm{IP}$, energy difference of interatomic-potential-relaxed
    structure evaluated (but not relaxed) using DFT is $\Delta E_{ac}^\mathrm{eD}$, and DFT-evaluated
    energy difference of DFT-relaxed structure starting from interatomic potential
    structure is $\Delta E_{ac}^\mathrm{rD}$.  Most atoms in MEAM structure have
    coordination $\geq 6$. Experimental defect density from
    Ref.~\onlinecite{nygren_j_appl_phys_1991} and energy from
    Ref.~\onlinecite{roorda_phys_rev_b_1991}.
    }
    \label{table:amorph_coord}
    \begin{tabular}{lrrrrrr}
        Model       & $c_3$  & $c_4$  & $c_5$   & $\Delta E_{ac}^\mathrm{IP}$ & $\Delta E_{ac}^{\mathrm{eD}}$ & $\Delta E_{ac}^{\mathrm{rD}}$ \\
        \hline
        lit. exper. &      &  $\geq 99$ &    & 0.137 \\
GAP & 1.4 & 98.1 & 0.5 & 0.15 & 0.14 & 0.13 \\
EDIP & 0.5 & 94.4 & 5.1 & 0.22 & 0.22 & 0.19 \\
Tersoff & 0.0 & 98.1 & 1.9 & 0.22 & 0.18 & 0.17 \\
MEAM & 0.0 & 0.0 & 2.8 & 0.14 & 0.65 & 0.28 \\
SW & 2.3 & 75.5 & 21.8 & 0.20 & 0.29 & 0.23 \\
ReaxFF & 0.0 & 86.1 & 13.9 & 0.35 & 0.35 & 0.25 \\

    \end{tabular}
\end{table}

\subsection{Phase diagram}
\begin{figure}
    \centerline{\includegraphics[width=\columnwidth]{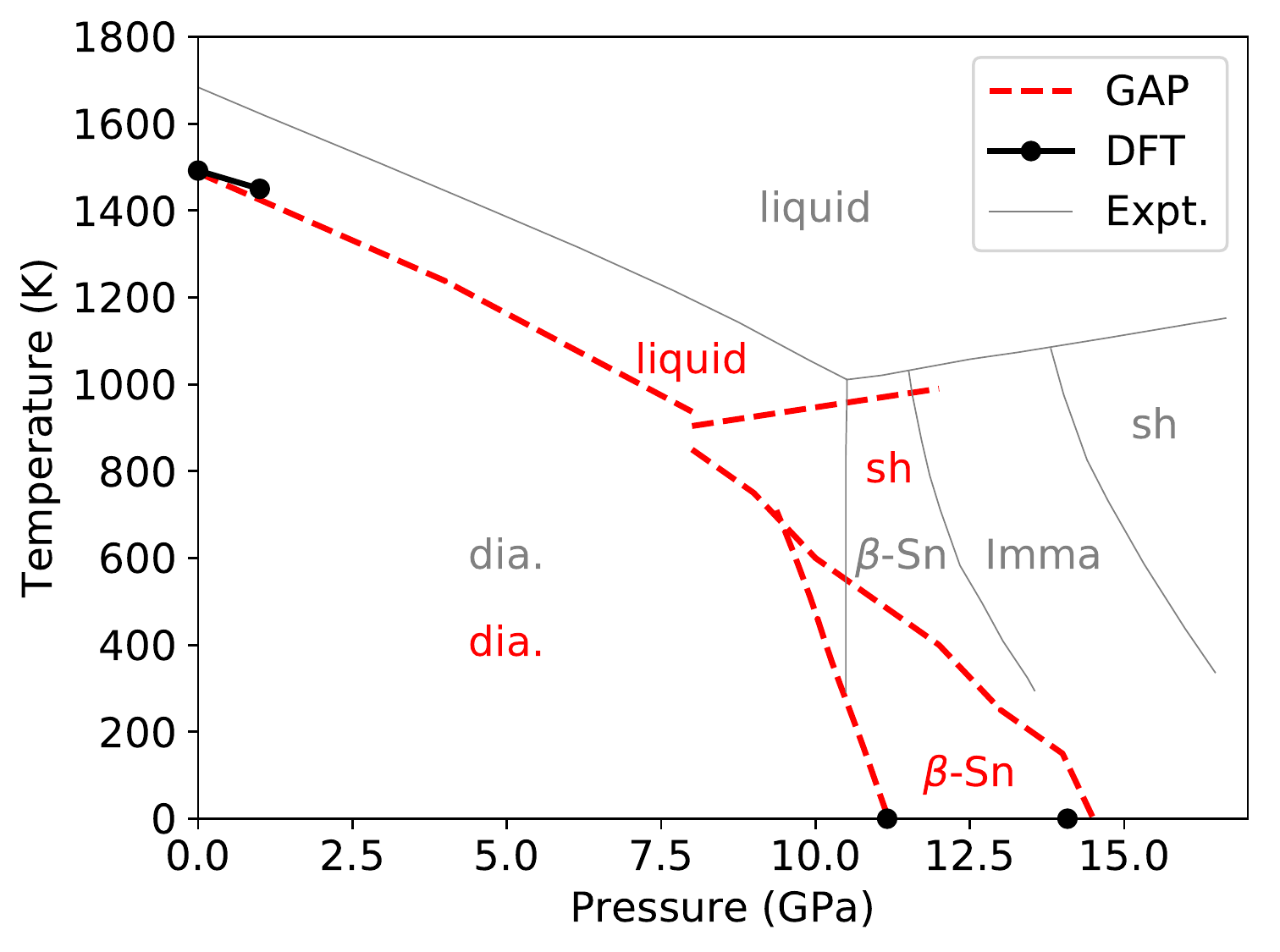}}
    \caption{Temperature-pressure phase diagram of silicon, computed with GAP (red), compared to available DFT results (black) and experimental phase transitions (grey). The finite temperature DFT result is from Ref.~\onlinecite{Alfe:2003gu}, and the experimental lines are from Ref.~\onlinecite{Voronin:2003dh}.}
    \label{fig:phase_diagram}
\end{figure}

The phase behaviour corresponding to an interatomic potential is a useful
benchmark: it not only informs the user about how realistic the model is, but
provides an indirect yet stringent test of the microscopic details of the PES.
The phase transitions result from a delicate balance between energetic and
entropic effects, and for finite temperature transitions
probe relatively high-energy configurations.  To
calculate the liquid-solid transition lines, we performed coexistence simulations for
the diamond and simple hexagonal structure at fixed
pressure and enthalpy, and measured the resulting average equilibrium temperature.~\cite{Morris:1994bi}
The diamond/liquid simulations contained
432 atoms and the pressure was fixed at the values of 0, 4 and 8~GPa, and
the simple hexagonal/liquid system contained 1024 atoms and the simulations
were carried out at 8 and 12~GPa.  To estimate the transition line between $\beta$-Sn and
simple hexagonal phases, we ran isothermic-isobaric molecular dynamics
simulations of both pure phases in a temperature range of 0-1000~K and pressure
range of 6-14~GPa, and observed the transition (which occurred in both directions in all cases)
by monitoring the Steinhardt bond-order parameters~\cite{Steinhardt:1983hz}. Finally, the
transition line between diamond and $\beta$-Sn structures was determined by calculating the
Gibbs free energy using the quasi-harmonic approximation (QHA). We also established that in these phases anharmonic
contributions to the free energy differences are negligible at 0~K. We used the
LAMMPS package for the MD simulations, and phonopy\cite{phonopy} for the phonon
calculations.  Figure~\ref{fig:phase_diagram} shows the calculated phase
diagram, compared to the published DFT results for the diamond/liquid melting
point\cite{Alfe:2003gu} and our own calculations with the Castep program for the
diamond/$\beta$-tin and $\beta$-tin/simple hexagonal transition pressures at
0~K. For comparison, we also show the experimentally determined phase
relations\cite{Voronin:2003dh}.  Note that the Imma phase is missing from the
calculated phase diagram. This is due to the fact that both our DFT calculations
and GAP model find the Imma phase to be metastable.

\subsection{Defects}

\subsubsection{Point defects}

Several point defects were represented in the fitting database (Table~\ref{table:sidatabase}), and their
formation energies would therefore be expected to be accurately reproduced
by the GAP.  Indeed, as Fig.~\ref{fig:big_fractional_error_plot} shows,
the relative error for the vacancy and three interstitial positions,
hexagonal, tetrahedral, and dumbbell, are all within at most 7\% of the
reference DFT values.  The only other potential that is close to this
level of accuracy is EDIP, with similar errors for all but the hexagonal
interstitial, where it is off by 14\%.  All the other potentials, as
well as DFTB, differ from our DFT calculations by tens of percent for
at least some of the defects.

Since point defects control properties such as diffusivity in bulk
silicon, their migration barriers are also of interest, and as
they represent bond breaking and formation processes, often present
a challenge for interatomic potentials.  Since the training database
configurations came from finite temperature MD, it
could in principle include configurations near the barrier, but since the
system spends relatively little time near the energy saddle point this is
actually unlikely~\cite{lamw2017q}.  However, the hexagonal and tetrahedral interstitials
are related by a short displacement, so one is typically a local minimum
and the other a saddle point along an interstitial diffusion pathway.
We find that GAP preserves the DFT ordering, although the energy
difference is underestimated, while the other potentials make much
larger errors, many reversing the relative order of the two high
symmetry geometries.  Two other related observables, the migration path of
the vacancy and the formation energy of the four-fold defect (the midpoint
of the concerted-exchange diffusion mechanism~\cite{Pandey:1986hc,Kaxiras:1993cb}),
which are not represented in the database, are discussed below in
Sec.~\ref{sec:validation:fourfold} and Sec.~\ref{sec:validation:vac-path}.

\subsubsection{Surfaces}

Surfaces are a class of defects that have particular importance for the behaviour of materials. Solids fail under tension by opening new surfaces, and it is on surfaces that reactions involving chemical species in the environment can
take place, where special functional layers can form e.g. by oxidation, and also where a crystal can grow under suitable conditions. Apart from useful applications, a rich complexity of bonding emerges on surfaces due to the subtle interplay of strain effects with the chemistry of dangling bonds. This makes surface formation energies, and particularly the energies and geometries of various reconstructions, a sensitive test of the accuracy of an interatomic potential.

\begin{figure}
\centerline{\includegraphics[width=\columnwidth]{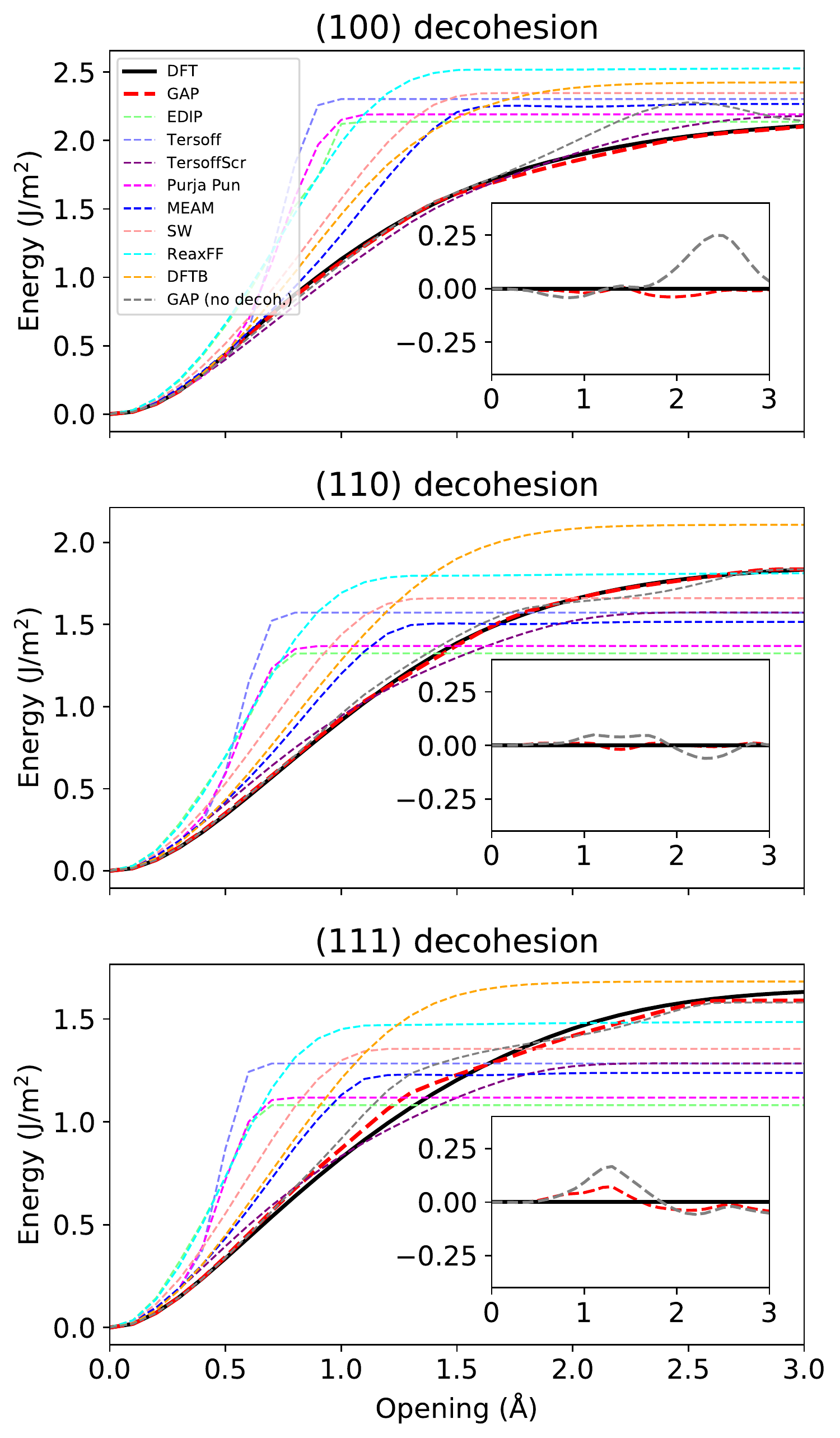}}
\caption{Decohesion energy of diamond-structure silicon along various directions (labelled according to the orientation of the opening surface).
Insets (same axes as main plot) show errors with respect to DFT for the current GAP (red) as well as for a previous version of the GAP (grey)
with a fitting database that did not include any
configurations along the separation path (or high energy crystal-lattice structures), but only final fully seprated surfaces.
}
\label{fig:decohesion}
\end{figure}

Figure~\ref{fig:decohesion} shows the energy as a function of separation
as a gap is opened up in a unit cell that is long in one direction and has the dimensions of the minimal surface unit cell in the orthogonal plane. For the purposes of this test, the atomic positions were not relaxed, but kept rigid relative to one another as the gap was opened. All analytical potentials apart from the screened Tersoff show far too short a range - they plateau much earlier than DFT, and in fact this was one of the motivating factors behind modifying the original Tersoff potential\cite{Pastewka2008,pastewka_phys_rev_b_2013}. 
The right hand side limit corresponds to the unrelaxed surface energy in each case, a property in which the potentials show about 30\% scatter. Note that in the case of the $(111)$ surface, DFT is believed to overestimate the surface energy\cite{Jaccodine:1963ks} and, e.g., Tersoff and its screened version were explicitly fit to reproduce the experimental value.

Note that the final version of the fitting database for the GAP
presented here includes configurations along the separation path, in
addition to fully separated surfaces.  An earlier version of
the GAP model~\cite{Bartok:2017hz} that did not include configurations
from the separation path correctly reproduced the {\em
fully separated} energy (since fully separated surfaces were included in the
fitting), but not the intermediate energies, as shown in the insets
in Fig.~\ref{fig:decohesion}.  The test results for the version of the
potential without the decohesion path configurations (as well as fewer
non-diamond crystal structures), listed in detail in the Supplemental
Information\cite{supplemental}, were very close to the final GAP values, with most tested
quantities differing by less than 1\%.
The only exceptions were the quantities directly related to configurations
newly added to the database, and a few other tests (described later in this subsection and in Section.~\ref{sec:diinter}) that were not explicitly fit
(two di-interstitial formation energies that changed by 3\% and 10\%, and
the (111) reconstructed surface energies that changed by 2\%-3\%).
This example shows that the flexibility of the GAP functional form makes
it possible to correct shortcomings by adding configurations to the database
without significantly affecting accuracy for other configurations.

\begin{figure}
\centerline{\includegraphics[width=4cm]{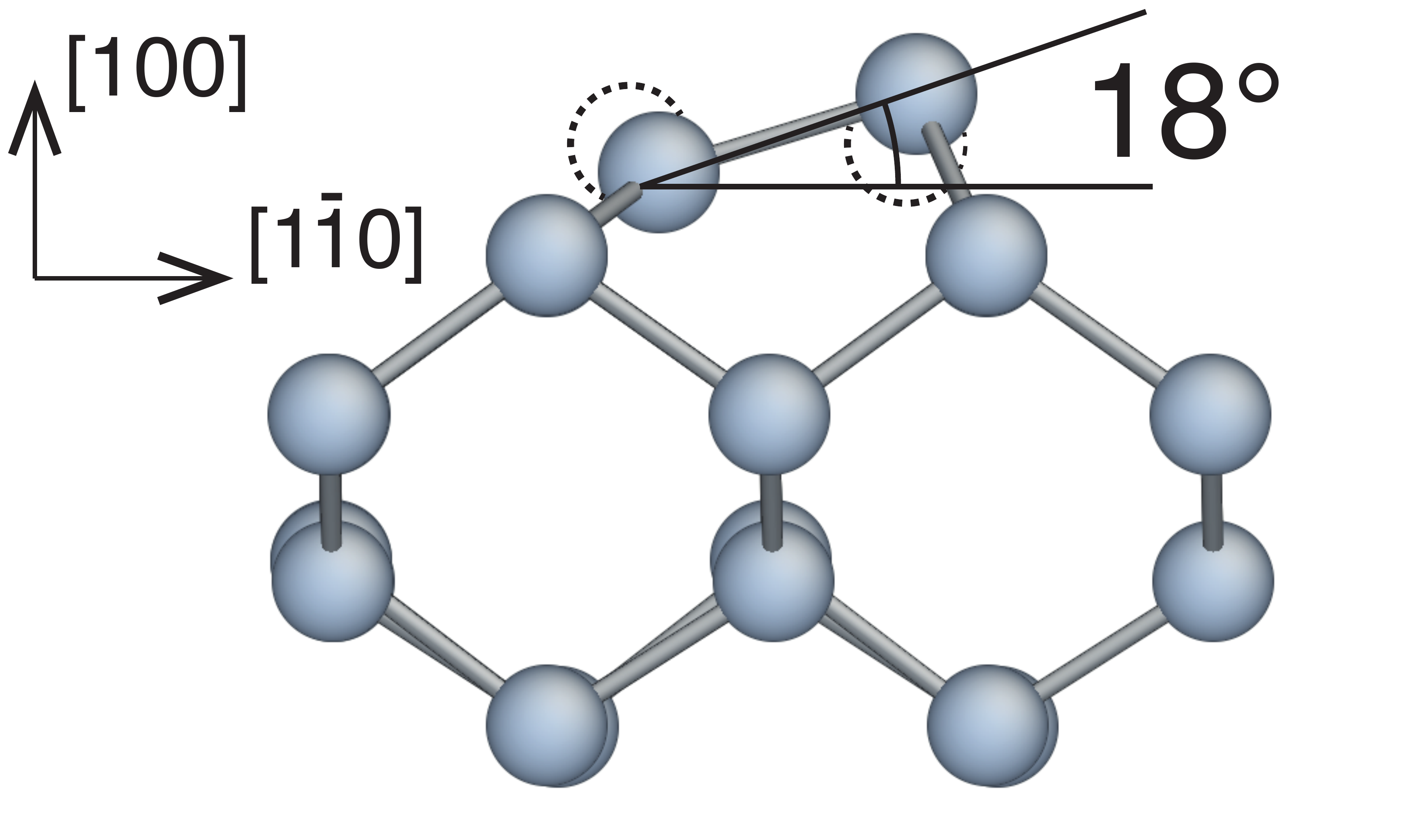}}
\caption{Geometry of the 2x1 reconstruction of the (100) surface, showing the tilting of the surface dimers. The dashed circles show the surface dimer in the untilted (but rebonded) position.}
\label{fig:surfacerecon100}
\end{figure}

Figure~\ref{fig:surfacerecon100} shows the geometry of the tilted-dimer
$2\times1$ reconstruction, one of the low energy configurations of the (100)
surface, which forms spontaneously from the as-cut surface.  In this
reconstruction the surface atoms dimerize to form additional bonds,
and the dimers tilt (by $18^\circ$ in our DFT calculations) due to a
Jahn-Teller effect\cite{Haneman:1987vr}, which would seem to require an
explicit description of the electrons.  In fact, none of the analytical
potentials reproduce the substantial tilting (zero tilt for all but EDIP,
which tilts by $4^\circ$).  Only GAP, with its relatively long range
and flexible form, captures the tilting in reasonable agreement with
DFT (-$2.5^\circ$ error).  The DFTB model, with its minimal description of
electronic structure also shows the breaking of symmetry with a similar error
on the resulting bond angle of about -$2.3^\circ$.

\begin{figure}
\centerline{\includegraphics[width=8.5cm]{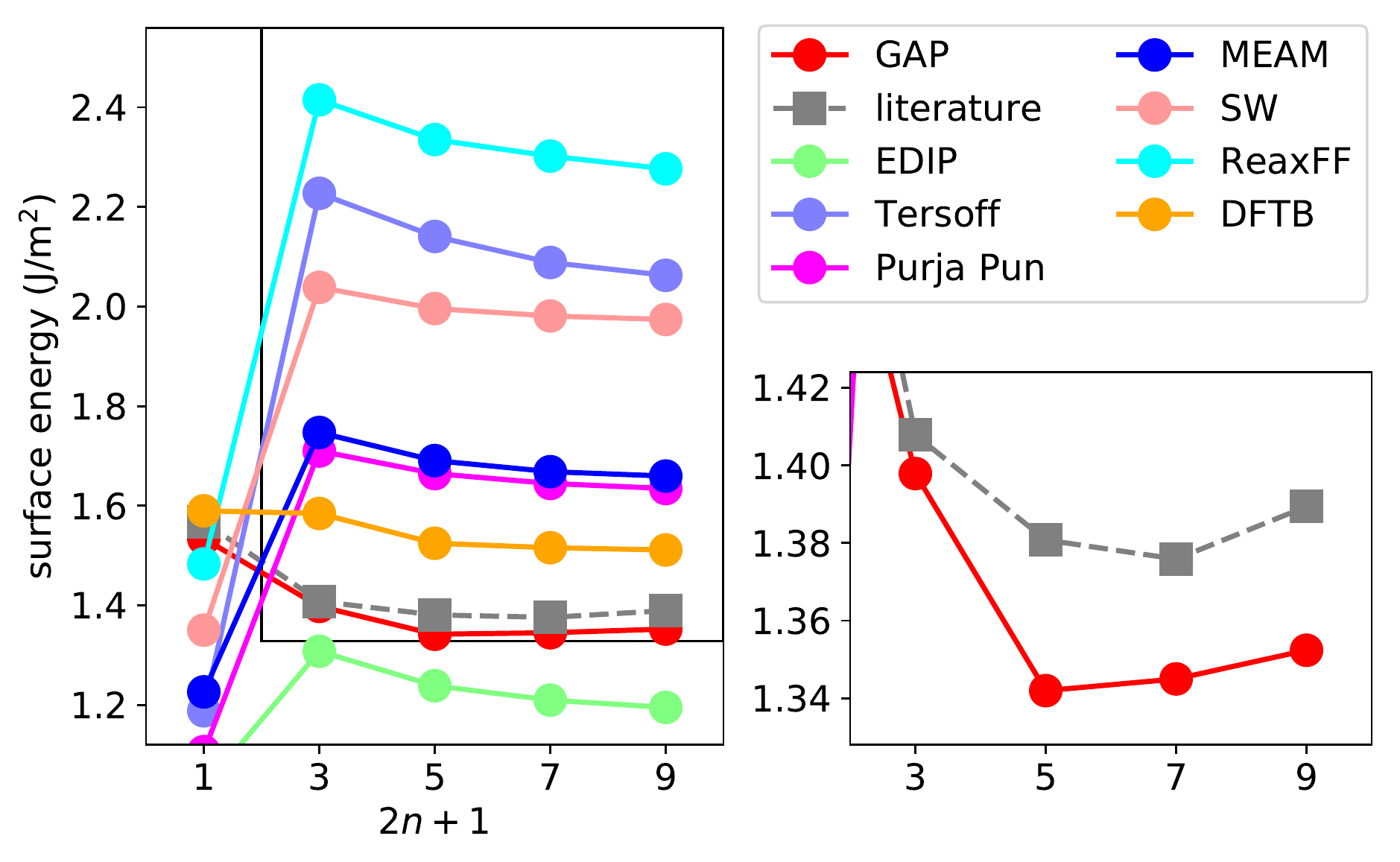}}
\caption{Formation energy of the dimer-adatom-stacking-fault (DAS)
reconstruction of the (111) surface for various surface unit cell
sizes $n=(3,5,7,9)$ computed with different models. The value shown
at $n=1$ corresponds not to a DAS reconstruction, but rather the
unreconstructed surface. The box on the lower right is a magnified
view that shows just the DFT and GAP results. }
\label{fig:surfacerecon111}
\end{figure}

\begin{figure}
\centerline{\includegraphics[width=4cm]{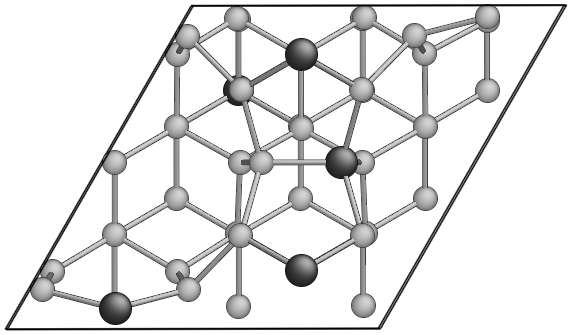}\includegraphics[width=4cm]{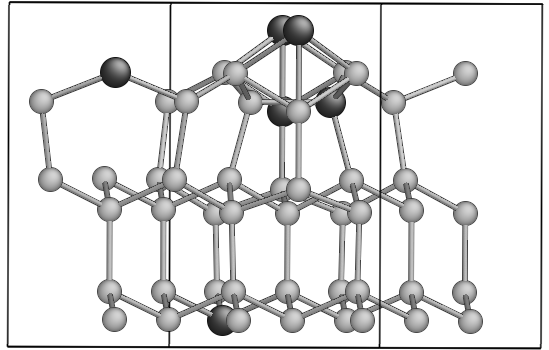}}
\caption{Two views of the DAS $3 \times 3$ (111) surface reconstruction configuration that is in the database with atoms, marked with dark grey, whose environments were selected to be amongst representative set for the purposes of defining the GAP model.
}
\label{fig:surfacerecon111_das3x3}
\end{figure}

The lowest energy configuration of the $(111)$ surface is the famous
$7\times 7$ dimer-adatom-stacking-fault (DAS) reconstruction,
already alluded to in the introduction. It is a rather complex
structure, involving a 2D super-lattice of 10-atom rings, connected
by dimerized dislocation cores that separating triangles of stacking
faults, half of which have extra atoms on top. A family of analogous
structures can be defined by varying the number of dimers, $n \ge 3$,
between the vertices of the super-lattice, leading to the designation
$(2n+1)\times(2n+1)$. As shown in Fig.~\ref{fig:surfacerecon111},
all analytical potentials predict these reconstructions to be higher
in energy than the unreconstructed surface (shown in place of $n=1$
for simplicity), and furthermore, within the family of DAS structures,
the energy goes down as $n$ goes up. Computing accurate DFT energies
is a nontrivial calculation, and its prediction that the $7\times7$
DAS structure is the lowest energy configuration was a significant early
triumph of DFT~\cite{Stich:1992hh}. Here our reference is a more recent
careful determination of the DFT energies\cite{Solares:2005gm}. The
DFTB model again stands out as qualitatively different from the
analytical models, but still failing to show quantitative agreement
with DFT. The GAP model, which includes in its training database
just a single configuration of the $3\times3$ DAS structure (shown in
Fig.~\ref{fig:surfacerecon111_das3x3}), gives energies with error
below 0.05~J/m$^2$ (much smaller than a meV/atom over the supercell),
correctly predicting the DAS family to be lower in energy than
the unreconstructed surface, and also giving an energy minimum.

The lowest energy structure for the present potential happens to be
for $n=5$, within 0.01~J/m$^2$ of the $7\times7$ structure. The energy differences are
much smaller than the target (and assumed) error in the GAP model,
and as such this level of detail is not robust: the earlier variant of the potential
fitted to a slightly different database (in ways unrelated to the (111)
surface) show the $7\times7$ DAS structure as the global minimum, as
shown before\cite{SciAdvpaper}. What {\em is} robust is the relationship
of the energies of the DAS family to other types of reconstructions,
and the upturn in energy for $n=9$. Significantly more data relevant to
these structures would be needed in order to robustly capture the finest
of relative energies within the DAS family.

Figure~\ref{fig:surfacerecon111_das3x3} shows which atoms were picked (automatically, by the CUR decomposition of the descriptor matrix, as mentioned above) to be part of the representative set: mostly those that are unique to the DAS family of reconstruction and do not appear elsewhere in the dataset, i.e. the adatom, the atom just below it, one of the dimer atoms in at the boundary of the stacking fault, and one atom on the 10-ring that surrounds the vertices of the surface unit cell.

\subsection{Crack Propagation}

The atomic-scale details of crack propagation have proved particularily
challenging to model, since  sufficent accuracy to describe bond
breaking processes must be combined with large model systems to
avoid unrealistic strain gradients~\cite{Bitzek2015}. Interatomic
potentials which provide an otherwise good description of the bulk
and elastic properties of silicon (e.g. the Stillinger-Weber and
Tersoff potentials) tend to overestimate the lattice trapping
barriers to brittle fracture, resulting in an overestimate of
the fracture toughness as well as an erroneously ductile material
response including features such as crack arrest and dislocation
emission~\cite{Swadener2002,Holland1998,Holland1998b}. Progress has
been made using reactive potentials such as ReaxFF~\cite{buehler_phys_rev_lett_2006}
or with hybrid quantum/classical approaches where an \emph{ab
initio} crack tip model is embedded within a larger classical model
system~\cite{Bernstein2003,Kermode2008,Gleizer2014}. The latter limits
the applicability to timescales accessible to DFT, making it extremely
challenging to study processes such as thermally activated crack
growth~\cite{Kermode2015}.

\begin{figure}
    \includegraphics[width=\columnwidth]{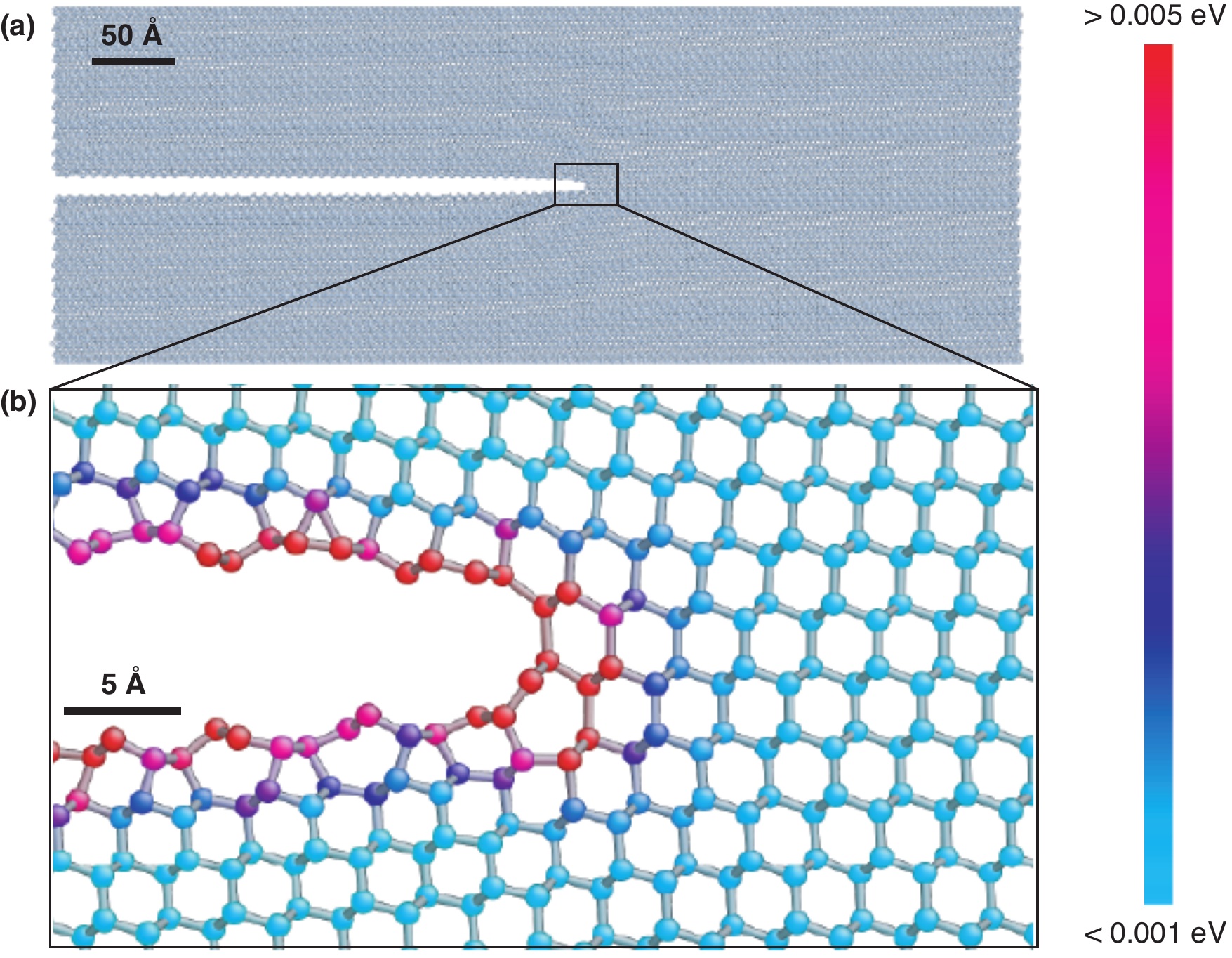}
    \caption{
    (a) Snapshot from a quasi-static simulation of fracture in the Si$(111)[1\bar{1}0]$ cleavage system at a strain energy release rate of $G=5.13$~J/m$^2$. Model system contains 23,496 atoms and has dimensions $600\times200\times3.86$~\AA$^3$. (b) Close up of the crack tip, which has undergone a crack tip reconstruction as previously reported in DFT-based hybrid simulations in Ref.~\onlinecite{Kermode2008}. Atoms are coloured by the predicted error per atom of the GAP model, from blue (low) to red (high).
    \label{fig:111-fracture}}
\end{figure}

To test the accuracy of our new GAP model for fracture, we considered
the well studied $(111)[1\bar{1}0]$ cleavage system, where fracture is
known to exhibit a low speed instability triggered by the formation of a
crack tip reconstruction~\cite{Kermode2008}. We performed simulations of
a 23,496 atom model system of dimensions $600\times200\times3.86$~\AA$^3$
using both molecular dynamics at 300~K with a range of strain rates between $10^{-6}$
and $10^{-4}$ fs$^{-1}$, and quasi-static strain increments followed by relaxation.
In all cases the trajectories obtained
were consistent with those expected from our earlier DFT-based hybrid
simulations as reported in Ref.~\onlinecite{Kermode2008}. The GAP model predicts
brittle fracture morphology with an atomically smooth fracture surface
and the occasional formation of crack tip reconstruction and subsequent
surface steps in the `downward' $[111]$ direction, in line with the
results of our previous study. A snapshot from a quasi-static simulation showing
the formation of the crack tip reconstruction at a strain energy release
rate of $G=5.13$~J/m$^2$ is illustrated in Fig.~\ref{fig:111-fracture}.
The atoms are coloured by the predicted error of the GAP model, showing
high confidence in the bulk and with larger predicted errors at the
crack tip; nevertheless, qualitatively correct surfaces and crack tip
reconstructions are obtained.

\section{Results: Validation}
\label{sec:validation}

In addition to the tests presented in the previous section,
we tested quantities and configurations that are physically important but do not
map so cleanly to particular geometries in the database.  The first
is random structure seach, which probes a very wide range of geometries,
bonding topologies, and energies.  The second is a test of the vibrational
properties (harmonic phonons and anhramonic Gr\"uneisen parameters)
of the diamond structure, which are only implicitly included in the
fit through the perturbed diamond configurations.  Finally two types of
defects were tested, a high-symmetry grain boundary and di-interstitials,
which have geometries related to, but clearly different than, the defects
in the fitting database.

\subsection{Random structure search}

\begin{figure}
    \centerline{\includegraphics[width=\columnwidth]{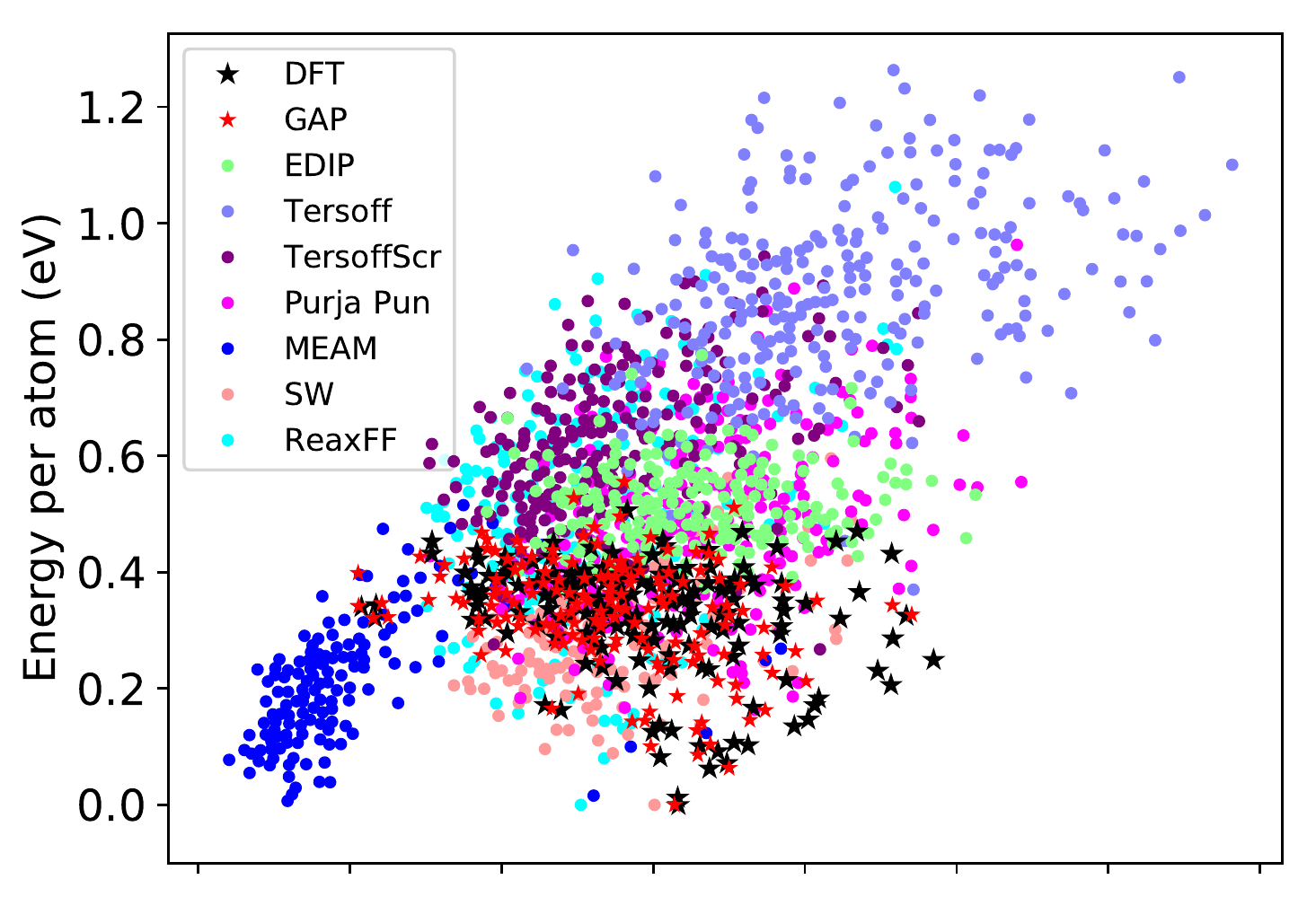}}
    \centerline{\includegraphics[width=\columnwidth]{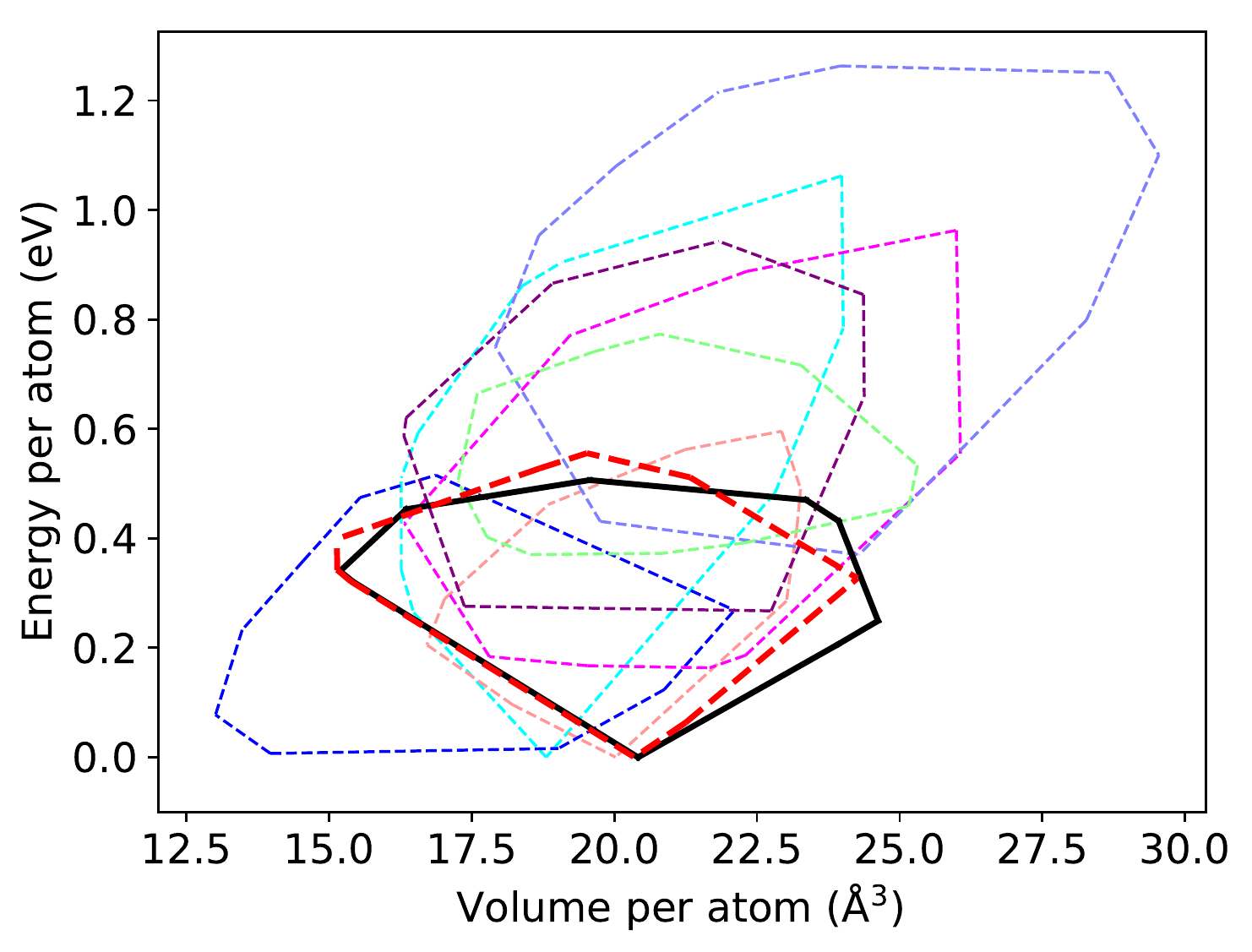}}
    \caption{Relaxed volumes and energies (relative to diamond structure) for random structure
        searches.  Top panel shows scatterplot with DFT (black stars), GAP (red stars), and various
        other interatomic potentials (various color circles).  Bottom panel shows convex
        hull surrounding all minima for each method with same x-axis and colors as top panel.  }
    \label{fig:RSS_E_V}
\end{figure}

The random structure search~\cite{Pickard2006,Pickard2011} (RSS) method provides a global
test of the potential energy surface, including not only regions
near the physically reasonable minima (i.e. typical bulk lattices
with small distortions and defects that vary only locally from the
bulk structure), but also much more distorted and correspondingly
higher energy configurations.  We carried out RSS using the various
interatomic potentials and DFT for 8 atom unit cells with constraints
on the initial shape (close to cubic) and interatomic distances
($>1.7$~\AA) to exclude unphysically close atoms, relaxed with the two-point
steepest descent~\cite{barzilai_ima_j_numer_anal_1988} method.  The resulting distribution
of configuration energy and volume are plotted in Fig.~\ref{fig:RSS_E_V}.
The GAP results show a similar distribution to DFT, with the
diamond structure at the correct volume, a few structures with energies up to
0.2~eV/atom higher, mostly at comparable or somewhat larger volumes (with
one or two exceptions at substantially smaller volume), and a large group at
more than 0.2~eV/atom higher at comparable or smaller volume.  None of the
interatomic potentials give a similar distribution, and some of the more
sophisticated ones give drastically different distributions, including
low energies for extremely small (25\% below diamond structure) volumes,
or unphysical local minima at very high energies ($> 0.4$~eV/atom).

\begin{figure}
    \centerline{\includegraphics[width=\columnwidth]{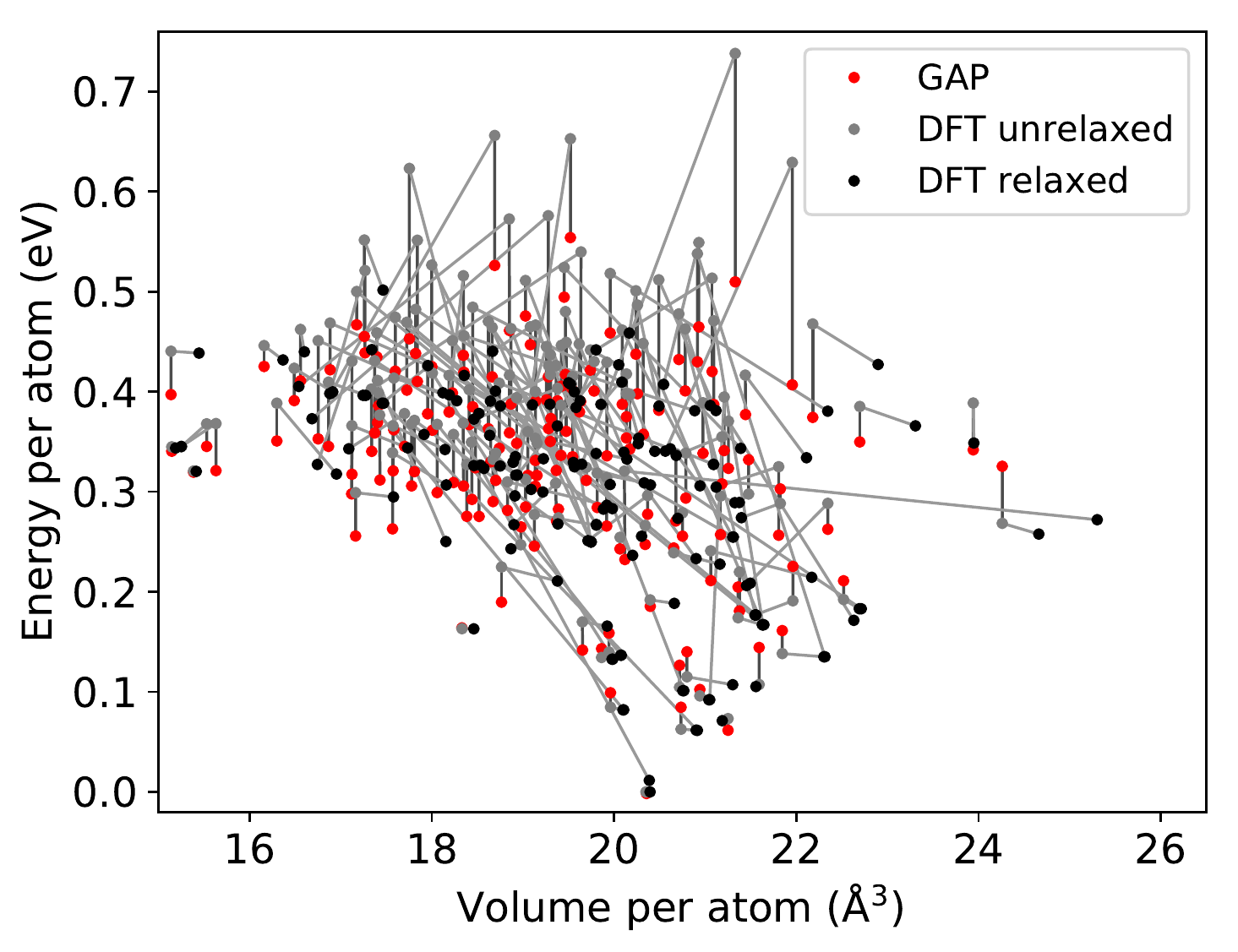}}
    \caption{Energies and volumes of relaxation of GAP RSS minima with
    DFT forces and stresses.  Red circles indicate GAP minima, grey circles indicate DFT
    energies of GAP minima configurations, and black circles indicate
    energy and volume of DFT relaxed configurations starting from the corresponding
    GAP minimum. }
    \label{fig:RSS_dogleg}
\end{figure}

\begin{figure}
    \centerline{\includegraphics[width=\columnwidth]{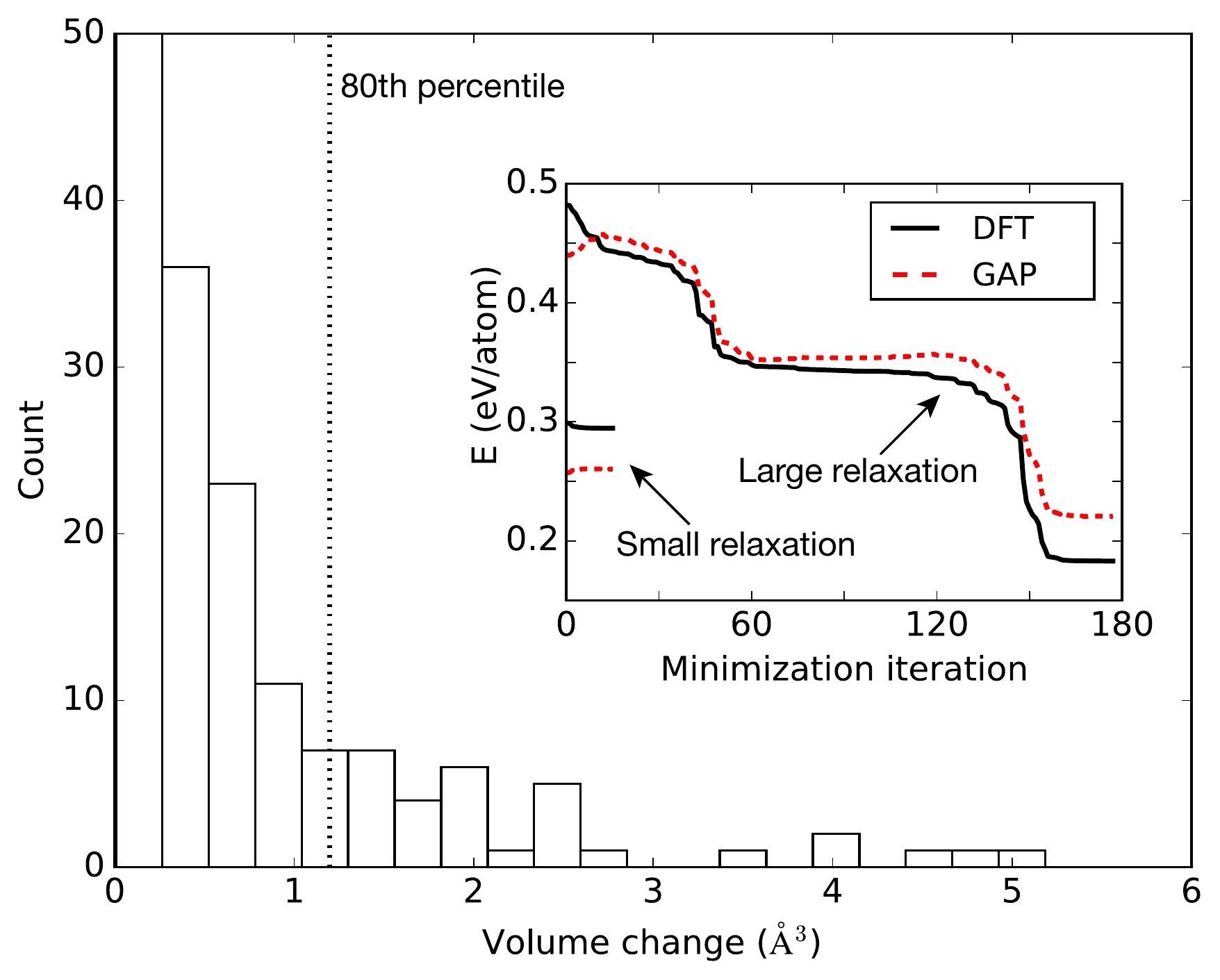}}
    \caption{Distribution of volume changes for relaxation of GAP minima
    using DFT.  Dotted vertical line indicates the 80'th percentile.
    Inset shows energy per atom for trajectories generated by DFT relaxation
    starting from GAP RSS minimum, computed with DFT (black solid line) and GAP (red
    dashed line), for one configuration with a small volume change and
    one with a large volume change.}
    \label{fig:RSS_rerelax_volume_changes_distrib}
\end{figure}

While the distribution of energies and volumes for the GAP relaxed
minima is similar to that of the DFT relaxed ones, that
does not necessarily mean that individual minima predicted by GAP
are also DFT minima.  To test this, we further relaxed
the GAP minima using DFT, and plotted the resulting positions
on the (E,V) plane in Fig.~\ref{fig:RSS_dogleg}.  The plot shows
that in many, although clearly not all, cases the GAP energy for
GAP minimum configurations is close to the DFT energy for the same
configuration, and further relaxation with DFT does not change the
volume or energy very much.  The distribution of volume changes, shown
in Fig.~\ref{fig:RSS_rerelax_volume_changes_distrib}, confirms that
most volume changes are small, with 80\% falling below 1~\AA$^3$/atom,
or about 5\% of the diamond structure volume.

A better understanding of the large volume changes that sometimes occur
during DFT relaxation comes from calculating the GAP energy for the DFT
relaxation trajectory configurations.  Two examples of these energy
variations are shown in the inset of Fig.~\ref{fig:RSS_rerelax_volume_changes_distrib}.  For the
configuration with a small overall volume change,
the DFT energy goes down a bit and quickly flattens as the DFT minimum
is reached, while the GAP energy on the same trajectory goes up a bit
(as it must, since the initial configuration is a GAP local minimum), and
also flattens.  On the other hand, for the configuration with the large volume change, the DFT energy does not flatten immediately,
but instead goes through a sequence of drops and flat regions as it
explores various near minima on the PES.  While the initial GAP local
minimum is not also a DFT local minimum, the barrier that separates it
from other minima on the GAP PES is small, and all of the configurations
that the DFT minimization trajectory goes through have very similar energies (within
0.05~eV/atom) with DFT and GAP.  This shows that while the GAP PES is
not perfect, differing in the positions and height of some small energy
barriers, the overall shape of the PES is in fact in good agreement
with DFT.

\subsection{Phonons and thermal expansion}

\begin{figure}
   \centerline{\includegraphics[width=\columnwidth]{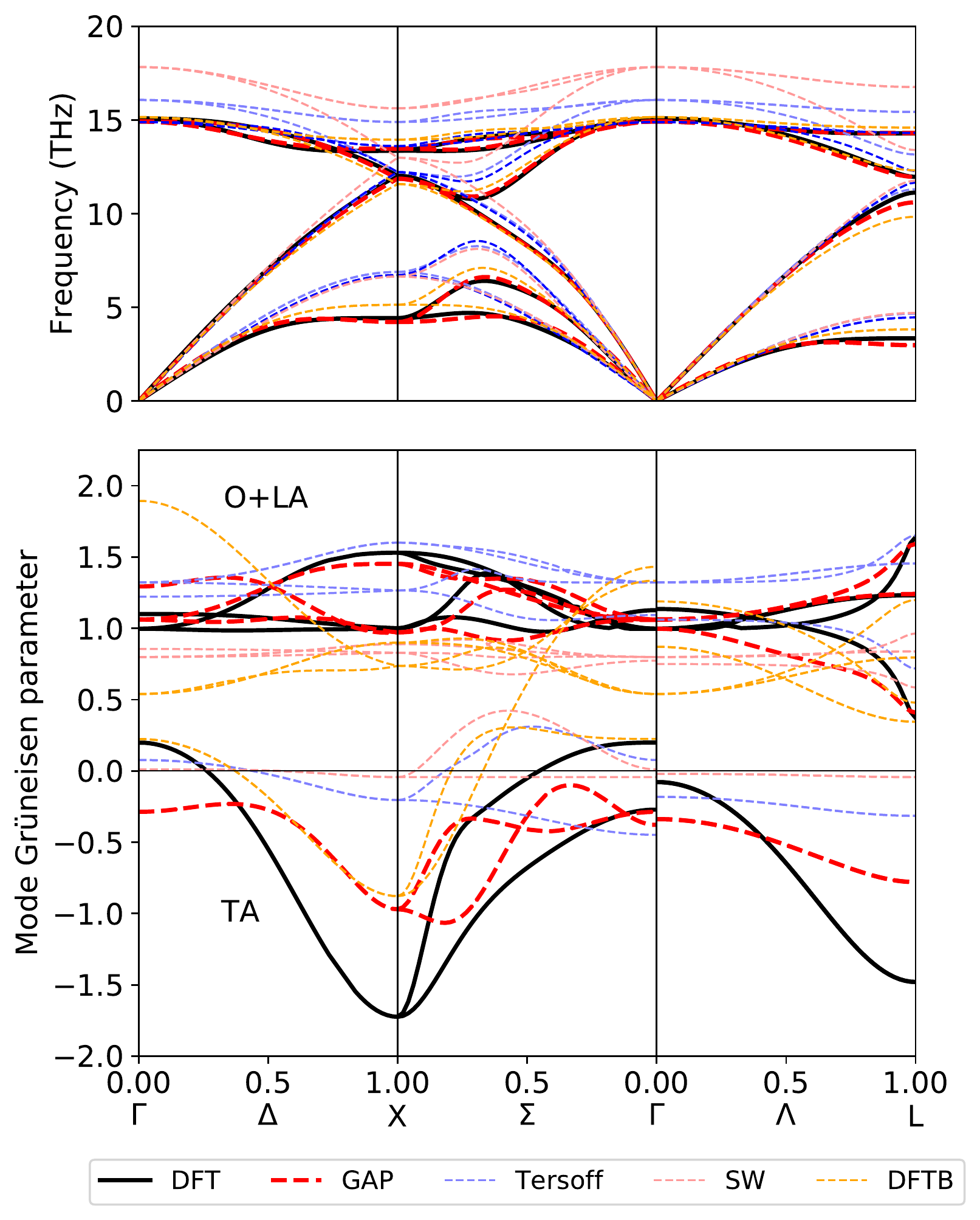}}
   \caption{Top panel: phonon dispersion curves of cubic diamond silicon at 0~GPa with various models.
   Bottom panel: dispersion curves of mode Gr\"uneisen parameters.}
   \label{fig:phonon}
\end{figure}

Vibrational properties probe the PES in the region close to the minima, and
influence the thermodynamic and transport behaviour of the material. We computed
the phonon and mode Gr\"uneisen dispersion curves in the cubic diamond structure with GAP, DFT,
DFTB and various interatomic potentials using phonopy\cite{phonopy}. The results are shown in
Fig.~\ref{fig:phonon}. Even though the phonon frequencies were not included in
the database explicitly, there is an excellent agreement between DFT and GAP.
The analytical interatomic potentials are generally in good qualitative
agreement for the phonon spectrum, although they overestimate the acoustic
branch zone-edge and all optical branch frequencies, while DFTB is
in significantly better agreement with DFT.  Not unexpectedly, the slope of
the dispersion curves at small k-vectors is well reproduced, corresponding to
the generally good agreement of the elastic constants of the examined potentials
with DFT.
GAP Results for the mode Gr\"uneisen parameter are more mixed, with the transverse acoustic
branch showing a large discrepancy, indicating that the force data of
near-equilibrium  crystalline configurations is not sufficient for the fitting
procedure to resolve the anharmonicity of the PES.  The analytical potentials and DFTB,
on the other hand, differ qualitatively from the DFT calculation for all branches.

\begin{figure}
   \centerline{\includegraphics[width=\columnwidth]{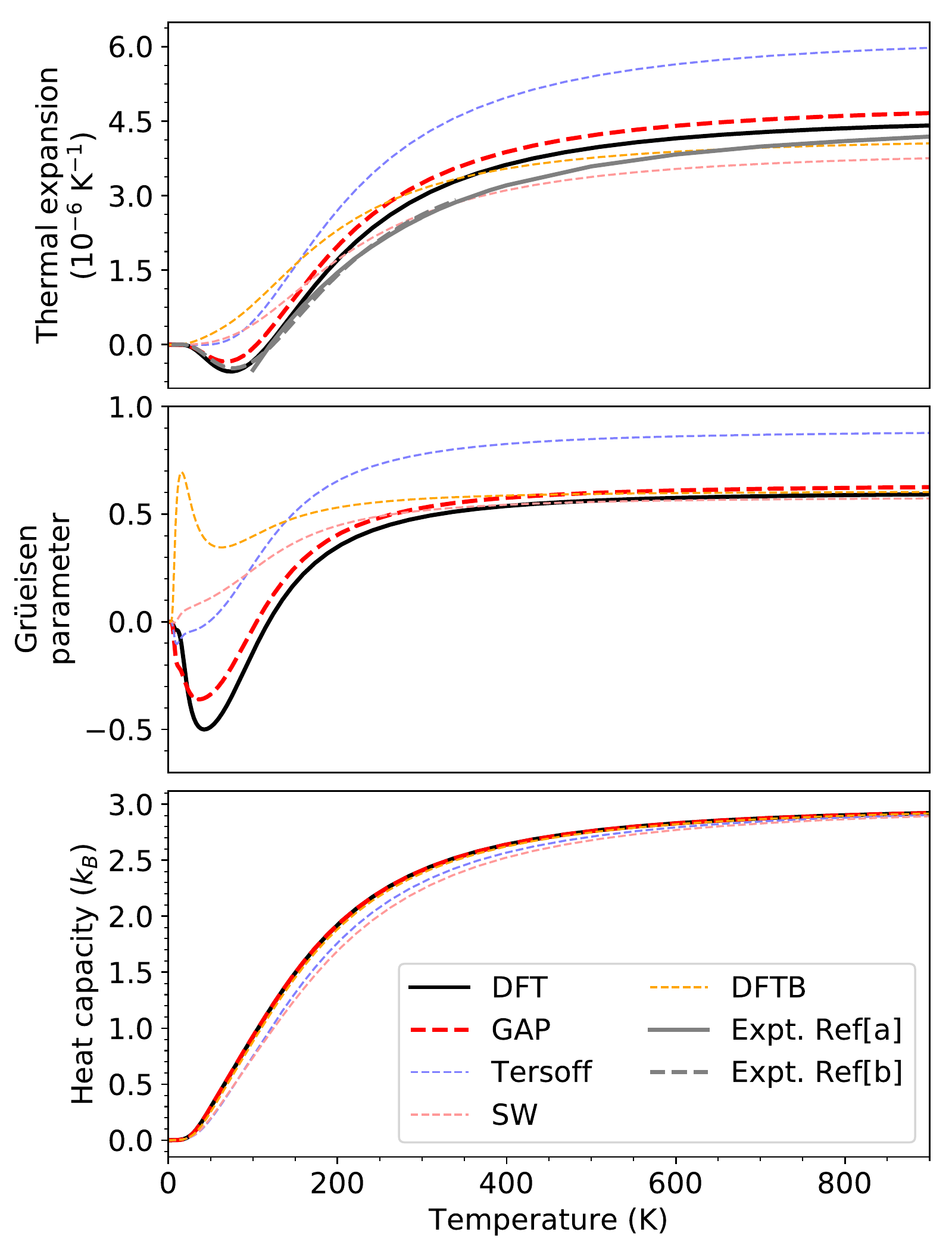}}
   \caption{Linear thermal expansion, Gr\"uneisen parameter, and specific
   heat of silicon computed with various models,
   using the quasi-harmonic approximation, compared to experimental results: \\
   a: Lyon et al\cite{Lyon:1977en}\\
   b: Okada and Tokumaru\cite{Okada:1984cz}.\\
   }
   \label{fig:thexp}
\end{figure}

Diamond structure silicon displays negative thermal expansion at low
temperatures\cite{Okada:1984cz} due to phonon entropic effects, and this behavior
is reproduced by DFT\cite{Biernacki:1989vs}. To benchmark this unusual feature and other
thermal properties of GAP, we calculated the thermal expansion and Gr\"uneisen
parameters using QHA and the heat capacity from the phonon
frequencies\cite{Mounet:2005we}. Figure~\ref{fig:thexp} shows the temperature
dependence of the linear thermal expansion, the Gr\"uneisen parameter and the
heat capacity of various models including GAP and DFT. Even though the mode
Gr\"uneisen parameters of DFT are not accurately reproduced by GAP, these
averaged thermodynamic properties show a reasonable agreement, including the
temperature region with negative thermal expansion.
Analytic potentials and DFTB
show a good agreement in the temperature dependence of the heat capacity, not
surprisingly, as phonon frequencies are generally similar.
However, the errors that the other tested models made in the mode Gr\"uneisen
parameters shown in the previous figures manifest themselves here in the form of
qualitative errors in the thermal expansion, where none reproduce the negative
values at low temperature.
At small temperatures, mostly low energy states are excited, and in particular, those in
the transverse acoustical (TA) branch. The hardening of phonons, or in other words, the strongly negative mode
Gr\"uneisen paramters, in the TA branch have been associated with the negative
thermal expansion of silicon~\cite{Biernacki:1989vs}. The shift to large positive
values of the TA mode Gr\"uneisen parameters of DFTB in the $\Lambda$ direction
is probably the reason for the spurious maximum in the Gr\"uneisen parameter.

\subsection{Generalized stacking faults}

\begin{figure}
    \centerline{\includegraphics[width=\columnwidth]{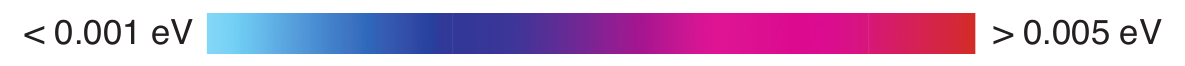}}
           \centerline{\includegraphics[width=\columnwidth]{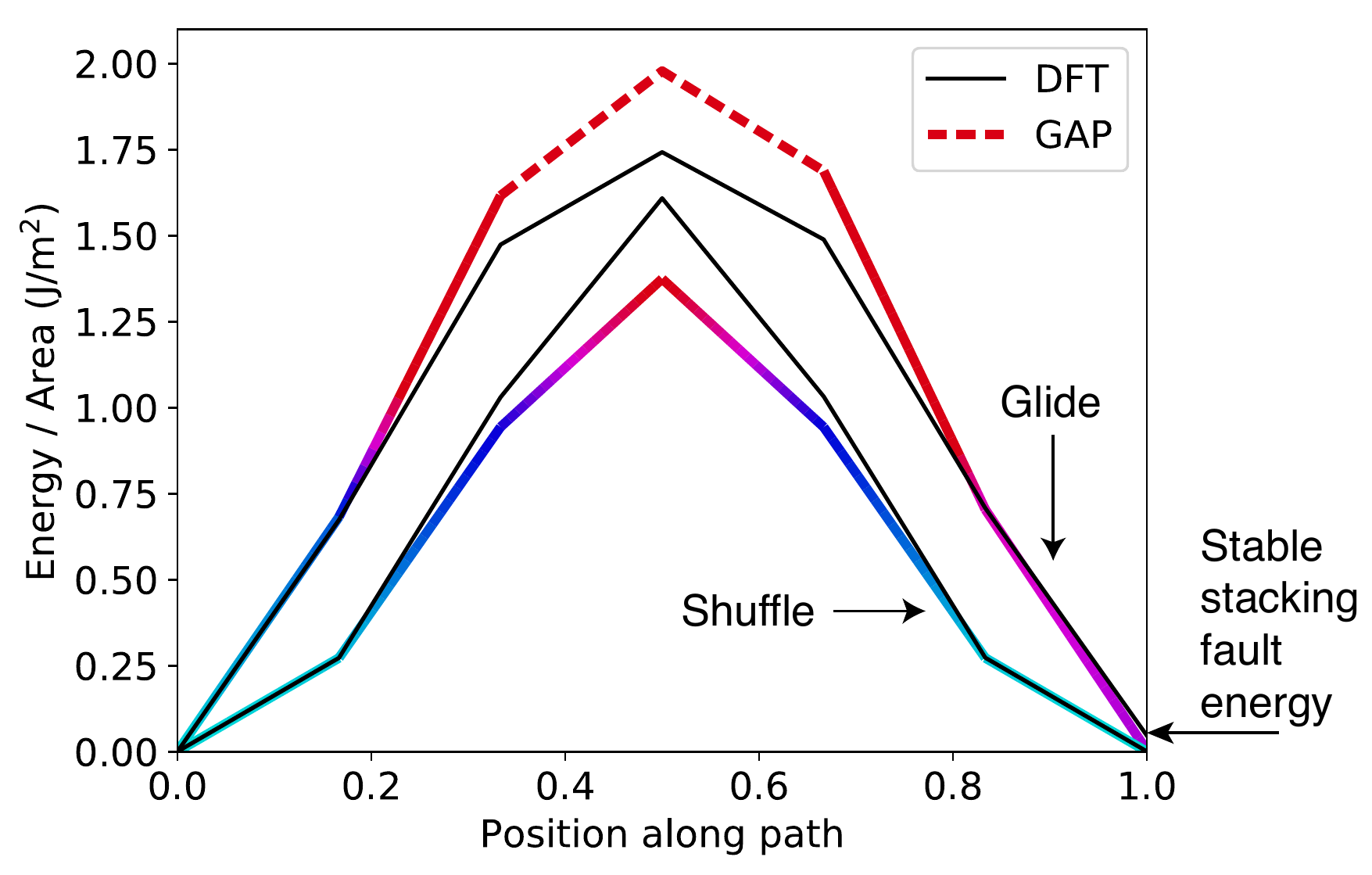}}
    \caption{
    Relaxed generalized stacking fault energies along minimum barrier
    energy path directions ($[112]$ for glide and $[110]$ for shuffle) computed with
    DFT (black solid lines) and GAP.
    The thick curve showing the GAP model energies is coloured according to
    the maximum per-atom predicted error of the GAP model,  and dashed where the predicted error exceeds the
    scale maximum of 5 meV/atom.
    The upper two curves correspond to glide plane and the lower two to the shuffle plane.
    }
    \label{fig:gsf_GAP}
\end{figure}

One type of planar defect that is important as a representative
of dislocation properties which control macroscopic behavior such
as plasticity and fracture, is the generalized stacking fault (GSF)
surface~\cite{vitek_phil_mag_1968}.
This generalization of the conventional stacking
fault, which has been studied extensively in silicon~\cite{Kaxiras:1993bla,Juan:1996co,deKoning:1998bx,Godet:2003cu},
is the energy as a function of arbitrary in-plane shift between
two blocks of otherwise undisturbed crystal.  We focus on the diamond
structure $(111)$ planes, where the GSF can be introduced in glide
planes (within a bilayer) or shuffle planes (between bilayers).
We calculated the energy along high symmetry paths on the GSF surface
connecting equivalent representations of the ideal crystal in a $1
\times 1$ $(111)$ surface cell with 9 bilayers (18 atoms) in the normal
direction, relaxing the atoms parallel to the surface normal.  In
each case we chose the minimum barrier energy path, $[112]$ for
the glide fault and $[110]$ for the shuffle fault.
The resulting energies along the paths for GAP and the reference DFT results
are shown in Fig.~\ref{fig:gsf_GAP}.  For both relaxed
shuffle and glide GSFs the GAP results are in good agreement with
DFT.  The predicted errors shown in the plot are significant near the peaks of both
paths, consistent with a relatively large disagreement with the
DFT reference values, and with the fact that neither type of configuration
was included in the fitting database.

\begin{figure}
    \centerline{\includegraphics[width=\columnwidth]{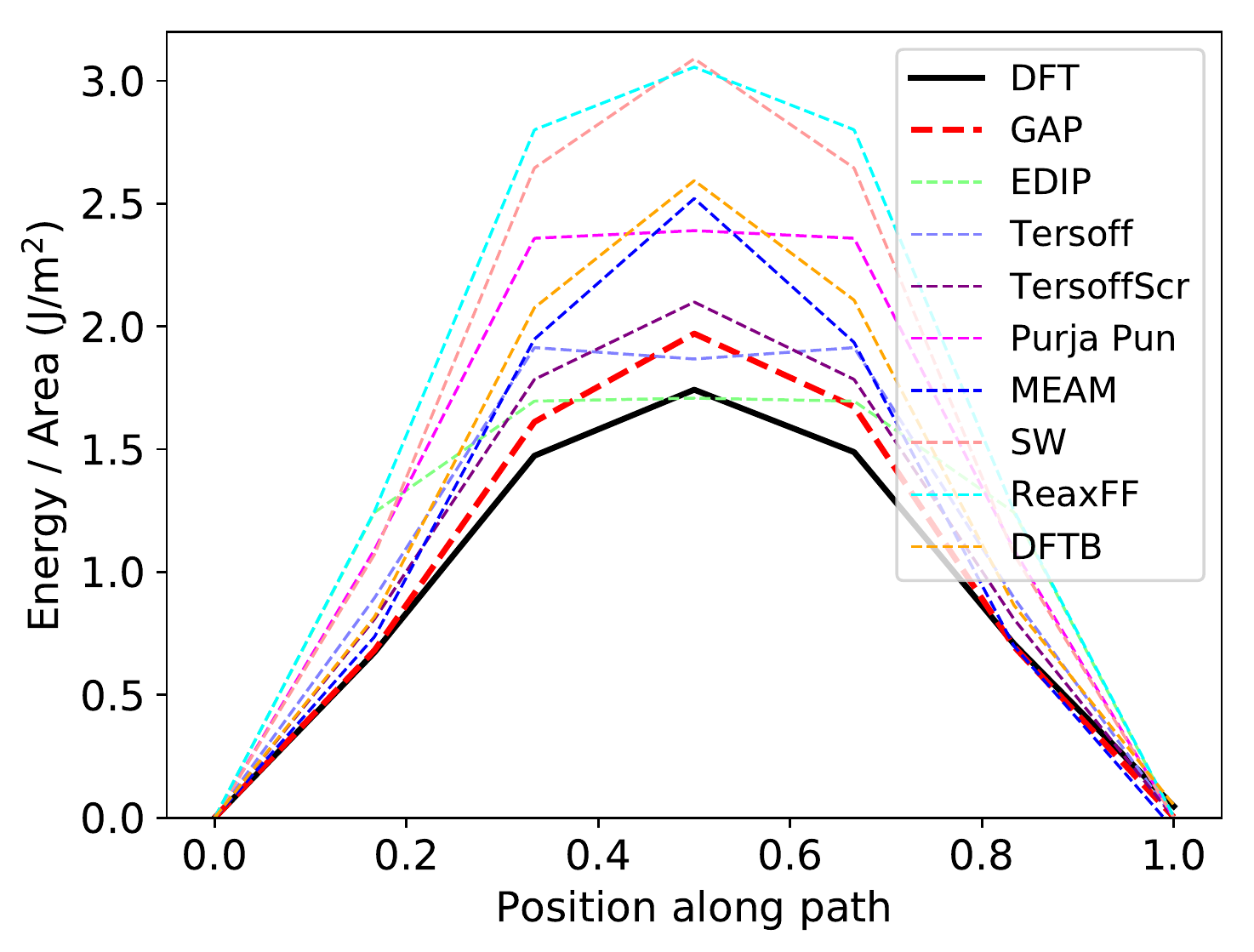}}
    \caption{Relaxed glide-plane generalized stacking fault energies along minimum
    barrier energy path $[112]$ direction computed with
    DFT (solid lines), GAP (red dashed lines with symbols), and other
    interatomic potentials (other color dashed lines).}
    \label{fig:gsf_glide_ALL}
\end{figure}

 A comparison of the energy along one of the paths, glide plane relaxed, for all potentials, is shown in Fig.~\ref{fig:gsf_glide_ALL}, and the corresponding fractional errors in the peak energy along all relaxed paths relative to DFT
for all the interatomic potentials are shown in Fig.~\ref{fig:big_fractional_error_plot}.  The results for GAP
show reasonable agreement with DFT, similar to the best of the other
interatomic potentials, and much better than most.

The final point on the glide plane GSF path is the conventional {\em stable}
stacking fault energy $\gamma_\mathrm{sf}$, which is listed in Table~\ref{table:sf}.
The DFT reference value is small and positive, indicating that the hexagonal
stacking is higher in energy than cubic diamond structure. To get a sense of the scale, note that
the glide curve does not quite reach zero at the right hand side of Fig.~\ref{fig:gsf_GAP}:
that mismatch corresponds to the stable stacking fault energy.  The GAP value
is positive but much too small, indicating that diamond structure is indeed
the lowest energy configuration, but underestimating the energy difference.
There are four atoms with non-diamond-like second neighbour environment,
and the DFT energy difference corresponds to a contribution of about 10~meV from each roughly in correspondence with the $\sim\!2.5$~meV/atom predicted error (purple colour). The elevated predicted error shows that GAP's range and flexibility can distinguish these environments, and the $\gamma_\mathrm{sf}$ value could probably be improved by extending the database.
While most potentials tested are short ranged and give exactly zero energy, ReaxFF has a similar value to GAP, while MEAM gives a qualitatively incorrect negative $\gamma_\mathrm{sf}$.  The DFTB model is the only one that accurately reproduces the DFT value.

\begin{table}[!htbp]
\caption{Stable stacking fault energy $\gamma_\mathrm{sf}$ for each model.
}
\label{table:sf}
\centering
\begin{tabular}{lr}
Model & $\gamma_\mathrm{sf}$ (J/m$^2$) \\ \hline
DFT & 0.047 \\
GAP & 0.002 \\
EDIP & 0.000 \\
Tersoff & 0.000 \\
TersoffScr & 0.001 \\
Purja Pun & 0.000 \\
MEAM & -0.046 \\
SW & 0.000 \\
ReaxFF & 0.004 \\
DFTB & 0.052 \\
\end{tabular}

\end{table}

\subsection{Grain boundary}

Another class of planar defects that was not included in the fitting database
are grain boundaries, which are the interfaces between identical
crystal lattices in different orientations.  As a simple example of
these structures we chose the $(112)~\Sigma 3$ tilt boundary of the
diamond structure, which can be represented by a relatively small unit
cell and can therefore be efficiently computed with DFT.  We computed the
energy per unit area of this grain boundary with the various interatomic
potentials and DFTB, as well as DFT, using a cell with 48 atoms,
which had a single interface unit cell and was about 27~{\AA} long normal to the boundary.
The resulting fractional errors relative to the DFT value are shown in
Fig.~\ref{fig:big_fractional_error_plot}, and the GAP force errors
for the DFT relaxed configuration are shown in Fig.~\ref{fig:pred-err}.
Despite the fact that the
grain boundary structure was not in the fitting database, the GAP energy
is in excellent agreement with DFT.  The difference between the DFT and GAP relaxed
geometries is also small, as indicated by the small magnitudes of the GAP
forces in the DFT relaxed geometry (Fig.~\ref{fig:pred-err}), and the
corresponding displacements (not shown) are nearly imperceptible.
The accuracy of the other interatomic potentials
varies considerably, with some also in very good agreement but others
with very large energy errors relative to the DFT reference.

\subsection{Four-fold defect}
\label{sec:validation:fourfold}

\begin{figure}
    \centerline{
        \begin{minipage}[t]{0.88\columnwidth}
            \vspace{0pt}
            \includegraphics[width=\columnwidth]{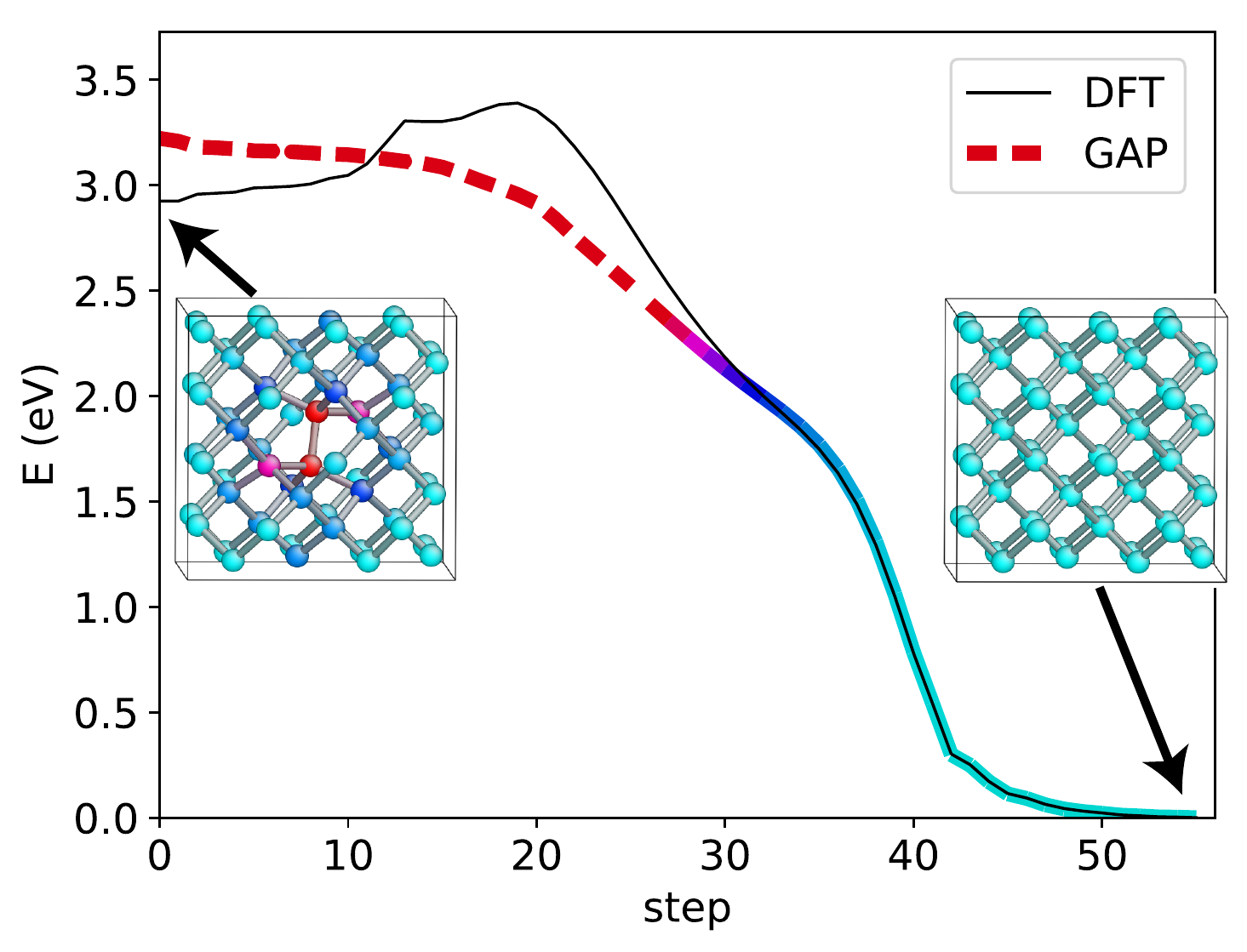}
        \end{minipage}
        \begin{minipage}[t]{0.15\columnwidth}
            \vspace{0pt}
            \includegraphics[width=\columnwidth]{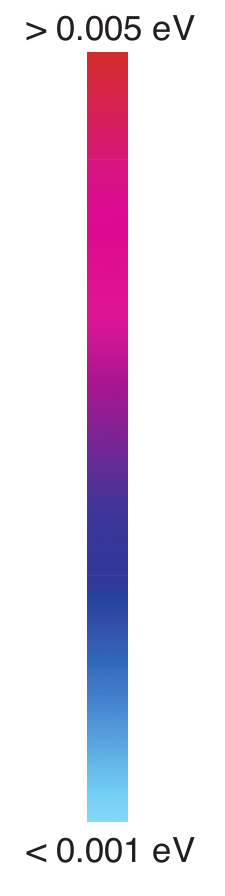}
        \end{minipage}
    }
\caption{Relaxation path of the GAP model showing the instability of
the four-fold defect. The left hand side of the plot corresponds to the
local minimum of the four-fold defect for the DFT model. The black curve
shows the energy of the configurations of this path evaluated with DFT
(this is {\em not} a DFT minimum energy path, but of course still shows a
barrier). The thick curve shows the GAP model energies,  coloured according
to the maximum per-atom predicted error of the GAP model, and  is dashed where the predicted
error exceeds the scale maximum of 5~meV/atom.
}
   \label{fig:fourfold}
\end{figure}

The point defect with the lowest formation energy in the diamond
structure of silicon is the so-called ``four-fold coordinated
defect''\cite{Goedecker:2002ba}, which is formed by a bond rotation followed
by reconnecting all broken bonds. The energy barrier for the reverse
process (i.e. annealing out this defect) is relatively small, and
the GAP model does not in fact stabilise this defect, as shown in
Fig.~\ref{fig:fourfold}. Indeed, the database does not contain anything
resembling the bond rotation process or the final defect structure,
and this is quantitatively shown by the predicted error. The energies
of the GAP model agree very well with those of DFT up to where the
predicted error (taken as the maximum over all atoms) is lower than about
3~meV/atom, and strongly deviate after that. Similarly to the planar
defects, the predicted error gives a good qualitative indication of
where the database is deficient and is in need of extension.

\subsection{Vacancy migration}
\label{sec:validation:vac-path}

\begin{figure}
  \centerline{\includegraphics[width=\columnwidth]{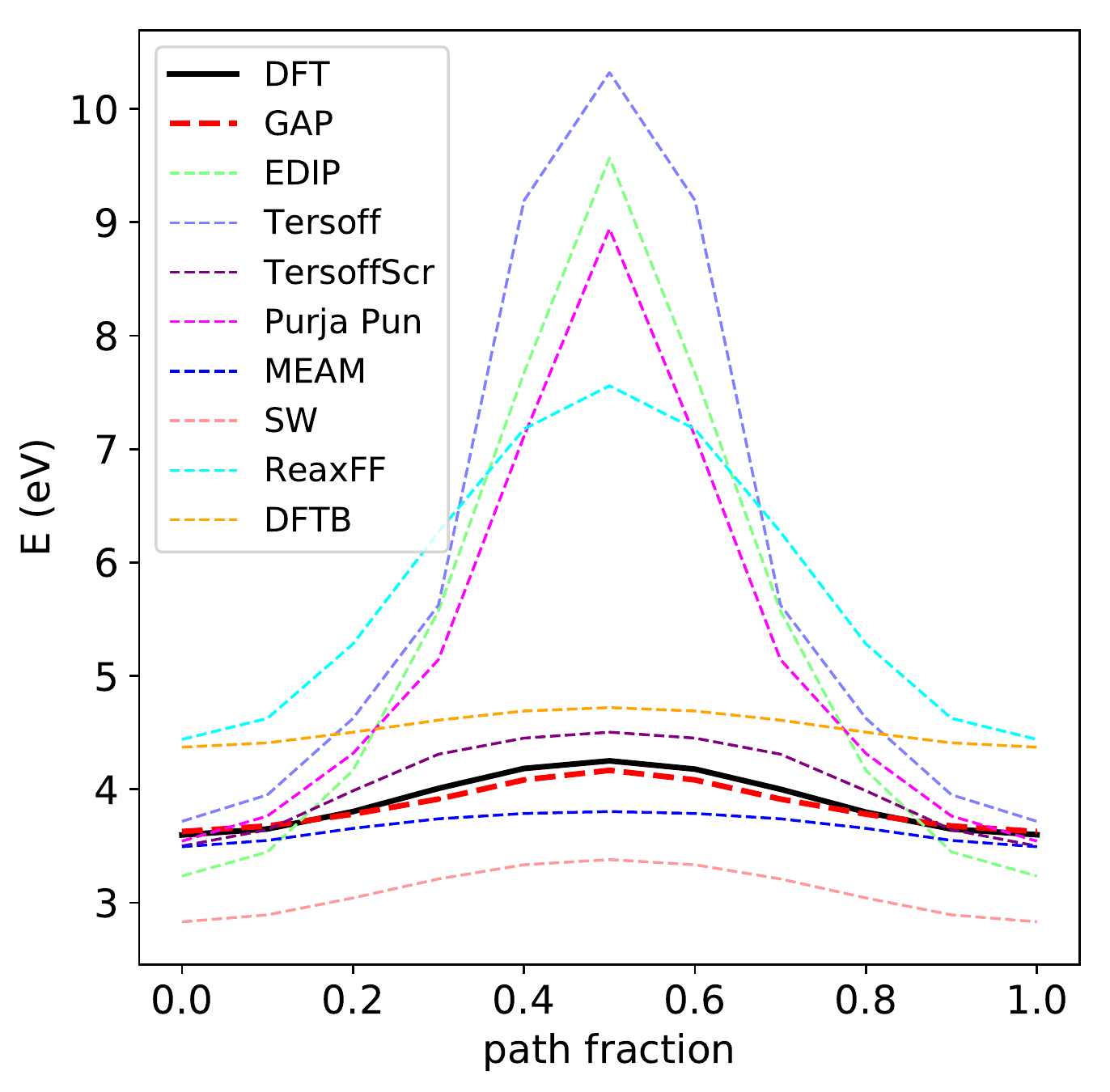}}
  \caption{Unrelaxed energy profiles for the migration of a vacancy, consisting of a series of
    linearly interpolated configurations between relaxed vacancy endpoints, in a 63 atom diamond cell for all models.
    Energies are calculated relative to a bulk crystal with the same number of atoms so that the endpoints indicate the
    relaxed vacancy formation energy. \label{fig:vacancy-path-all}}
\end{figure}

We compared the migration paths for vacancies in 63 atom diamond structure cells predicted
by the various models, as a test of their ability to describe bond breaking
processes. The endpoints were relaxed with preconditioned
LBFGS\cite{Packwood2016} to a maximum force tolerance of $10^{-3}$~eV/\AA{},
and the path was calculated as a linear interpolation between the two
relaxed endpoints.
The intermediate configurations were not relaxed
(as in, for example, the nudged elastic band method\cite{Henkelman2000}), because features
in the PES of many of the potentials led to ill-behaved paths, similar
to the inconsistencies previously noted for the Tersoff potential~\cite{Posselt:2008jr}.
The results shown in Fig.~\ref{fig:vacancy-path-all} indicate the wide variability in the quality of
the predictions from the interatomic potentials in comparison to DFT, with many
of the models significantly over or underestimating both the formation energy and the
migration barrier for vacancies.

\begin{figure}
  \centerline{\includegraphics[width=\columnwidth]{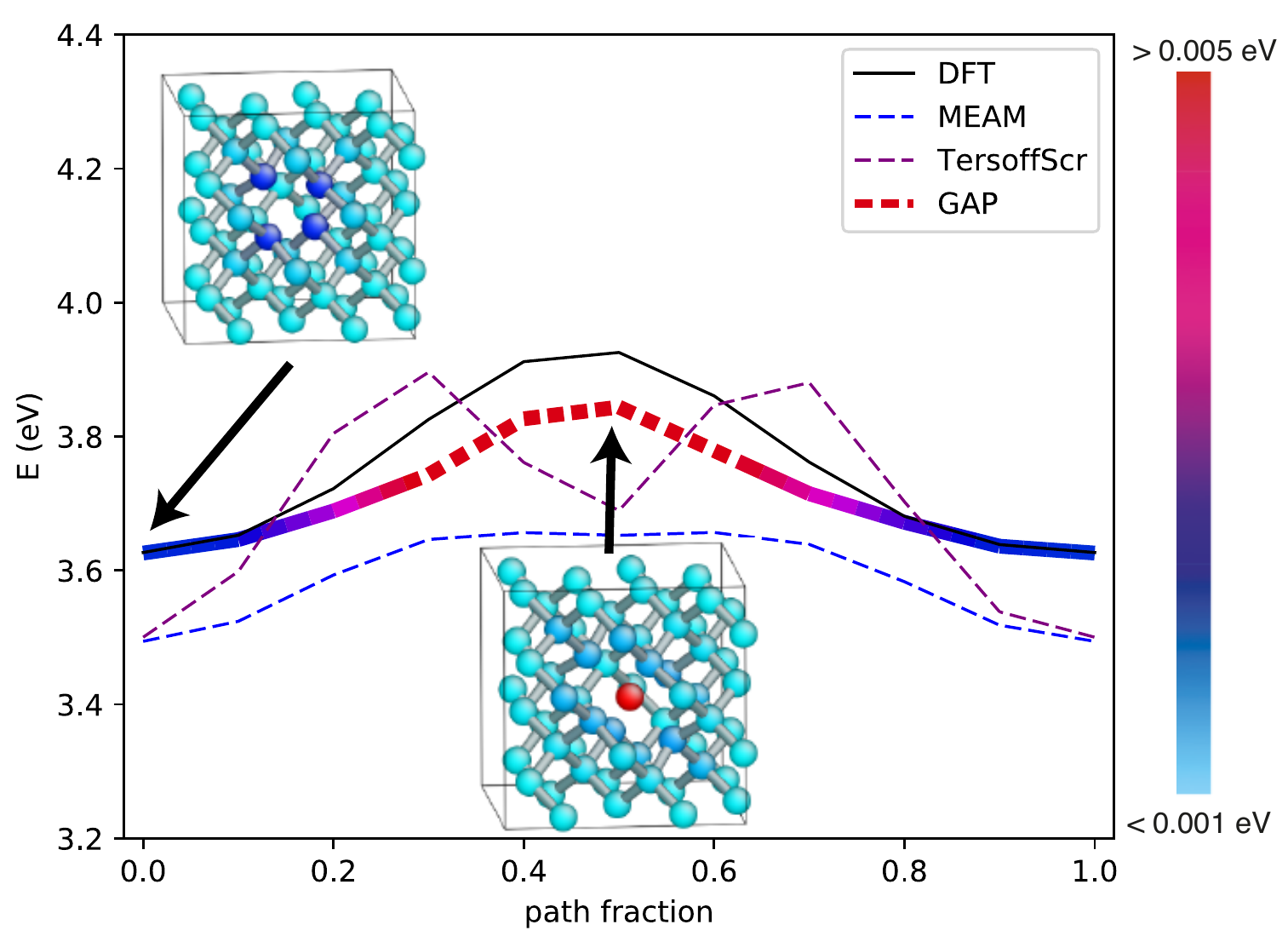}}

  \begin{tabular}{l|r|r|r|r}
    Model & Formation / eV & Error & Barrier / eV & Error \\
    \hline
    DFT & 3.63 & --- & 0.30 & --- \\
    MEAM & 3.49 & -3.7\% & 0.16 & -45.6\% \\
    TersoffScr & 3.50 & -3.5\% & 0.40 & 32.5\% \\
    GAP & 3.63 & -0.0\% & 0.22 & -27.0\% \\
  \end{tabular}

  \caption{Comparison of DFT, GAP, MEAM and TersoffScr models for the migration of a vacancy.
           Lines show NEB minimum energy pathways for a 63 atom cell, with the thick GAP line
           and inset images coloured by the predicted error, becoming dashed where
           the maximum error exceeds 0.005~eV/\AA{}, showing the model is more confident at the minima than the saddle.
           \label{fig:vacancy-path-gap-dft}
}
\end{figure}

For GAP, MEAM and TersoffScr, which produce formation
energies and barriers close to DFT,
the minimum energy path (MEP) was determined using the nudged
elastic band (NEB)\cite{Henkelman2000} algorithm as implemented in ASE
using 9 intermediate
images between minima. The results shown in Fig.~\ref{fig:vacancy-path-gap-dft} show
that GAP produces the most accurate MEP in comparison to DFT, albeit
with an underestimate of the barrier. TersoffScr predicts a local
minimum at the split vacancy configuration. The insets show the
per-atom GAP predicted errors at one of the minima and close to the
saddle point; the model is more confident near the minima since it has
been trained on similar configurations, while saddle-like
configurations were not present in the training data. The predicted
error here again provides a useful guide to the expected reliability of
the model, with good agreement with DFT where it is low and
decreasing agreement where it exceeds 3~meV per atom.

\subsection{Di-interstitials}
\label{sec:diinter}

\begin{figure}
\centerline{\includegraphics[width=\columnwidth]{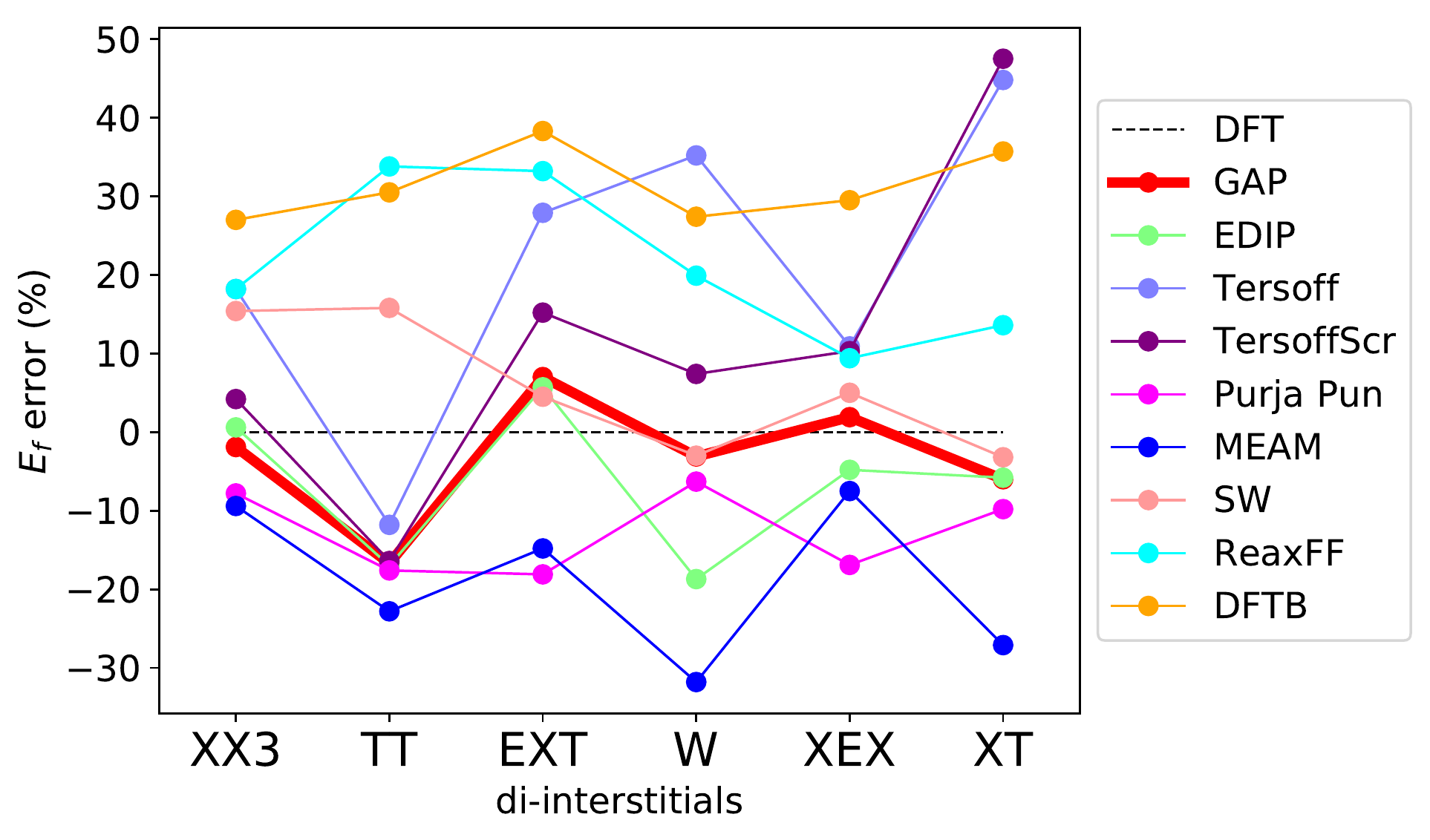}}
\caption{Percentage error in the formation energy of di-interstitials (I$_2$) in various configurations for a variety of interatomic potentials, relative to DFT.}
\label{fig:dii}
\end{figure}

Although configurations including simple point defects, such as the
mono-vacancy and the interstitial were part of our training database,
the di-interstitial provides an interesting test case of transferability
to new defect types. The atomic neighbour environments involved in the
di-interstitial are clearly different from anything that was explicitly
included in the database. Figure~\ref{fig:dii} shows the percentage error
that various interatomic potentials make in the formation energy of the
diinterstital for six different conformations. We used 66 atoms in the
unit cell, including the two extra  that were added to a conventional
64 atom cubic unit cell. The starting positions\cite{Borodin_dii}
were relaxed with each potential, as well as DFT, and the final energies
from each local minima compared.

\begin{figure*}
\begin{tabular}{m{0.3in} m{0.15\textwidth} m{0.15\textwidth} m{0.15\textwidth} m{0.15\textwidth} m{0.15\textwidth}  m{0.15\textwidth}}
    GAP &
    \includegraphics[width=0.15\textwidth]{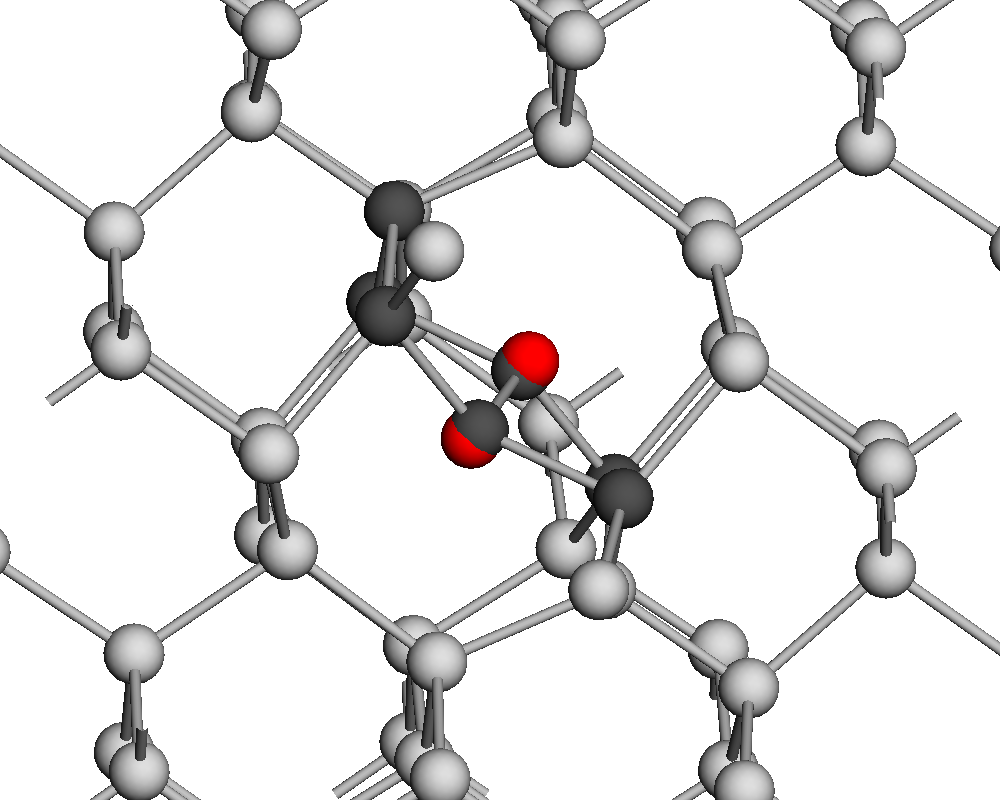} &
    \includegraphics[width=0.15\textwidth]{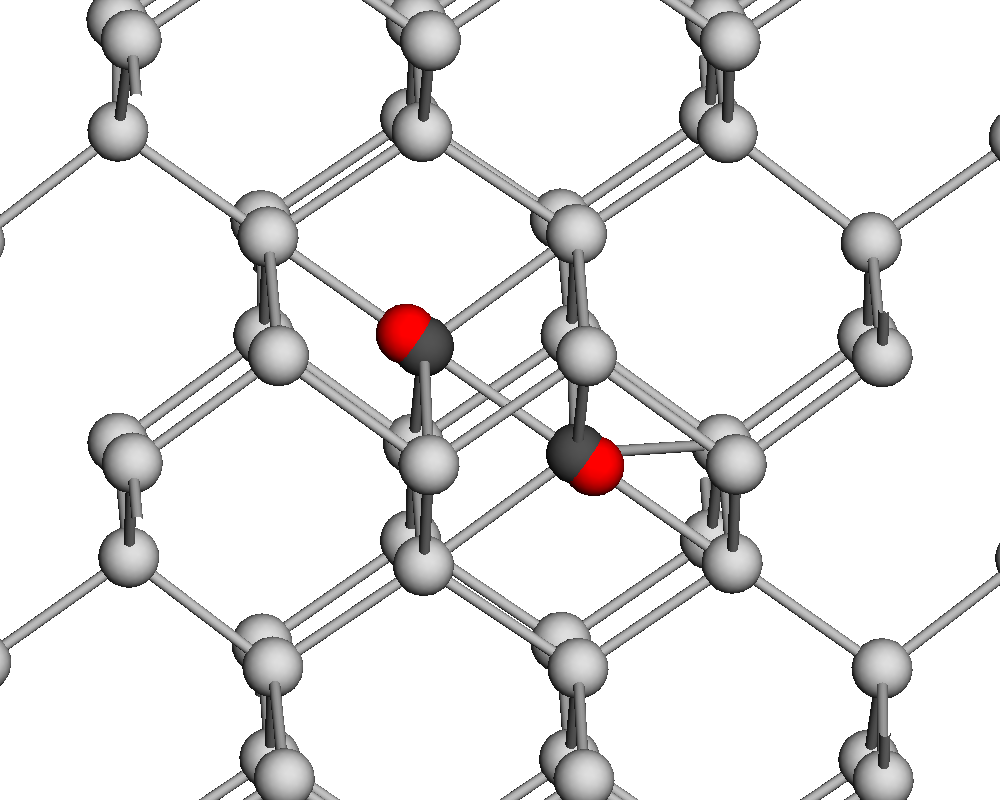} &
    \includegraphics[width=0.15\textwidth]{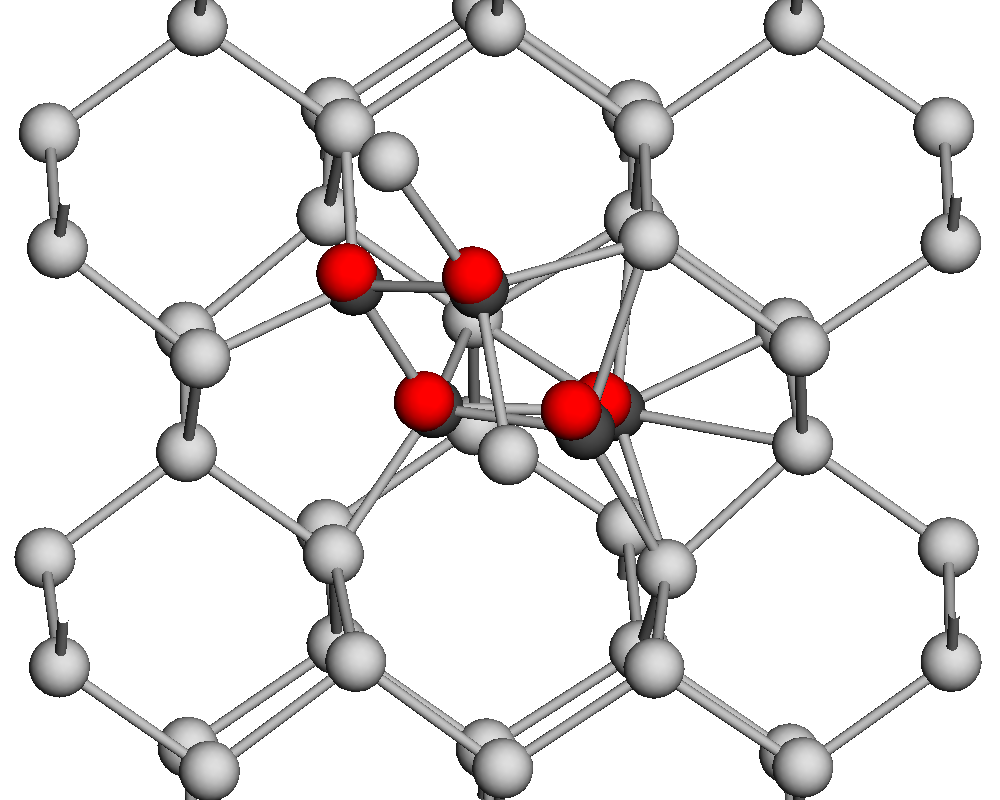} &
    \includegraphics[width=0.15\textwidth]{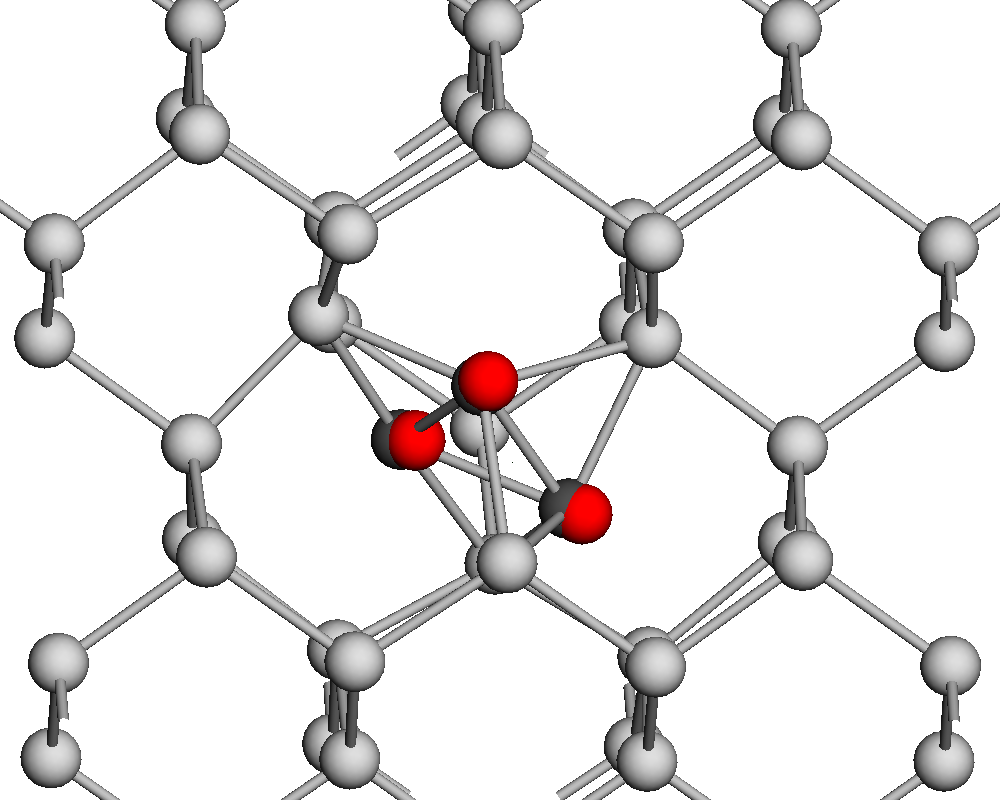} &
    \includegraphics[width=0.15\textwidth]{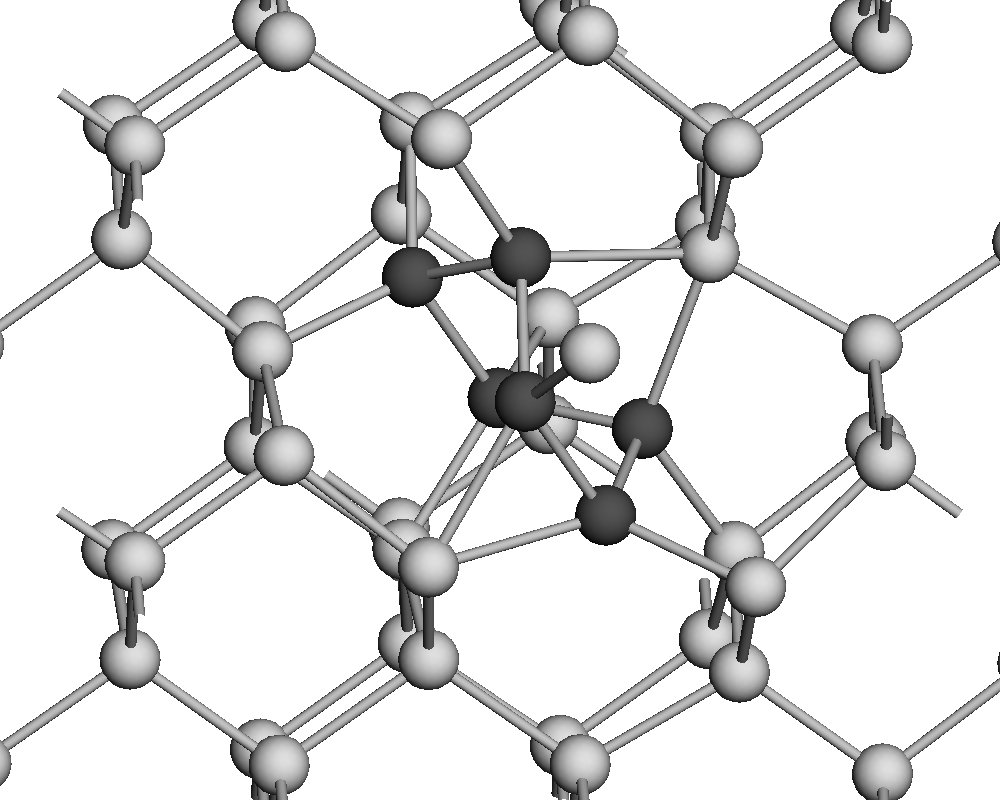} &
    \includegraphics[width=0.15\textwidth]{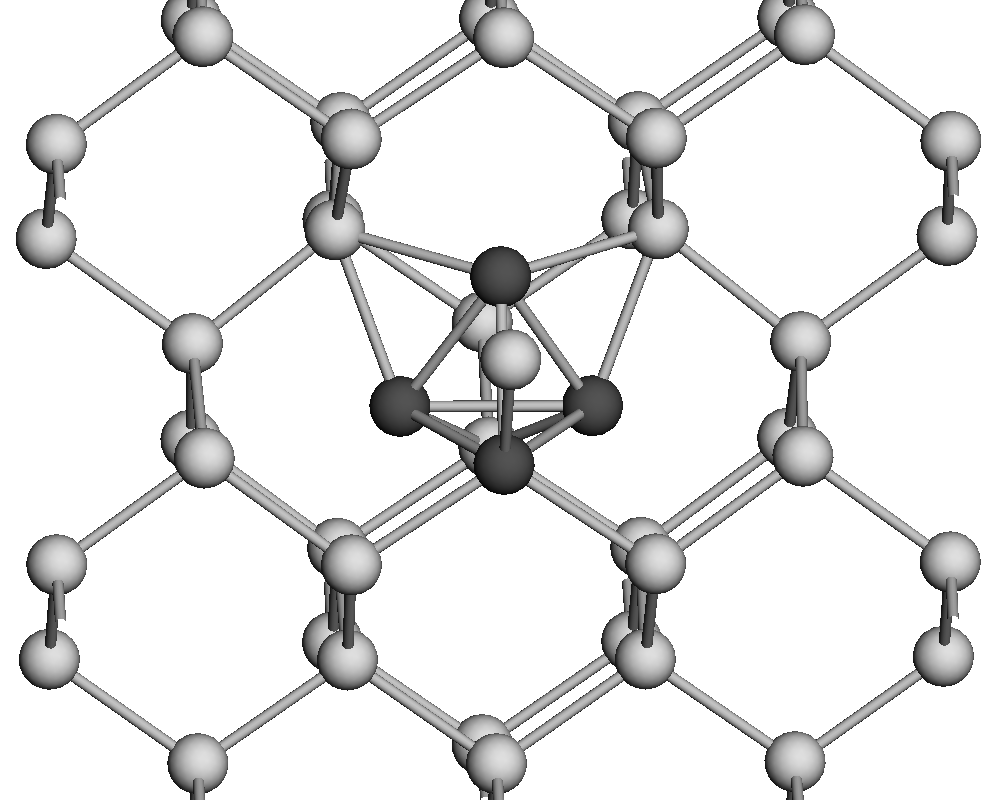} \\ \\
    SW &
    \includegraphics[width=0.15\textwidth]{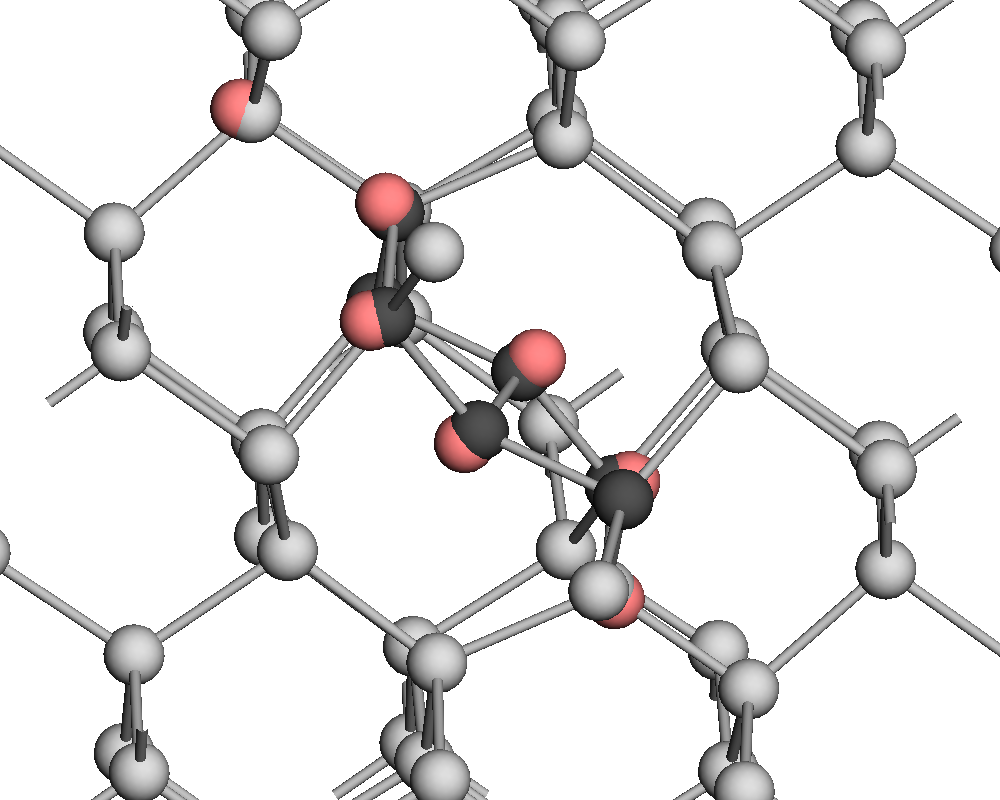} &
    \includegraphics[width=0.15\textwidth]{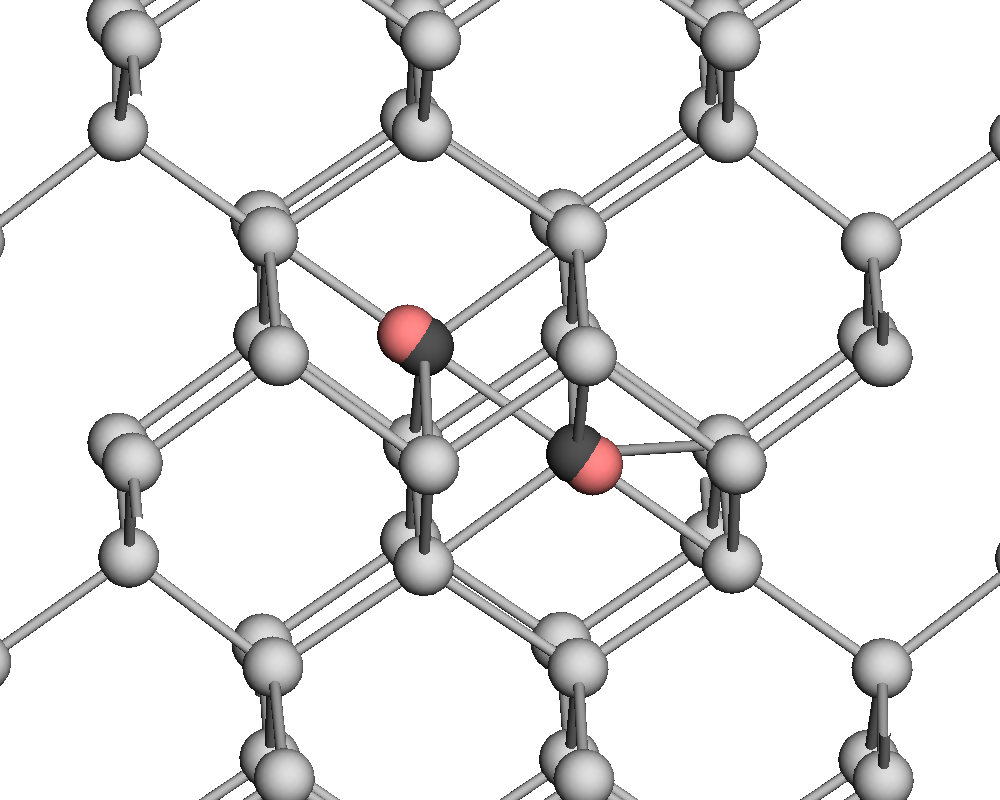} &
    \includegraphics[width=0.15\textwidth]{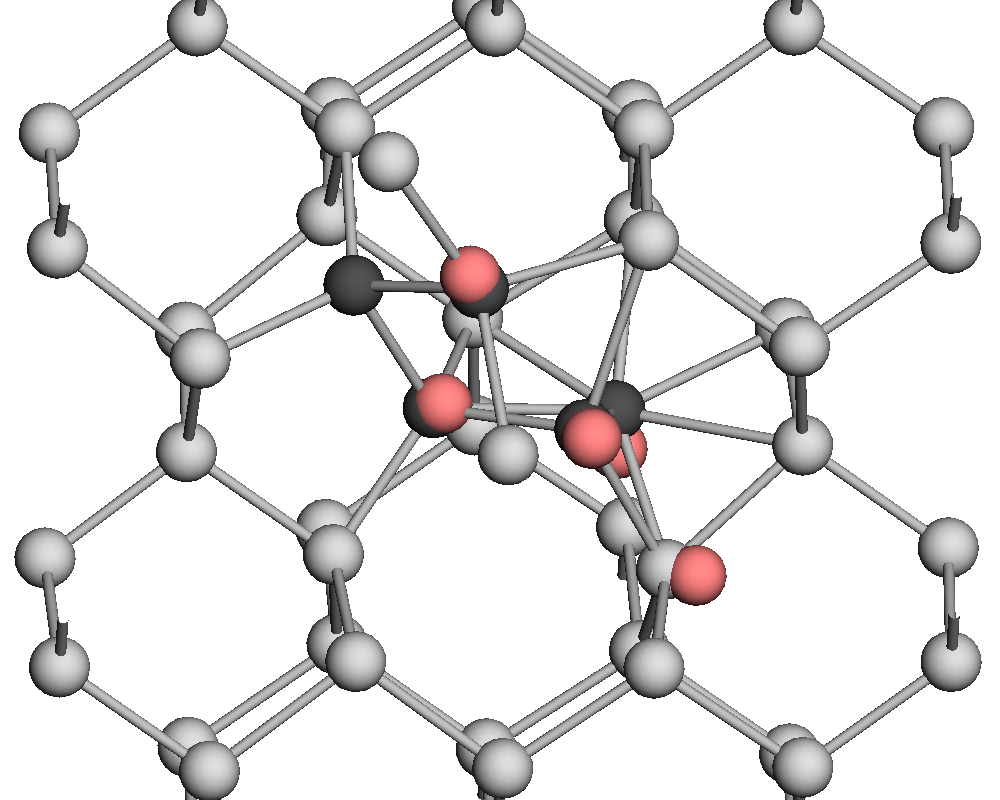} &
    \includegraphics[width=0.15\textwidth]{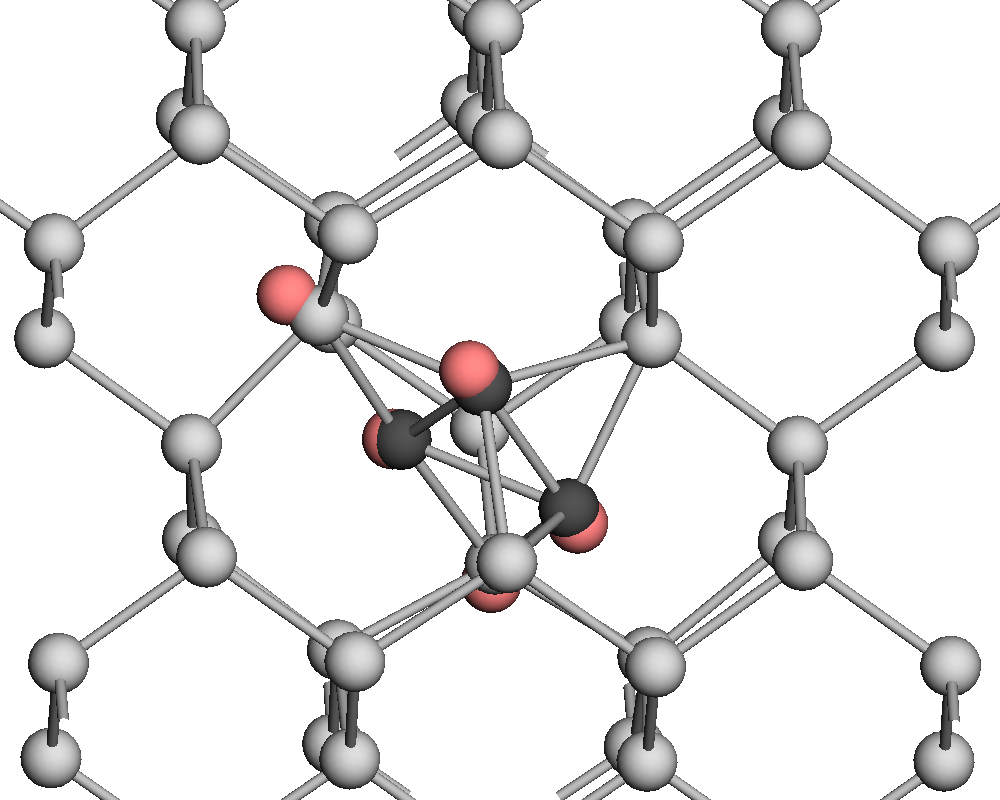} &
    \includegraphics[width=0.15\textwidth]{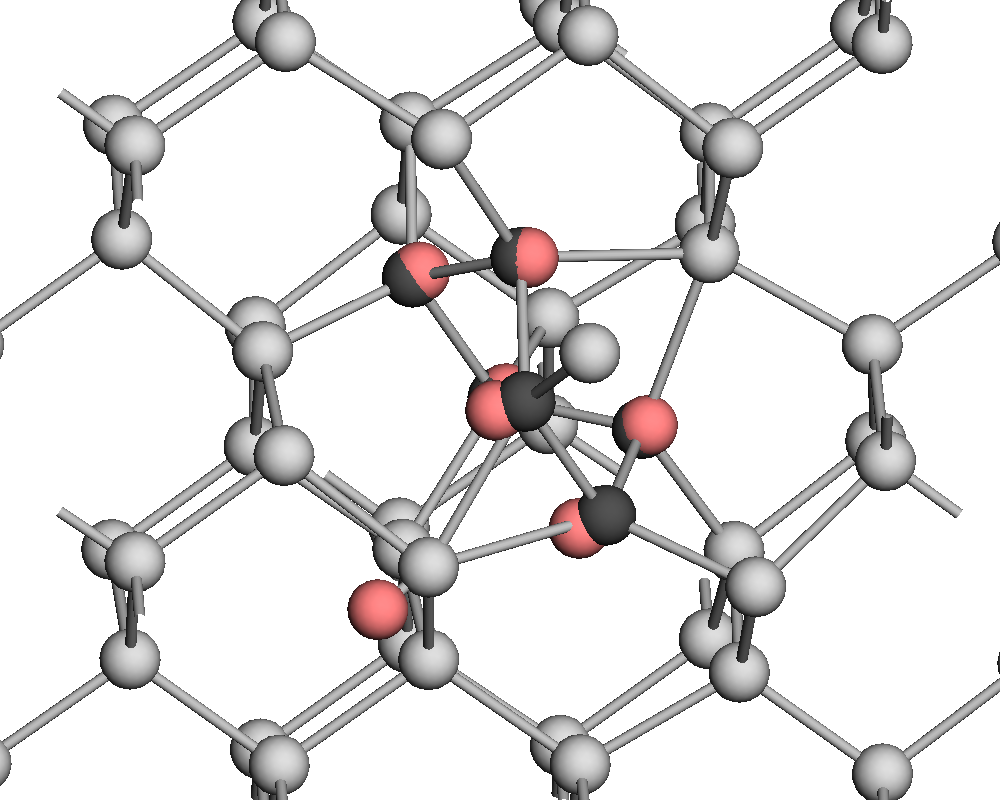} &
    \includegraphics[width=0.15\textwidth]{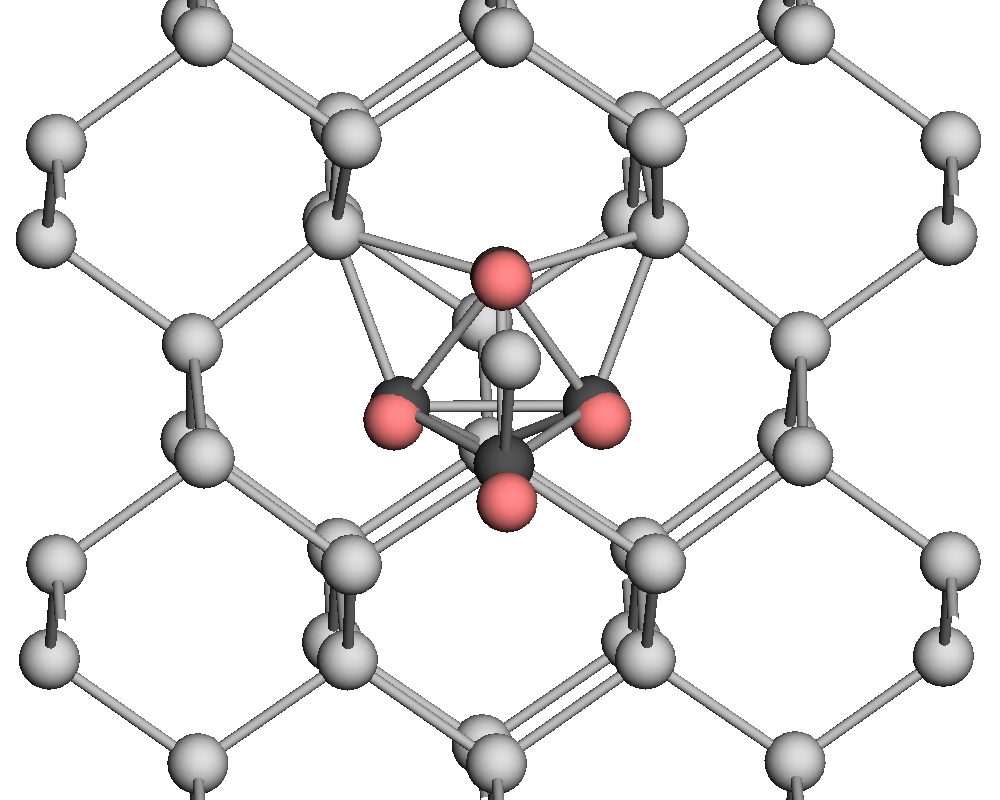} \\ \\
       &
    \multicolumn{1}{c}{XX3} & \multicolumn{1}{c}{ TT } & \multicolumn{1}{c}{ EXT } & \multicolumn{1}{c}{ W } & \multicolumn{1}{c}{ XEX } & \multicolumn{1}{c}{ XT }
\end{tabular}

\caption{Visualization of the atomic configurations of the relaxed di-interstitial structures with GAP (top row) and Stillinger-Weber (bottom row).  Dark grey spheres show reference DFT-relaxed positions of defect atoms (those that are significantly different from perfect lattice position or topology), light grey spheres show reference DFT-relaxed positions of other atoms, and colored spheres show interatomic-potential-relaxed positions of defect atoms that are more than 0.1~{\AA} from the corresponding DFT atom position.
}
\label{fig:dii_viz}
\end{figure*}

The results show that most potentials struggle with this property. EDIP for example, which performs relatively well for the mono-vacancy and the single interstitial makes up to 20\% error here. The Stillinger-Weber model on the other hand, which made errors of over 50\% for the single point defects, looks rather better here. MEAM, ReaxFF and DFTB perform poorly, similarly to the case of single point defects. It is also clear from the plot, that all potentials struggle with the TT configuration (even the best result is over 10\%), including GAP (which otherwise has errors less than about 6\%). Figure~\ref{fig:dii_viz} shows the corresponding relaxed geometries using the GAP and Stillinger-Weber model, as well as DFT, with coloured markings for significant deviations. The Stillinger-Weber model, despite its competitive energy accuracy, shows many more distorted geometries.

\section{Conclusion}
\label{sec:conclusion}

The benefit of the non-parametric approach for creating
interatomic potentials, as presented here, is first and foremost its accuracy in matching the target potential energy surface.
This enhanced accuracy is not just a quantitative improvement,
but actually leads to a {\em qualitatively} better potential:
the GAP model for silicon presented here provides a uniformly high
accuracy across a wide range of properties and systems including bulk
structures, point, and plane defects tested, while maintaining useful
properties of conventional interatomic potentials: locality and linear
scaling computational cost.  This achievement requires an accurate
description of the energetics of a wide range of configurations, including
both fully bonded systems as well as bond breaking.  The ability of
the Si GAP to accurately describe both the energy and forces during the
bond breaking process, including surface decohesion, unstable stacking
fault minimum energy paths, and point defect migration barriers, is an
especially important point which has been challenging for interatomic
potentials to achieve due to their limited variational freedom and
short range.  Such a comprehensive description of silicon has never before
been achieved, despite many efforts, with analytical potentials.
The probabilistic nature
of the Gaussian process technique allows for uncertainty quantification
(and similar measures are possible to obtain using other non-parametric
fitting techniques), and this is useful in assessing when
configurations are encountered that are too far from the training
set and are likely to have large errors.  Initial efforts here
strongly suggest that accuracy can be further improved for specific properties
by adding relevant configurations to the database without compromising the accuracy
for other properties.

There are of course limitations of the specific potential
presented here. The major ones are that (i) we focused on only ambient
pressure here, and a comprehensive silicon potential would be expected
to reproduce the large variety of high pressure phases at finite temperatures; (ii) we limited ourselves to considering an elemental material. While it is true that
increasing the number of different elements increases the dimensionality
and therefore the complexity of the configuration space, several recent
works found that including different atomic species do not
qualitatively change the difficulty of the problem.~\cite{De:2016iaa,Artrith:2017fc} Another consequence of
using the elemental silicon condensed phase example is that no long range interactions beyond
5~\AA\ (such as Coulomb or van der Waals) are needed. Properly including
long range interaction and integrating it with a high dimensional fit of the short range
interactions is still an outstanding problem. Using a multi-scale
description to capture these is an alternative approach\cite{Hirn2017}. Another deficiency
of the present potential is that the training database was assembled ``by
hand'', using an {\em ad hoc} iterative process. It would be desirable
to establish protocols that allow the essentially automated construction
of databases suitable for predicting and studying specified macroscopic
phenomena.  There is every hope that the built-in uncertainty
quantification can be used in the future to build much better databases
and design algorithms that automatically select novel configurations
encountered during a simulation for inclusion or even to generate new atomic configurations
that are optimised to improve the database.

Finally, we are not making any claims about the optimality
of the SOAP kernel and the corresponding basis functions. In
particular, our implementation has a computational cost of around
100~ms/atom, and it is certainly possible that there are basis functions
that are cheaper to calculate and better suited to the problem, so that
fewer of them might be enough to achieve the same accuracy.\cite{Thompson:2015dw,Shapeev:2016kn}

We believe that this potential, perhaps extended by the
addition of particular geometries of interest or by a re-evaluation of
the reference database with more accurate methods, will enable a new
and more quantitative approach to simulations of structural properties
of silicon.

We are well aware that the merits of the silicon
potential presented here will not satisfy all possible audiences: while
it is undeniable that the potential is far more accurate and transferable
than any before it, its remaining shortcomings are not completely trivial, and
only further work will conclusively show that they are easily overcome.  Nevertheless,
we hope that the present success in building a generally applicable
potential will allow this to serve as a template for building such models
for other materials, enabling scientifically and technologically
relevant simulations that have thus far been limited by the tradeoffs
between accuracy and computational cost.

The potential is available for anyone to use, and is provided in the
form of an XML file for the QUIP code~\cite{QUIP} as supplementary
material.  In addition to usage directly with QUIP, it can be used with
the LAMMPS\cite{lammps} software with the ``\verb|pair_style quip|'' command,
as well as from ASE\cite{Larsen2017} through QUIP's \texttt{quippy}
Python module. The datafile includes a copy of the training database
structures and associated DFT data.  The potential used throughout this paper has the unique label
{\tt GAP\_2017\_6\_17\_60\_4\_3\_56\_165}.

\begin{acknowledgments}
NB's work was supported by the Office of Naval Research through the U.~S.
Naval Research Laboratory's core basic research program.
JRK ackowledges funding from the EPSRC under grant number EP/P002188/1 and the Royal Society under grant number RG160691.
ABP received a Research Fellowship from Magdalene College, Cambridge between 2010 and 2013,
and later he was supported by a Leverhulme Early Career Fellowship and the Isaac Newton Trust until 2016.
ABP also acknowledges support from the Collaborative Computational Project for NMR Crystallography (CCP-NC) and UKCP Consortium,
both funded by the Engineering and Physical Sciences Research Council (EPSRC) (EP/M022501/1 and EP/P022561/1, respectively)
An award of computer time was provided by the Innovative and Novel Computational Impact on Theory and Experiment (INCITE) program. This research used resources of the Argonne Leadership Computing Facility, which is a DOE Office of Science User Facility supported under Contract DE-AC02-06CH11357.
We are grateful for computational support from the UK national high performance computing service, ARCHER, for which access was obtained via the UKCP consortium and funded by EPSRC grant ref EP/P022561/1 and EP/K014560/1.
Additional computing facilities were provided by the Scientific Computing Research Technology Platform of the University of Warwick.
The authors gratefully acknowledge useful discussion with L. Pastewka, V. Deringer, C. J. Pickard, and M. C. Payne.
\end{acknowledgments}

\bibliography{silicon,Si_GAP_ML_review,Rupp_qmml}

\end{document}